\newlength{\absize}
\documentclass[12pt]{article}
\usepackage{epsf}
\usepackage{hyperref}
\usepackage{latexsym}
\usepackage{amssymb}
\usepackage{setspace}
\renewcommand{\arraystretch}{2}

\newcommand{\figsize}{\small}

\input prepictex
\input pictex
\input postpictex
\newdimen\tdim
\tdim=\unitlength
\def\stpltsmbl{\setplotsymbol ({\small .})}

\def\tarrow{\arrow <5\tdim> [.3,.6]}

\newbox\sru
\setbox\sru=\hbox{\beginpicture
\setcoordinatesystem units <\tdim,\tdim>
\stpltsmbl
\setquadratic
\plot
  0.0   0.0
  4.8   1.5
  7.5   5.0
  7.3   8.5
  5.0  10.0
  2.7   8.5
  2.5   5.0
  5.2   1.5
 10.0   0.0
/
\endpicture}
\def\springru #1 #2 *#3 /{\multiput {\copy\sru}  at
#1 #2 *#3 10 0 /}

\renewcommand{\bar}{\overline}

\newcommand{\spur}[1]{\!\not\! #1 \,}
\newcommand{\un}{\mathcal{U}}

\newcommand{\cA}{\mathcal{A}}
\newcommand{\cV}{\mathcal{V}}
\newcommand{\cL}{\mathcal{L}}
\newcommand{\cO}{\mathcal{O}}
\newcommand{\cB}{\mathcal{B}}

\newcommand{\cM}{\mathcal{M}}
\newcommand{\pd}{\partial}
\newcommand{\vev}[1]{\langle 0|\mbox{T}#1|0\rangle}
\renewcommand{\slash}[1]{#1\!\!\!/}

\newcommand{\be}{\begin{equation}}
\newcommand{\ee}{\end{equation}}
\newcommand{\bea}{\begin{eqnarray}}
\newcommand{\eea}{\end{eqnarray}}
\newcommand{\nn}{\nonumber}
\newcommand{\tall}{\frac{}{}}
\newcommand{\comment}[1]{}

\begin{document}

\thispagestyle{empty}
\pagestyle{empty}
\newcommand{\starttext}{\newpage\normalsize
 \pagestyle{plain}
 \setlength{\baselineskip}{3ex}\par
 \setcounter{footnote}{0}
 \renewcommand{\thefootnote}{\arabic{footnote}}
 }
\newcommand{\preprint}[1]{\begin{flushright}
 \setlength{\baselineskip}{3ex}#1\end{flushright}}
\renewcommand{\title}[1]{\begin{center}\LARGE
 #1\end{center}\par}
\renewcommand{\author}[1]{\vspace{2ex}{\large\begin{center}
 \setlength{\baselineskip}{3ex}#1\par\end{center}}}
\renewcommand{\thanks}[1]{\footnote{#1}}
\renewcommand{\abstract}[1]{\vspace{2ex}\normalsize\begin{center}
 \centerline{\bf Abstract}\par\vspace{2ex}\parbox{\absize}{#1
 \setlength{\baselineskip}{2.5ex}\par}
 \end{center}}

\title{Unparticle self-interactions}
\author{
 Howard~Georgi,\thanks{\noindent \tt georgi@physics.harvard.edu}
 Yevgeny~Kats,\thanks{\noindent \tt kats@physics.harvard.edu}
 \\ \medskip
Center for the Fundamental Laws of Nature\\
Jefferson Physical Laboratory \\
Harvard University \\
Cambridge, MA 02138
 }
\date{\today}
\abstract{We develop techniques for studying the effects of self-interactions in the conformal sector of an unparticle model. Their physics is encoded in the higher $n$-point functions of the conformal theory. We study inclusive processes and argue that the inclusive production of unparticle stuff in standard model processes due to the unparticle self-interactions can be decomposed using the conformal partial wave expansion and its generalizations into a sum over contributions from the production of various kinds of unparticle stuff, corresponding to different primary conformal operators. Such processes typically involve the production of unparticle stuff associated with operators other than those to which the standard model couples directly. Thus just as interactions between particles allow scattering processes to produce new particles in the final state, so unparticle self-interactions cause the production of various kinds of unparticle stuff.  We discuss both inclusive and exclusive methods for computing these processes.  The resulting picture, we believe, is a step towards understanding what unparticle stuff ``looks like'' because it is quite analogous to way we describe the production and scattering of ordinary particles in quantum field theory, with the primary conformal operators playing the role of particles and the coefficients in the conformal partial wave expansion (and its generalization to include more fields) playing the role of amplitudes. We exemplify our methods in the 2D toy model that we discussed previously in which the Banks-Zaks theory is exactly solvable.}

\newpage
\tableofcontents
\starttext

\section{Introduction}
\setcounter{equation}{0}

Since the original formulation of unparticle physics~\cite{Georgi:2007ek,Georgi:2007si} as an effective field theory in which the standard model couples only at high energies to a ``Banks-Zaks''~\cite{Belavin:1974gu,Caswell:1974gg,Banks:1981nn} sector that is scale-invariant at low energies, this idea has been explored both theoretically and phenomenologically in many papers. We start by mentioning what we find the most interesting theoretical developments.

The way unparticle physics arises from weakly coupled Banks-Zaks-like theories was analyzed explicitly in several examples in~\cite{Strassler:2008bv,Grinstein:2008qk}. Strongly coupled Banks-Zaks sectors can include theories in the conformal window of supersymmetric QCD~\cite{Seiberg:1994pq} as discussed in~\cite{Fox:2007sy,Nakayama:2007qu,Strassler:2008bv}, or other supersymmetric or non-supersymmetric gauge theories~\cite{Ryttov:2007sr,Ryttov:2007cx,Sannino:2008ha}. Quite generally (no counterexamples are known), unitary scale-invariant theories have the full conformal invariance~\cite{Polchinski:1987dy,Dorigoni:2009ra}. This imposes lower bounds on operator dimensions and dictates the tensor structure of the unparticle propagators~\cite{Grinstein:2008qk,Mack:1975je,Todorov-Mintchev-Petkova}.  The dimension of a primary\footnote{A primary operator is one that behaves covariantly under conformal transformations. Any operator of a definite scaling dimension that is not a derivative of another operator is primary. In 2D CFTs these operators are often referred to as ``quasi-primary,'' while the term ``primary'' refers to the properties of operators with respect to the full Virasoro algebra.} vector operator $\cO^\mu$ must be as large as $d_\un \geq 3$. This would usually suppress the possible interactions with the standard model both in the absolute magnitude and relative to the accompanying standard model contact terms~\cite{Grinstein:2008qk}. Similarly, for an antisymmetric tensor $\cO^{\mu\nu}$ $d_\un \geq 2$ and for a symmetric traceless tensor $\cO^{\mu\nu}$ $d_\un \geq 4$. However, scalar operators can start from $d_\un \geq 1$. Spin-$\frac{1}{2}$ operators must have $d_\un \geq 3/2$, and they mostly become relevant if it is possible to describe unparticle stuff charged under the standard model gauge interactions (otherwise, the fermionic unparticle operator can only couple to the standard model fields similarly to a singlet neutrino.) Conformal invariance also introduces a specific $d_\un$-dependence into the tensor structure of the 2-point functions in momentum space~\cite{Grinstein:2008qk,Todorov-Mintchev-Petkova}. It is also interesting to mention the work~\cite{Hofman:2008ar} that used the conformal symmetry to study energy-momentum correlations in unparticle stuff.  In particular, they find that the total momentum flux in a particular direction is always proportional to the energy flux, like for a massless particle, and derive the angle-dependent correlation functions of energy and charge.

The possibility that the coupling of an unparticle operator to the Higgs would break the low energy conformal symmetry of the Banks-Zaks sector was discussed in~\cite{Fox:2007sy}. An interaction of the form $(1/M_\un^{d_{UV}-2})|H|^2\cO_{UV}$ that flows in the IR to $(\Lambda_\un^{d_{UV}-d_\un}/M_\un^{d_{UV}-2})|H|^2\cO_\un$ becomes a relevant operator in the conformal sector (for $d_\un < 4$) introducing a scale $\Lambda_{\not\un}^{4-d_\un} \sim \left(\Lambda_\un/M_\un\right)^{d_{UV}-d_\un}M_\un^{2-d_\un}\,v^2$, where $v$ is the Higgs vev. This can break the conformal symmetry at energies $E \lesssim \Lambda_{\not\un}$ (while preserving unparticle behavior for $\Lambda_{\not\un} \ll E \ll \Lambda_\un$). The consequences of the breaking depend on the particular realization of the Banks-Zaks sector, and in many cases rich and surprising phenomenology is expected (see~\cite{Bander:2007nd,Kikuchi:2007qd,Delgado:2007dx} and especially the hidden valley picture of~\cite{Strassler:2008bv}).

The fact that some conformal field theories are dual to gravitational theories in anti-de Sitter space with one extra dimension (the AdS/CFT correspondence~\cite{Maldacena:1997re,Gubser:1998bc,Witten:1998qj} or its more phenomenological variants, such as the Randall-Sundrum models~\cite{RS2,RS1}) can be useful for studying various aspects of unparticle physics from a different perspective. For example, some works on unparticle phenomenology involving internal scalar unparticle propagators obtained divergent results for $d_\un > 2$~\cite{Cheung:2007zza,Liao:2007bx,Zhou:2007zq}. Looking from the AdS perspective, the authors of~\cite{Cacciapaglia:2008ns} (following an earlier work~\cite{PerezVictoria:2001pa}) explained that for $d_\un > 2$ (or $d_\un > 5/2$ for fermions) the propagator includes UV-dependent terms that do not decouple when the UV cutoff is removed. These terms cancel the divergences, and in fact dominate much of the physics in this range of $d_\un$. As pointed out already in~\cite{PerezVictoria:2001pa} (see also~\cite{PerezVictoria:2008pd}), these counterterms are needed in order for the two-point function to be a well-defined distribution. In the context of unparticle physics, these are just the standard model contact terms generated from integrating out the high energy physics that couples the standard model to the Banks-Zaks sector. An explicit example of this was presented in~\cite{Grinstein:2008qk}. The renormalization group running of such terms has been analyzed from the holographic perspective in~\cite{Ho:2009zv}. On the other hand, the UV completion provided by the RS2 model~\cite{RS2} has been analyzed in~\cite{Friedland:2009iy}. The AdS/CFT correspondence was also used in~\cite{Cacciapaglia:2008ns} for analyzing how the unparticle stuff can couple to gauge interactions, similarly to their earlier suggestion of gauging a non-local action~\cite{Cacciapaglia:2007jq} (see also~\cite{Liao:2008dd,Galloway:2008sq,Licht:2008km,Ilderton:2008ab}). Other uses of the AdS picture included describing Banks-Zaks sectors in which the conformal invariance is broken in the IR~\cite{Stephanov:2007ry,Strassler:2008bv,Cacciapaglia:2008ns} and analyzing the possibility suggested in~\cite{Stancato:2008mp} that ``unhiggs'' is responsible for the electroweak symmetry breaking~\cite{Falkowski:2008yr}.

The disadvantage of the AdS/CFT-based approaches is the difficulty to explicitly specify the 4D Lagrangian description of the physics. The original AdS/CFT setup can describe the Banks-Zaks sector in its low-energy limit, the CFT. However, since the number of examples from string theory is limited, typically one needs to pick the field content and the Lagrangian in the AdS space ``by hand'',\footnote{It is widely believed that knowing how (or whether) such a choice can actually be realized in string theory is not essential in order for the correspondence to work.} without knowing what 4D theory is being described. Many CFTs do not have weakly coupled AdS duals at all. Those that do are typically large-$N$ gauge theories with large 't Hooft coupling $\lambda$ (see also~\cite{Kovtun:2008kw}). There is no special reason to believe that the conformal sectors in our world should belong to this class of theories. Some properties may work out anyway, as in AdS/QCD~\cite{Erlich:2005qh,DaRold:2005zs}, but it can often be hard to separate a real effect from an artifact of an uncontrolled approximation. Certain features of unparticle physics that are missed by assuming $N\to\infty$, $\lambda\to\infty$ were discussed in~\cite{Strassler:2008bv}. Furthermore, while knowing correlation functions of CFTs is important for unparticle physics, one would like to include the coupling to the standard model as well and describe the UV completion of the combined theory. In order to add these non-conformal ingredients, the AdS space needs to be modified in the IR. For example, the RS2 model~\cite{RS2} cuts off the AdS space by a brane. As a result, the CFT gets cut off in the UV due to interactions with the boundary values of the AdS fields (gravity and others) which become dynamical fields in the 4D theory~\cite{Gubser:1999vj,ArkaniHamed:2000ds,PerezVictoria:2001pa}. Unparticle physics aspects of an RS2 model with a massive vector field in the bulk were analyzed in~\cite{Friedland:2009iy}. In general, different ways of modifying the AdS space in the IR and regularizing its physics will correspond to different choices of the coupling to the standard model in the UV. But unfortunately the relation between the two sides of this extended correspondence is far from being straightforward. Therefore, while AdS-based models can provide very useful guidance and examples, their ability to describe realistic unparticle physics scenarios is limited.

The understanding of unparticle physics is incomplete without taking into account the self-interactions of the low energy conformal sector. The main goal of the present paper is to contribute to this understanding. So far most works have focused on the $2$-point function of the unparticle operator $\cO$ to which the standard model couples, with a only few excursions~\cite{Strassler:2008bv,Feng:2008ae} into the more complicated higher $n$-point functions which contain the information about the interactions.

Scale invariance requires the $2$-point function of $\cO$ of dimension $d$ to have the form\footnote{For simplicity of presentation, we assumed here that the operator is a Lorentz scalar.}
\be
\vev{\cO(x)\cO(0)} \propto \frac{1}{\left(-x^2 + i\epsilon\right)^d} \propto \int\frac{d^Dp}{(2\pi)^D}\,e^{-ipx}\left(-p^2 - i\epsilon\right)^{d - D/2}
\ee
where $D$ is the spacetime dimension. There are two ways in which this $2$-point function can appear in physical processes. First, it can appear as an internal line in a process that includes two standard model--unparticle interaction vertices. The momentum-space expression for the $2$-point function makes the calculation straightforward. This results in interesting effects due to the non-zero imaginary part at all $p^2$~\cite{Georgi:2007si,Cheung:2007zza}. The other physical effect is the production of unparticle stuff. The corresponding phase space can be determined either from scale invariance~\cite{Georgi:2007ek} or by computing the imaginary part of the 2-point function~\cite{Georgi:2007si}:
\be
\Phi \propto \left(p^2\right)^{d - D/2}\,\theta(p^0)\,\theta(p^2)
\ee
These two types of processes describe the physics at the leading order in the standard model--unparticle coupling.

Higher order processes require additional machinery since they will depend on $3$- and higher $n$-point functions of $\cO$. Furthermore, we will argue that it is useful to not restrict our attention solely to the operator $\cO$ that couples to the standard model sector. Instead we will use more of the power of conformal invariance and consider the primary operators $\cO_j$ of dimension $d_j$ of the conformal field theory.  Conformal invariance requires the $2$-point functions to have the form\footnote{If there is more than one operator with a given dimension and tensor structure, we can choose linear combinations to get the $\delta_{jk}$. Again, for simplicity of presentation, we assumed here that the operators are Lorentz scalars, but most of what we say can be easily generalized to higher tensors. Various general properties of conformal theories are discussed in~\cite{Ginsparg:1988ui,Osborn:1993cr,Fradkin:1978pp,Fradkin:1997df,Todorov-Mintchev-Petkova}.}
\be
\vev{\cO_j(x)\cO_k(0)} \propto \frac{\delta_{jk}}{\left(-x^2 + i\epsilon\right)^d} \propto \delta_{jk}\int\frac{d^Dp}{(2\pi)^D}\,e^{-ipx}\left(-p^2 - i\epsilon\right)^{d - D/2}
\label{conformal2point}\ee
The $\delta_{jk}$ in (\ref{conformal2point}) allows us to identify different primary operators with different kinds of unparticle stuff. For each $\cO_j$, there is a unique phase space given by
\be
\Phi_j \propto \left(p^2\right)^{d_j - D/2}\,\theta(p^0)\,\theta(p^2)
\label{phikphasespace}
\ee

The form of the $3$-point function is also fixed by conformal invariance. In particular (for example), for primary scalars $\cO_i$ with dimensions $d_i$:
\be
\vev{\cO_1(x_1)\cO_2(x_2)\cO_3(x_3)} \propto \frac{1}{\left(x_{12}^2\right)^{(d_1 + d_2 - d_3)/2}
\left(x_{13}^2\right)^{(d_1 - d_2 + d_3)/2}
\left(x_{23}^2\right)^{(-d_1 + d_2 + d_3)/2}}
\label{3pt-scalar}
\ee
where $x_{ij} \equiv x_i - x_j$. This was used in~\cite{Feng:2008ae} (in the case where all $\cO_i$ are the same) to study processes that involve $3$ {\bf internal} standard model--unparticle vertices. Higher $n$-point functions are highly constrained by conformal invariance, but not completely determined.

Our goal in this paper is to understand how to use this structure to analyze processes in which unparticle stuff is produced as an {\bf outgoing} state. We will discuss two approaches. We will show how to analyze inclusive processes by looking at the discontinuities across physical cuts in the $n$-functions of $\cO$. And we will argue that these discontinuities are related to a sum over primary conformal operators of squared ``amplitudes'' for the production of the corresponding unparticle stuff. These amplitudes, in turn, are determined by the coefficient functions in the conformal partial wave expansion. The resulting picture, we believe, is a step towards understanding what unparticle stuff ``looks like'' because it is quite analogous to way we describe the production and scattering of ordinary particles in quantum field theory, with the primary conformal operators playing the role of particles and the coefficients in the conformal partial wave expansion (and its generalization to include more fields) playing the role of amplitudes.

{\figsize\begin{figure}[htb]
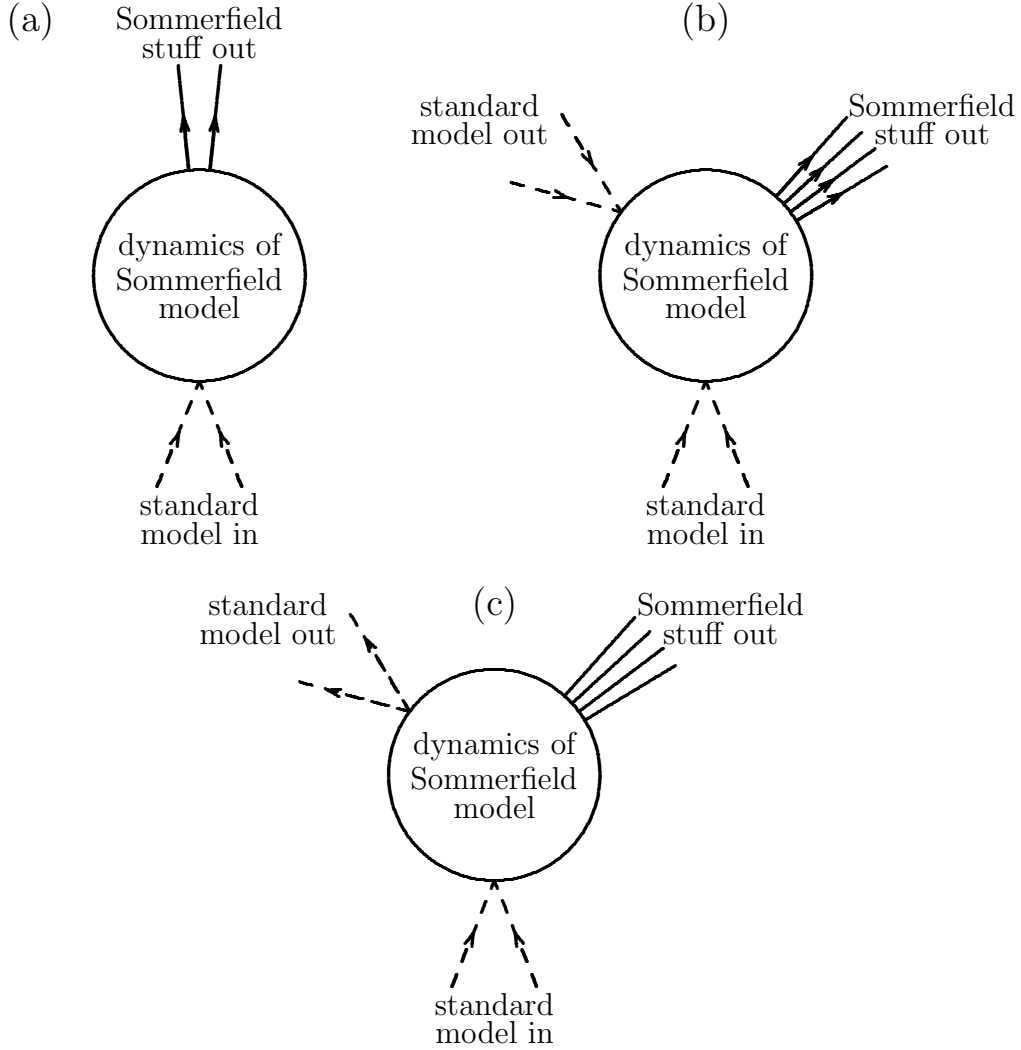

$$\beginpicture
\setcoordinatesystem units <0.8\tdim,0.8\tdim>
\stpltsmbl
\circulararc 360 degrees from 50 0 center at 0 0
\put {\stack{dynamics of,Sommerfield,model}} at 0 0
\startrotation by .995 -.1 about 0 0
\tarrow from 0 50 to 0 75
\plot 0 50 0 100 /
\stoprotation
\startrotation by .995 .1 about 0 0
\tarrow from 0 50 to 0 75
\plot 0 50 0 100 /
\stoprotation
\put {\stack{Sommerfield,stuff out}} at 0 115
\setdashes
\plot -20 -100 0 -50 /
\plot 20 -100 0 -50 /
\tarrow from -20 -100 to -10 -75
\tarrow from 20 -100 to 10 -75
\put {\stack{standard,model in}} at 0 -115
\linethickness=0pt
\putrule from -100 0 to 100 0
\put{\large(a)} at -80 120
\endpicture
\beginpicture
\setcoordinatesystem units <0.8\tdim,0.8\tdim>
\stpltsmbl
\circulararc 360 degrees from 50 0 center at 0 0
\put {\stack{dynamics of,Sommerfield,model}} at 0 0
\startrotation by .995 -.1 about 0 0
\tarrow from 40 30 to 60 45
\plot 40 30 80 60 /
\stoprotation
\tarrow from 40 30 to 60 45
\plot 40 30 80 60 /
\startrotation by .995 .1 about 0 0
\tarrow from 40 30 to 60 45
\plot 40 30 80 60 /
\stoprotation
\startrotation by .98 .198 about 0 0
\tarrow from 40 30 to 60 45
\plot 40 30 80 60 /
\stoprotation
\setdashes
\plot -20 -100 0 -50 /
\plot 20 -100 0 -50 /
\tarrow from -20 -100 to -10 -75
\tarrow from 20 -100 to 10 -75
\put {\stack{standard,model in}} at 0 -115
\startrotation by -.6 -.8 about 0 0
\plot -20 -100 0 -50 /
\plot 20 -100 0 -50 /
\tarrow from -20 -100 to -10 -75
\tarrow from 20 -100 to 10 -75
\put {\stack{standard,model out}} at 5 -130
\stoprotation
\startrotation by -.6 .8 about 0 0
\put {\stack{Sommerfield,stuff out}} at -5 -130
\stoprotation
\linethickness=0pt
\putrule from -100 0 to 100 0
\put{\large(b)} at 0 120
\endpicture$$
$$\beginpicture
\setcoordinatesystem units <0.8\tdim,0.8\tdim>
\stpltsmbl
\circulararc 360 degrees from 50 0 center at 0 0
\put {\stack{dynamics of,Sommerfield,model}} at 0 0
\startrotation by .995 -.1 about 0 0
\plot 40 30 80 60 /
\stoprotation
\plot 40 30 80 60 /
\startrotation by .995 .1 about 0 0
\plot 40 30 80 60 /
\stoprotation
\startrotation by .98 .198 about 0 0
\plot 40 30 80 60 /
\stoprotation
\setdashes
\plot -20 -100 0 -50 /
\plot 20 -100 0 -50 /
\tarrow from -20 -100 to -10 -75
\tarrow from 20 -100 to 10 -75
\put {\stack{standard,model in}} at 0 -115
\startrotation by -.6 -.8 about 0 0
\plot -20 -100 0 -50 /
\plot 20 -100 0 -50 /
\tarrow from 0 -50 to -15 -87
\tarrow from 0 -50 to 15 -87
\put {\stack{standard,model out}} at 5 -130
\stoprotation
\startrotation by -.6 .8 about 0 0
\put {\stack{Sommerfield,stuff out}} at -5 -130
\stoprotation
\linethickness=0pt
\putrule from -100 0 to 100 0
\putrule from 0 0 to 0 100
\put{\large(c)} at 0 80
\endpicture$$
\caption{\figsize\sf\label{fig-1}(a) A disappearance process. (b) A
missing charge process. (c) A missing energy process. The arrows indicate the flow of the global $U(1)$ charge.}\end{figure}}

We will present the basic ideas, which are valid for any conformal theory in any number of dimensions, in section~\ref{sec-self-int}, and dedicate most of the rest of the paper to testing them on the 2D example of unparticle physics that we discussed in~\cite{Georgi:2008pq}. The Banks-Zaks sector in that example is the Sommerfield model of massless fermions coupled to a massive vector field. This model is exactly solvable and flows to the Thirring model in the infrared, as we will discuss in section~\ref{sec-model}. We couple the Sommerfield model to a toy standard model, which is simply a massive scalar carrying a global $U(1)$ charge. In the infrared, the resulting interaction flows to a coupling of two charged scalars to an unparticle operator with a fractional anomalous dimension. In section~\ref{correlation} we apply the operator product expansion to the solution of the Sommerfield model to find the exact $2n$-point functions of the unparticle operator. We discuss the mathematical structures that appear in the solution and certain infrared issues involved in looking at them in momentum space. Before studying self-interactions, we briefly review in section~\ref{sec-disappearance} the simplest unparticle process shown in figure~\ref{fig-1}a in which two toy standard model scalars ``disappear'' into unparticle stuff. Because we have the exact solution for the Banks-Zaks sector correlation functions, we can see precisely how the system makes the transition from the low-energy unparticle physics to the high-energy physics of free particles. The answer, as we discussed in~\cite{Georgi:2008pq}, is rather simple. The ``spectrum'' of the model that we can see in that process consists of unparticle stuff and massive bosons. As the incoming energy of the standard model particles is increased, the unparticle stuff is always there but more and more massive bosons are emitted and the combination becomes more and more like the free-fermion production cross-section. In sections~\ref{sec-mc} and~\ref{sec-me} we analyze the missing charge and missing energy processes shown in figures~\ref{fig-1}b and~\ref{fig-1}c, respectively. These {\bf inclusive} processes are mediated by the conformal sector self-interactions. We show both inclusive and exclusive methods of calculation. The spectrum again includes unparticle stuff and massive bosons, but the unparticle stuff corresponds to operators other than those to which the standard model couples directly (there are also massless excitations that correspond to operators of integer dimensions). We comment on the massive bosons in section~\ref{sec-comments-massive}. In section~\ref{conclusions} we summarize the conclusions.

\newpage
\section{Unparticle self-interactions\label{sec-self-int}}
\setcounter{equation}{0}

A useful concept in conformal field theories is the conformal partial-wave expansion~\cite{Ferrara:1972xe,Ferrara:1973vz,Dobrev:1975ru,Mack:1976pa,Fradkin:1978pp,Fradkin:1997df} (see also~\cite{D'Hoker:1999jp,Arutyunov:2000ku,Dolan:2000ut} and references therein in the context of AdS/CFT), which is a generalization of the operator product expansion (OPE). For any two operators $\cO_1(x_1)$ and $\cO_2(x_2)$ with an arbitrary separation between them, it is possible to write
\be
\mbox{T}\cO_1(x_1)\cO_2(x_2)|0\rangle
= \sum_k \int d^Dx\; iQ_k(x|x_1,x_2)\,\cO_k(x)|0\rangle
\label{CPWE}
\ee
where $\mbox{T}$ denotes time-ordering, $\cO_k$ are the various primary operators in the theory (unlike in the OPE, their derivatives need not be included separately) and the coefficients $iQ_k(x|x_1,x_2)$ are the 3-point functions of $\cO_1(x_1)$, $\cO_2(x_2)$ and $\cO_k(x)$, with the $\cO_k$ leg amputated, namely
\be
\int d^Dx\;\vev{\cO_k(x')\cO_k(x)}\, iQ_k(x|x_1,x_2) = \vev{\cO_k(x')\cO_1(x_1)\cO_2(x_2)}
\label{amputated}
\ee
If $\cO_1$ and $\cO_2$ are scalars, the operators $\cO_k$ are completely symmetric traceless tensors~\cite{Mack:1976pa} whose amputated 3-point functions $Q_k(x|x_1,x_2)$ are fully determined, up to a constant prefactor, by the dimensions of the three operators and the tensor rank of $\cO_k$.

Amputated 3-point functions are exactly what is needed for computing processes that produce unparticle stuff corresponding to the operator whose leg is amputated. This follows because using (\ref{CPWE}) twice we can write
\bea
&&\vev{\cO^*_2(x_2)\cO^*_1(x_1)\cO_1(y_1)\cO_2(y_2)} \nn\\
&&= \sum_k \int d^Dx\,d^Dy\; Q_k^\ast(x|x_1,x_2)\, \vev{\cO_k^\ast(x)\cO_k(y)}\, Q_k(y|y_1,y_2)
\label{4pt-CPWE}
\eea
For example, suppose we had the coupling
\be
\cL_{\rm int} \propto \phi^2\cO
\label{Lint}
\ee
where $\phi$ is a standard model field and $\cO$ is an unparticle operator. According to (\ref{4pt-CPWE}) with $\cO_1 = \cO_2 = \cO$, by taking the discontinuity across the cut of the 4-point function of $\cO$ (figure~\ref{fig-4pt-cut}), which corresponds to the inclusive process
\be
\phi + \phi \to \phi + \phi + \mbox{unparticle stuff}
\label{prodinc}
\ee
we obtain the sum of cross-section for the processes
\be
\phi + \phi \to \phi + \phi + \{\mbox{$\cO_k$ stuff}\}
\label{prodOk}
\ee
where $\cO_k$ are the various primary operators in the theory (that do not necessarily couple to the standard model directly). The amplitudes $\cM$ of these processes are the amputated 3-point functions $Q_k$ (times factors coming from the standard model), while cutting the $\cO_k$ propagators in (\ref{4pt-CPWE}) gives the $\cO_k$ phase spaces in (\ref{phikphasespace}).

{\figsize\begin{figure}[htb]
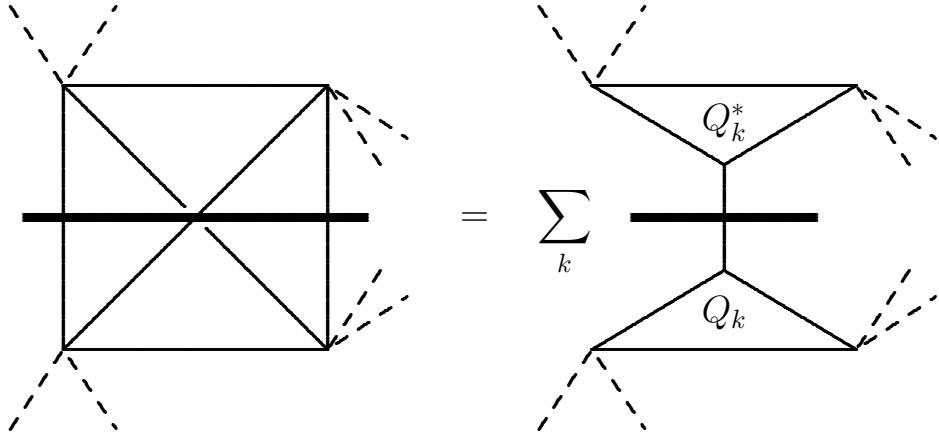

$$\beginpicture
\setcoordinatesystem units <1\tdim,1\tdim>
\stpltsmbl
\plot -50 -50 50 50 -50 50 -50 -50 50 -50 50 50 /
\plot -50 50 -4 4 /
\plot 4 -4 50 -50 /
\plot -65 0 65 0 /
\plot -65 1 65 1 /
\plot -65 -1 65 -1 /
\setdashes
\plot -70 80 -50 50 -30 80 /
\plot -70 -80 -50 -50 -30 -80 /
\plot 70 -20 50 -50 80 -30 /
\plot 70 20 50 50 80 30 /
\linethickness=0pt
\putrule from 0 -90 to 0 90
\putrule from -100 0 to 100 0
\endpicture
\beginpicture
\setcoordinatesystem units <1\tdim,1\tdim>
\stpltsmbl
\plot 50 50 -50 50 0 20 50 50 /
\plot 50 -50 -50 -50 0 -20 50 -50 /
\plot 0 20 0 -20 /
\plot -35 0 35 0 /
\plot -35 1 35 1 /
\plot -35 -1 35 -1 /
\setdashes
\plot -70 80 -50 50 -30 80 /
\plot -70 -80 -50 -50 -30 -80 /
\plot 70 -20 50 -50 80 -30 /
\plot 70 20 50 50 80 30 /
\put {\large$Q_k^\ast$} [c] at 0 36
\put {\large$Q_k$} [c] at 0 -36
\put {\large$\displaystyle =\quad \sum_k$} [c] at -75 -5
\linethickness=0pt
\putrule from 0 -90 to 0 90
\putrule from -100 0 to 100 0
\endpicture$$
\caption{\figsize\sf\label{fig-4pt-cut}On the left: the unparticle 4-point function (schematically) with a cut over the unparticle stuff for computing the cross-section of the process (\ref{prodinc}), inclusive with respect to the unparticle stuff. The dashed lines are the standard model particles $\phi$. Each vertex is the interaction (\ref{Lint}). (One should include also diagrams in which the standard model particles are attached to the 4-point function in other possible ways, as we will do in section~\ref{sec-mc}.) On the right: representation (\ref{4pt-CPWE}) of the 4-point function as a sum of terms, each with two amputated 3-point functions $Q_k$ connected by an $\cO_k$ propagator. The cut through the $\cO_k$ propagator allows us to interpret the terms in the sum as related to the cross-sections of the exclusive processes (\ref{prodOk}).}\end{figure}}

{\figsize\begin{figure}[htb]
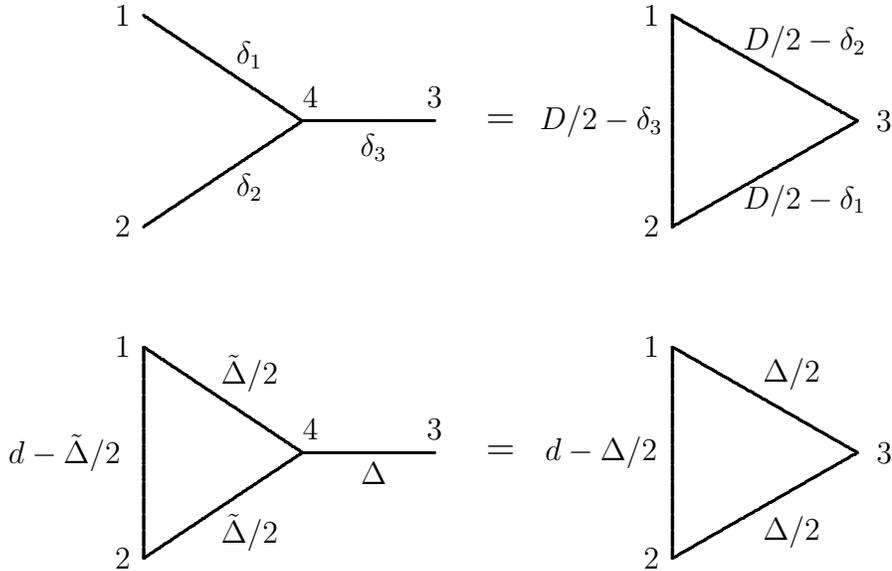

\bea
\beginpicture
\setcoordinatesystem units <1\tdim,1\tdim>
\stpltsmbl
\plot -60 40 0 0 /
\plot -60 -40 0 0 /
\plot 0 0 50 0 /
\put {$1$} [c] at -68 40
\put {$2$} [c] at -68 -40
\put {$4$} [c] at 3 9
\put {$3$} [c] at 50 9
\put {$\delta_1$} [c] at -20 25
\put {$\delta_2$} [c] at -20 -25
\put {$\delta_3$} [c] at 27 -10
\put {\large$=$} [c] at 75 0
\plot 140 40 210 0 /
\plot 140 -40 210 0 /
\plot 140 40 140 -40 /
\put {$1$} [c] at 132 40
\put {$2$} [c] at 132 -40
\put {$3$} [c] at 220 0
\put {$D/2 - \delta_3$} [c] at 113 0
\put {$D/2 - \delta_2$} [c] at 190 30
\put {$D/2 - \delta_1$} [c] at 190 -30
\linethickness=0pt
\putrule from 0 -50 to 0 50
\putrule from -80 0 to 230 0
\endpicture\nn\\\nn\\\nn\\
\beginpicture
\setcoordinatesystem units <1\tdim,1\tdim>
\stpltsmbl
\plot -60 40 0 0 /
\plot -60 -40 0 0 /
\plot 0 0 50 0 /
\plot -60 40 -60 -40 /
\put {$1$} [c] at -68 40
\put {$2$} [c] at -68 -40
\put {$4$} [c] at 3 9
\put {$3$} [c] at 50 9
\put {$\tilde\Delta/2$} [c] at -20 30
\put {$\tilde\Delta/2$} [c] at -20 -30
\put {$\Delta$} [c] at 27 -8
\put {$d - \tilde\Delta/2$} [c] at -90 0
\put {\large$=$} [c] at 75 0
\plot 140 40 210 0 /
\plot 140 -40 210 0 /
\plot 140 40 140 -40 /
\put {$1$} [c] at 132 40
\put {$2$} [c] at 132 -40
\put {$3$} [c] at 220 0
\put {$d - \Delta/2$} [c] at 113 0
\put {$\Delta/2$} [c] at 185 30
\put {$\Delta/2$} [c] at 185 -30
\linethickness=0pt
\putrule from 0 -50 to 0 50
\putrule from -80 0 to 230 0
\endpicture\nn
\eea
\caption{\figsize\sf\label{fig-D'EPP}The D'EPP formula in its original form (\ref{D'EPP}) (top) and in the form (\ref{amputation}) that is useful for amputation (bottom).}\end{figure}}

There are powerful tools in conformal field theory for working out these amplitudes explicitly. The amputated 3-point function is formally proportional to the ordinary 3-point function of $\cO_1$ and $\cO_2$ with a ``shadow operator'' corresponding to the third operator $\cO_k$. The shadow operator has the same tensor rank $\ell$ as $\cO_k$, but dimension $\tilde\Delta = D - \Delta$, where $\Delta$ is the dimension of $\cO_k$ and $D$ is the spacetime dimension. This method is based on the D'EPP formula~\cite{DEPP} (see also~\cite{Ferrara:1972xe,Todorov-Mintchev-Petkova,Fradkin:1997df,Diaz:2006nm}), which says that for $\delta_1 + \delta_2 + \delta_3 = D$ (in Euclidean space)
\be
\int d^Dx_4\, \frac{\Gamma(\delta_1)\,\Gamma(\delta_2)\,\Gamma(\delta_3)} {\left(x_{14}^2\right)^{\delta_1} \left(x_{24}^2\right)^{\delta_2} \left(x_{43}^2\right)^{\delta_3}} = \pi^{D/2}\,\frac{\Gamma\left(\frac{D}{2}-\delta_3\right) \Gamma\left(\frac{D}{2}-\delta_2\right) \Gamma\left(\frac{D}{2}-\delta_1\right)}{\left(x_{12}^2\right)^{D/2 - \delta_3} \left(x_{13}^2\right)^{D/2 - \delta_2} \left(x_{23}^2\right)^{D/2 - \delta_1}}
\label{D'EPP}
\ee
For solving (\ref{amputated}) we can use this formula as
\be
\frac{1}{\left(x_{12}^2\right)^{d - \tilde\Delta/2}}\int d^Dx_4\, \frac{[\Gamma(\tilde\Delta/2)]^2\, \Gamma(\Delta)}{\left(x_{14}^2\right)^{\tilde\Delta/2} \left(x_{24}^2\right)^{\tilde\Delta/2} \left(x_{43}^2\right)^{\Delta}}
= \pi^{D/2}\, \frac{\Gamma(D/2-\Delta)\, [\Gamma(\Delta/2)]^2}
{\left(x_{12}^2\right)^{d - \Delta/2}
\left(x_{13}^2\right)^{\Delta/2} \left(x_{23}^2\right)^{\Delta/2}}
\label{amputation}
\ee
This is described graphically in figure~\ref{fig-D'EPP}. Since the 3-point function (\ref{3pt-scalar}) of two scalars $\cO_1$ and $\cO_2$ of dimension $d$ with a third scalar\footnote{A generalization for higher tensors exists as well. We will consider a vector operator in section~\ref{sec-me-amputated}.} $\cO_k$ of dimension $\Delta$, and the two-point function of $\cO_k$, are
\be
\vev{\cO_1(x_1)\,\cO_2(x_2)\,\cO_k(x_3)} =
\frac{C_3}{\left(-x_{12}^2+i\epsilon\right)^{d - \Delta/2}
\left(-x_{13}^2+i\epsilon\right)^{\Delta/2}
\left(-x_{23}^2+i\epsilon\right)^{\Delta/2}}
\label{3pt-scalar-2same}
\ee
\be
\vev{\cO_k(x)\cO_k(0)} = \frac{C_2}{\left(-x^2+i\epsilon\right)^\Delta}
\ee
where $C_3$ and $C_2$ are constants, we obtain the amputated 3-point function to be
\bea
&&Q_k(x_4|x_1,x_2) = \frac{[\Gamma(\tilde\Delta/2)]^2\,\Gamma(\Delta)\,C_3} {\pi^{D/2}\,[\Gamma(\Delta/2)]^2\,\Gamma(D/2-\Delta)\,C_2}\nn\\
&&\qquad\qquad\qquad\quad\times \frac{1}{\left(-x_{12}^2 + i\epsilon\right)^{d - \tilde\Delta/2} \left(-x_{14}^2 + i\epsilon\right)^{\tilde\Delta/2} \left(-x_{24}^2 + i\epsilon\right)^{\tilde\Delta/2}}
\eea
In momentum space it gives the amplitude
\be
\cM = \frac{2^{D-2d+\Delta}\,\Gamma(\Delta)\,\Gamma(D-d-\Delta/2)\,C_3} {\Gamma(D/2-\Delta)\, \Gamma(d+\Delta/2-D/2)\,C_2}\, I(P,Q)
\ee
(times the standard model factors), where
\be
I(P,Q) \equiv
\int \frac{d^Dk}{\left(-k^2-i\epsilon\right)^{D - d - \Delta/2} \left(-(P-k)^2-i\epsilon\right)^{\Delta/2}\left(-(k-(P-Q))^2-i\epsilon\right)^{\Delta/2}}
\label{IPQ}
\ee
where $P$ is the momentum incoming from the standard model at $x_1$ and $Q$ is the unparticle momentum outgoing at $x_4$.
The phase space of $\cO_k$ is
\bea
\Phi(Q) &=& -\frac{\pi^{D/2}\,\sin(\pi(\Delta-D/2))\,\Gamma(D/2-\Delta)\,C_2} {2^{2\Delta-D-1}\,\Gamma(\Delta)} \left(Q^2\right)^{\Delta-D/2} \theta(Q^0)\,\theta(Q^2)\nn\\
&=& \frac{\pi^{D/2+1}\,C_2} {2^{2\Delta-D-1}\,\Gamma(\Delta)\, \Gamma(\Delta - D/2 + 1)} \left(Q^2\right)^{\Delta-D/2} \theta(Q^0)\,\theta(Q^2)
\eea
Note that $\Phi$ is positive if $C_2 > 0$ and $\Delta \geq D/2 - 1$, which is exactly the well-known bound on the dimension of a scalar operator~\cite{Todorov-Mintchev-Petkova,Fradkin:1997df}. For the limiting case $\Delta = D/2 - 1$, use
\be
\lim_{\epsilon\to0^+}\frac{\epsilon\,\theta(Q^2)}{(Q^2)^{1-\epsilon}} = \delta(Q^2)
\ee
with $\epsilon = \Delta - (D/2 - 1)$ to obtain
\bea
\Phi(Q) = \frac{8\pi^{D/2+1}\,C_2} {\Gamma(D/2-1)}\; \theta(Q^0)\,\delta(Q^2)
\eea

We thus have
\bea
&&|\cM|^2\,\Phi = \frac{2^{3D-4d+1}\,\pi^{D/2+1}\,\Gamma(\Delta)\,[\Gamma(D-d-\Delta/2)]^2\,C_3^2} {[\Gamma(D/2-\Delta)]^2\, [\Gamma(d+\Delta/2-D/2)]^2\, \Gamma(\Delta - D/2 + 1)\, C_2}\nn\\
&&\qquad\qquad\quad \times \left|I(P,Q)\right|^2 \left(Q^2\right)^{\Delta-D/2} \theta(Q^0)\,\theta(Q^2)
\eea
We can combine the denominators in (\ref{IPQ}) by using
\be
\prod_j\frac{\Gamma(n_j)}{A^{n_j}}
= \int_0^1
\frac{\Gamma\left(\sum_k n_k\right)}
{\left(\sum_j\alpha_jA_j\right)^{\sum_k n_k}}\,
\delta\left(\sum_j\alpha_j - 1\right)
\prod_j\alpha_j^{n_j-1}\,d\alpha_j
\ee
which gives
\bea
|\cM|^2\,\Phi &=& \frac{2^{3D-4d+1}\,\pi^{D/2+1}\, \Gamma(\Delta)\,[\Gamma(D-d+\Delta/2)]^2\,C_3^2} {[\Gamma(D/2-\Delta)]^2\, [\Gamma(d+\Delta/2-D/2)]^2\,[\Gamma(\Delta/2)]^4\, \Gamma(\Delta - D/2 + 1)\,C_2}\nn\\
&&\times \left|\tilde I\left(P,Q\right)\right|^2 \left(Q^2\right)^{\Delta-D/2} \theta(Q^0)\,\theta(Q^2)
\label{sigma-contrib-int}
\eea
where
\bea
\tilde I(P,Q) &\equiv& \int_0^1 d\alpha_1\, d\alpha_2\, d\alpha_3\, \delta\left(\sum_{j=1}^3\alpha_j - 1\right)
\alpha_1^{D-d-\Delta/2-1}\,\alpha_2^{\Delta/2-1}\,\alpha_3^{\Delta/2-1}
\nn\\
&&\times \int \frac{d^Dk}{\left(k^2 + g(P,Q) + i\epsilon\right)^{D-d+\Delta/2}}
\eea
where we shifted the integration variable $k$ and defined
\be
g(P,Q) \equiv \alpha_1(1-\alpha_1)P^2 - 2\alpha_1\alpha_3 QP + \alpha_3(1-\alpha_3)Q^2
\ee

In the following sections we will do the inclusive and exclusive calculations explicitly in a 2D toy model, the Sommerfield model, where we consider the processes whose inclusive descriptions are given in figures~\ref{fig-1}b and c.

\section{2D toy model of unparticle physics\label{sec-model}}
\setcounter{equation}{0}

\subsection{Sommerfield model of a Banks-Zaks sector\label{sec-sommerfield}}

In this paper we apply the ideas of section~2 to analyze the unparticle stuff that appears in the Sommerfield model~\cite{Sommerfield:1964,Brown:1963,Thirring-Wess:1964,Dubin-Tarski:1967,Hagen:1967},
that is the Schwinger model~\cite{Schwinger:1962tp} with an additional mass term for the vector boson:\footnote{Our conventions, as in~\cite{Georgi:2008pq}, are:
$
g^{00}=-g^{11}=1,\,
\epsilon^{01}=-\epsilon^{10}=-\epsilon_{01}=\epsilon_{10}=1
$. From the defining properties $\{\gamma^\mu,\gamma^\nu\} = 2g^{\mu\nu}$ and $\gamma^5 = -\frac{1}{2}\epsilon_{\mu\nu}\gamma^\mu\gamma^\nu$, it follows that
$\gamma^\mu\gamma^5=-\epsilon^{\mu\nu}\gamma_\nu$ and $\gamma^\mu\gamma^\nu=g^{\mu\nu}+\epsilon^{\mu\nu}\gamma^5$, and we will use the representation
$
\gamma^0=
\pmatrix{
0&1\cr
1&0\cr
},\;
\gamma^1=
\pmatrix{
0&-1\cr
1&0\cr
},\;
\gamma^5=\gamma^0\gamma^1=
\pmatrix{
1&0\cr
0&-1\cr
}\,$. Then the components $\psi_1$ and $\psi_2$ describe a right-moving and left-moving fermion, respectively.}
\be
\cL =
\bar\psi\,(i\spur\pd - e\,\slash A)\,\psi
-\frac{1}{4}F^{\mu\nu}F_{\mu\nu}
+\frac{m_0^2}{2}A^\mu A_\mu
\label{Sommerfield-model}
\ee
We are interested in this theory since, like the Schwinger model, it is an exactly solvable model that becomes scale-invariant at low energies, and (unlike the Schwinger model) has fractional anomalous dimensions. At low energies, below the vector boson mass, the theory reduces to the Thirring model, that is the theory of a fermion with a quartic self-interaction which is scale-invariant at all energies~\cite{Thirring:1958in,Wilson:1970pq,Georgi:1971sk}.

In order to solve the model, it is convenient to decompose $A^\mu$ as\footnote{In 2D, an arbitrary vector $A^\mu$ can be expanded in terms of any non-null vector $k^\mu$ as
$$
A^\mu=
\frac{k^\mu k^\rho-\epsilon^{\mu\nu}k_\nu\,\epsilon^{\rho\sigma}k_\sigma}
{k^2}\, A_\rho
$$
In the Schwinger model ($m_0=0$), $\cV$ would be the unphysical longitudinal polarization that can be set to zero in the Lorenz gauge.}
\be
A^\mu = \pd^\mu \cV + \epsilon^{\mu\nu}\pd_\nu\cA
\label{A-decomposition}
\ee
The Lagrangian becomes
\be
\cL = i\bar\psi\spur\pd\psi -
e\bar\psi\gamma_\mu\psi\left(\pd^\mu\cV +
\epsilon^{\mu\nu}\pd_\nu\cA\right) + \frac{1}{2}\cA\,\Box^2\cA +
\frac{m_0^2}{2}\left(\pd_\mu\cV\pd^\mu\cV -
\pd_\mu\cA\pd^\mu\cA\right)
\label{Sommerfield-AV}
\ee
If we change the fermionic variable to
\be
\Psi = e^{ie\left(\cV + \cA\gamma^5\right)}\psi
\label{psi-redef}
\ee
the fermion becomes free:
\be
\cL = i\bar\Psi\spur\pd\Psi + \frac{m_0^2}{2}\pd_\mu\cV\pd^\mu\cV + \frac{1}{2}\cA\,\Box^2\cA - \frac{m^2}{2}\pd_\mu\cA\pd^\mu\cA
\label{Sommerfield-redefined}
\ee
In the last term of (\ref{Sommerfield-redefined}) $m_0^2$ has been replaced
by
\be
m^2 = m_0^2 + \frac{e^2}{\pi}
\ee
in order to account for the fact that the path integral measure is not
invariant under the $\cA$ part of (\ref{psi-redef})~\cite{Roskies:1980jh}.\footnote{The same
effect gives mass $e/\sqrt\pi$ to the gauge boson in the Schwinger
model. See also~\cite{Georgi:1971iu}.}

We can now calculate fermionic $n$-point functions as
\bea
\vev{\psi_\alpha(x)\ldots\psi^\ast_\beta(y)\ldots}
&=& \vev{e^{-ie\left(\cV(x)+\cA(x)\gamma^5_{x}\right)}\ldots
e^{ie\left(\cV(y)+\cA(y)\gamma^5_{y}\right)}\ldots} \nn\\
&&\times \vev{\Psi_\alpha(x)\ldots\Psi^\ast_\beta(y)\ldots}
\label{exact-fermionic}
\eea
where the subscript in $\gamma^5_x$ specifies that it acts on the spinor at point $x$ (in our representation, $\gamma^5\psi_1 = +\psi_1$, $\gamma^5\psi_2 = -\psi_2$.) Using Wick's theorem and the observation that
\bea
-i\int d^2x\,e^{ipx}\,\vev{\cV(x)\cV(0)} &=& \frac{1}{m_0^2\, p^2} \\
-i\int d^2x\,e^{ipx}\,\vev{\cA(x)\cA(0)} &=& \frac{1}{(p^2)^2 - m^2 p^2} = \frac{1}{m^2}\left(\frac{1}{p^2-m^2}
-\frac{1}{p^2}\right)
\label{AVprop}
\eea
we obtain that the $n$-point function is given by the corresponding free $n$-point function multiplied by
\bea
\prod_{j>i}C_0(x_i-x_j)^{\eta_{ij}}\,C(x_i-x_j)^{\eta_{ij}\kappa_{ij}}
\label{bosonic-factor}
\eea
where $i$, $j$ run over the $n$ points and
\bea
C_0(x) &=& \exp\left[i\frac{e^2}{m_0^2}\left[D(x) -
D(0)\right]\right]
\propto \left(-x^2+i\epsilon\right)^{-e^2/4\pi m_0^2} \\
C(x) &=& \exp\left[i\frac{e^2}{m^2}\left[(\Delta(x) - \Delta(0)) - (D(x) - D(0))\right]\right] \nn\\
&=&\exp\left[\frac{e^2}{2\pi m^2}
\left[K_0\left(m\sqrt{-x^2 + i\epsilon}\right) + \ln\left(\xi m\sqrt{-x^2 +
i\epsilon}\right)\right]\right]
\label{C(x)}
\eea
with
\be
\xi  = \frac{e^{\gamma_E}}{2}
\label{xi}
\ee
where we used\footnote{$K_0$ is the modified Bessel function of the
second kind and $x_0$ is an arbitrary constant that will cancel out in the
following. For $y \to \infty$, $K_0(y) \sim \sqrt\frac{\pi}{2y}\,e^{-y} \to
0$. For $y \to 0$, $K_0(y) = -\ln(y/2) - \gamma_E + \cO(y^2)$, where
$\gamma_E = -\Gamma'(1) \simeq 0.577$ is Euler's constant. Note that
$\Delta(0)-D(0)=(i/2\pi)\ln\left(e^{\gamma_E}x_0m/2\right)
=(i/2\pi)\ln\left(\xi x_0m\right)$ in $C(x)$ is finite. The $D(0)$ term in $C_0(x)$ does not have any effect since any $n$-point function has $-n/2$ such factors so their presence is equivalent to changing the normalization of the fermion field.}
\bea
&&\Delta(x) = \int\frac{d^2p}{(2\pi)^2}\frac{e^{-ipx}}{p^2 - m^2 + i\epsilon}
= -\frac{i}{2\pi}K_0\left(m\sqrt{-x^2 + i\epsilon}\right)
\label{massive-propagator} \\
&&D(x) = \int\frac{d^2p}{(2\pi)^2}\frac{e^{-ipx}}{p^2 + i\epsilon} = \frac{i}{4\pi}\ln\left(\frac{-x^2+i\epsilon}{x_0^2}\right)
\eea
and $\eta_{ij}$ and $\kappa_{ij}$ are sign factors that depend on whether the operators at the two points are the fields or their adjoints and their handedness:
\be
\eta_{ij} = \left\{\matrix{
  +1\,, & \psi \mbox{ and } \psi^\ast \cr
  -1\,, & (\psi \mbox{ and } \psi) \mbox{ or } (\psi^\ast \mbox{ and } \psi^\ast)}\right.
\qquad\quad
\kappa_{ij} \equiv \gamma^5_i\gamma^5_j = \left\{\matrix{
  +1\,, & \alpha = \beta \cr
  -1\,, & \alpha \neq \beta }\right.
\ee
Similar expressions for the correlation functions have been obtained by various methods in the past~\cite{Lowenstein:1971fc,Segre:1974sm,Hagen:1994eb}.

In the short-distance limit, $(x_i-x_j)^2 \ll 1/m^2$, $C(x)\to1$ and one obtains free-fermion behavior.\footnote{$C_0(x)$ does not contribute a fractional power to the correlation functions of fermion bilinears.} In the large-distance limit, $(x_i-x_j)^2 \gg 1/m^2$, $K_0$ does not contribute, and $C(x)$ is just a power of $x^2$:
\be
C(x) \to \left[(\xi m)^2(-x^2+i\epsilon)\right]^{e^2/4\pi m^2}
\label{C(x)-un}
\ee
leading to scale-invariant behavior with fractional anomalous
dimensions. This is the unparticle regime. Note that $m$  plays the role of $\Lambda_\un$ from~\cite{Georgi:2007ek}.

Additional properties of the Sommerfield model are described in appendix~\ref{sec-operators}.

\subsection{Coupling to the ``standard model''\label{sec-sm}}

We assume that the very high energy theory includes the interaction
\bea
\cL_{\rm int} &=& \frac{\mu}{2}\,
\left[\bar\psi(1+\gamma_5)\chi\,\phi^\ast
+
\bar\psi(1-\gamma_5)\chi\,\phi\right]
+ \mbox{h.c.}\nn\\
&=& \mu\,
\left(\psi_2^\ast\,\chi_1\,\phi^\ast
+
\psi_1^\ast\,\chi_2\,\phi\right)
+ \mbox{h.c.}
\label{highenergy}
\eea
that couples the fermion $\psi$ of the Banks-Zaks sector to a neutral
complex scalar $\phi$ with mass $m_\phi \ll m$ that plays the role of a standard model field. The interaction is mediated by the heavy fermion
$\chi$ with mass $M \gg m,\, \mu^2/m$ and the same coupling to $A^\mu$ as $\psi$.

The interaction preserves a global $U(1)$ symmetry with charge $+1$ for
$\phi^\ast$ and $\psi_1^\ast$ and charge $-1$ for $\psi_2^\ast$. Integrating out $\chi$ we obtain\footnote{In a 4D unparticle theory, the interaction corresponding to (\ref{lowenergy}) would typically be nonrenormalizable.}
\be
\cL_{\rm int} =
\frac{h}{2}
\left(\cO\,{\phi^\ast}^2 + \cO^\ast\phi^2\right)\,,\qquad
h \equiv \frac{2\mu^2}{M}
\label{lowenergy}
\ee
where the composite operator
\be
\cO\equiv \bar\psi\,\textstyle\frac{1}{2}\displaystyle\left(1+\gamma^5\right)\psi = \psi_2^\ast\psi_1
\label{O}
\ee
(that has charge $-2$ under the global $U(1)$ symmetry) will have a fractional anomalous dimension at low energies.

Since this symmetry acts chirally on the fermionic field one may be concerned about anomalies. Indeed, the axial current $j^{5\mu} = \bar\psi\gamma^\mu\gamma^5\psi$ has a non-zero divergence~\cite{Georgi:1971iu}, and in the Schwinger model the symmetry transformation gives rise to an additional term in the Lagrangian proportional to
\be
\cL_\theta \,\propto\, \pd^\mu j^5_\mu = -\frac{e}{\pi}\epsilon^{\mu\nu}\pd_\mu A_\nu = -\frac{e}{2\pi}\epsilon^{\mu\nu} F_{\mu\nu}
\ee
While this term is a total derivative, it cannot be eliminated in the Schwinger model due to instanton configurations in Euclidean space that do not decay fast enough at infinity and for which $\int d^2x\,\cL_\theta$ does not vanish~\cite{Callan:1976je}. This leads to the existence of degenerate $\theta$ vacua~\cite{Coleman:1975pw,Coleman:1976uz} that differ in the value of the symmetry-breaking condensate $\langle\cO\rangle \propto e^{i\theta}$. But in the Sommerfield model (unlike the Schwinger model that has gauge invariance) we can define a conserved axial current as
\be
\bar j^{5\mu} = j^{5\mu} + \frac{e}{\pi}\epsilon^{\mu\nu} A_\nu
\ee
Then, since a (non-anomalous) symmetry cannot be broken spontaneously in two dimensions~\cite{Coleman:1973ci}, there is no condensate. Note also that a non-vanishing condensate of an operator with a non-zero dimension would be in conflict with scale invariance. In the Schwinger model, the dimension of $\cO$ at the IR fixed point is $0$, but we will see that this is not the case in the Sommerfield model.

\section{Correlation functions of the unparticle operator\label{correlation}}
\setcounter{equation}{0}

The vertices in (\ref{highenergy}) that couple the standard model field $\phi$ to the fermions of the Sommerfield model involve also the heavy field $\chi$. Since $\chi$ effectively propagates over distances of order $1/M$ which are much shorter than the distances we want to consider, our effective interactions will involve two $\phi\,$s coupling to the leading operator in the operator product expansion (OPE)~\cite{Wilson:1969zs,Wilson:1970pq} of a product of two $\psi\,$s. In particular, we will be interested in
\be
\mbox{T}\psi_\beta^\ast(x_2)\psi_\alpha(x_1)
= c(x_2-x_1)\,\psi_\beta^\ast\psi_\alpha(x_2)+\cdots\qquad
(\alpha\neq\beta)
\label{OPE-5}
\ee
that defines the operator $\cO = \psi_2^\ast\psi_1$ from (\ref{O}) and its conjugate $\cO^\ast = \psi_1^\ast \psi_2$. Using this OPE and the exact solution for the fermionic correlation functions from section~\ref{sec-sommerfield} we will determine the correlation functions of $\cO$ and $\cO^\ast$. The coefficient function $c(x_2 - x_1)$ will not play any role since there is also a $\chi$ propagator connected between $x_1$ and $x_2$. Evaluating the convolution of the two at zero external momentum, that is all we need for energies much below $M$, would give a constant.

\subsection{2-point function}

{\figsize\begin{figure}[htb]
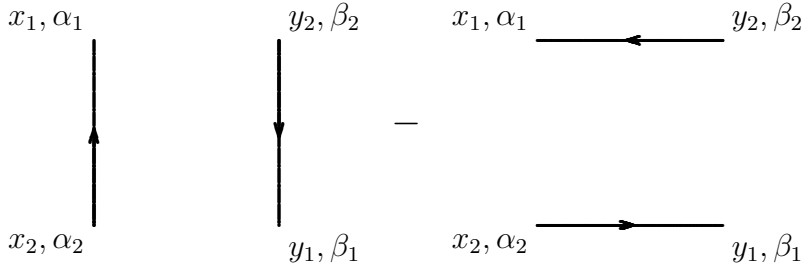

$$\beginpicture
\setcoordinatesystem units <0.7\tdim,0.7\tdim>
\put {$x_1,\alpha_1$} [rb] at -55 55
\put {$x_2,\alpha_2$} [rt] at -55 -55
\put {$y_2,\beta_2$} [lb] at 55 55
\put {$y_1,\beta_1$} [lt] at 55 -55
\stpltsmbl
\tarrow from -50 -50 to -50 2
\tarrow from 50 50 to 50 -2
\plot 50 50 50 -50 /
\plot -50 50 -50 -50 /
\endpicture
\quad\mbox{{\large$-$}}\quad
\beginpicture
\setcoordinatesystem units <0.7\tdim,0.7\tdim>
\put {$x_1,\alpha_1$} [rb] at -55 55
\put {$x_2,\alpha_2$} [rt] at -55 -55
\put {$y_2,\beta_2$} [lb] at 55 55
\put {$y_1,\beta_1$} [lt] at 55 -55
\stpltsmbl
\tarrow from -50 -50 to 2 -50
\tarrow from 50 50 to -2 50
\plot 50 -50 -50 -50 /
\plot 50 50 -50 50 /
\endpicture$$
\caption{\figsize\sf\label{fig-2}The free fermion skeleton corresponding to the 4-point function (\ref{4pt-def}).}\end{figure}}

{\figsize\begin{figure}[htb]
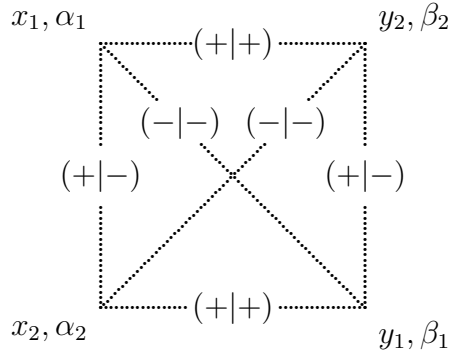

$$\beginpicture
\setcoordinatesystem units <\tdim,\tdim>
\put {$x_1,\alpha_1$} [rb] at -55 55
\put {$x_2,\alpha_2$} [rt] at -55 -55
\put {$y_2,\beta_2$} [lb] at 55 55
\put {$y_1,\beta_1$} [lt] at 55 -55
\stpltsmbl
\setdots <2pt>
\plot -50 -50 -16 -50 /
\plot 50 -50 16 -50 /
\plot 50 50 16 50 /
\plot -16 50 -50 50 /
\plot 50 50 50 10 /
\plot 50 -50 50 -10 /
\plot -50 50 -50 10 /
\plot -50 -50 -50 -10 /
\plot 50 50 28 28 /
\plot 12 12 -50 -50 /
\plot -50 50 -28 28 /
\plot -12 12 50 -50 /
\put {$(-|-)$} at 20 20
\put {$(-|-)$} at -20 20
\put {$(+|+)$} at 0 50
\put {$(+|+)$} at 0 -50
\put {$(+|-)$} at 50 0
\put {$(+|-)$} at -50 0
\linethickness=0pt
\putrule from -110 0 to 110 0
\endpicture$$
\caption{\figsize\sf\label{fig-signs}The sign factors $(\eta_{ij},\,\kappa_{ij})$ for the case $\alpha_1=\beta_2$, $\alpha_2=\beta_1$ ($\alpha_1\neq\alpha_2$).}\end{figure}}

Consider the fermionic 4-point function
\be
G^{(4)} = \vev{\psi^\ast_{\alpha_2}(x_2)\,\psi_{\alpha_1}(x_1)\,
\psi^\ast_{\beta_2}(y_2)\,\psi_{\beta_1}(y_1)}
\label{4pt-def}
\ee
We can express it as the free-fermion skeleton in figure~\ref{fig-2}, that is
\be
G^{(4)}_{\rm free} =
iS_0^{\alpha_1}(x_1-x_2)\, iS_0^{\beta_1}(y_1-y_2) \,\delta_{\alpha_1\alpha_2}\,\delta_{\beta_1\beta_2}
- iS_0^{\alpha_1}(x_1-y_2)\, iS_0^{\beta_1}(y_1-x_2) \,\delta_{\alpha_1\beta_2}\,\delta_{\beta_1\alpha_2}
\label{4pt-free}
\ee
multiplied by the bosonic factors (\ref{bosonic-factor}), where
$\delta_{\alpha\beta}\,S_0^\alpha(x)$ is the free-fermion propagator,\footnote{We define the propagator $S^\alpha(x)$ as
$
\delta_{\alpha\beta}\,S^\alpha(x)
\equiv -i\,\vev{\psi_\alpha(x)\,\psi^\ast_\beta(0)}
$.}
\be
S_0^1(x)
=\int \frac{d^2p}{(2\pi)^2}\,e^{-ipx}\,
\frac{p^+}{p^2+i\epsilon}
=-\frac{1}{2\pi}\frac{x^+}{x^2-i\epsilon}
\label{free-fermion-1}
\ee
\be
S_0^2(x)
=\int \frac{d^2p}{(2\pi)^2}\,e^{-ipx}\,
\frac{p^-}{p^2+i\epsilon}
=-\frac{1}{2\pi}\frac{x^-}{x^2-i\epsilon}
\label{free-fermion-2}
\ee
where we use the lightcone coordinates\footnote{Some properties of the lightcone coordinates are listed in appendix~\ref{sec-lightcone}.}
\be
x^\pm = x^0 \pm x^1
\label{lightcone-def}
\ee
For the $\cO$ and $\cO^\ast$ operators, we are interested in the case
\be
\alpha_1=\beta_2\,,\quad \alpha_2=\beta_1\qquad
(\alpha_1\neq\alpha_2)
\ee
Only the second term of (\ref{4pt-free}) survives, and the corresponding $\eta$ and $\kappa$ factors of (\ref{bosonic-factor}) are shown in figure~\ref{fig-signs}. In the limit $x_1\to x_2 \equiv x$, $y_1\to y_2 \equiv y$
this gives
\be
G^{(4)} =
\frac{C_0(x_1-x_2)\,C_0(y_1-y_2)}{C(x_1-x_2)\,C(y_1-y_2)}\, C(x-y)^4\, iS_0^1(x-y)\,iS_0^2(x-y)
\label{4pt-5}
\ee
On the other hand, using the OPE (\ref{OPE-5}), we can write (\ref{4pt-def}) in the same limit as
\be
G^{(4)} = c(x_2-x_1)\, c(y_2-y_1)\, \vev{\cO(x)\,\cO^\ast(y)}
\label{4pt-OPE-5}
\ee
Comparing (\ref{4pt-5}) and (\ref{4pt-OPE-5}) we see that we can take $c(x) = C_0(x)/C(x)$, and
\be
i\Delta_\cO(x) \equiv \vev{\cO(x)\,\cO^\ast(0)}
= C(x)^4\, iS_0^1(x)\, iS_0^2(x)
= \frac{C(x)^4}{(2\pi)^2\left(-x^2 + i\epsilon\right)}
\label{O-prop}
\ee

At distances large compared to $1/m$, using (\ref{C(x)-un}),
we see that the 2-point function is proportional to
an unparticle propagator\footnote{Here and below, we incorporate a
dimensional factor of $1/(\xi m)^{2a}$ in the unparticle propagator so
that it has the same engineering dimension as the $\cO$ propagator.}
\be
i\Delta_{\cO}(x) \;\to\; i\Delta_\un(x) =
\frac{1}{(2\pi)^2(\xi m)^{2a}\left(-x^2 + i\epsilon\right)^{1+a}}
\label{un-prop-position}
\ee
where
\be
a \equiv -\frac{e^2}{\pi m^2} = -\frac{1}{1+\pi m_0^2/e^2}
\label{anom-dim-5}
\ee
denotes the anomalous dimension of the composite operator $\cO \equiv \psi_2^\ast\psi_1$ (its total dimension is $d_\un = 1+a$). For $0<m_0<\infty$, $a$ is fractional, which leads to unparticle behavior.\footnote{For $m_0 = 0$ we obtain $a = -1$, i.e., $d_\un = 0$, which is the Schwinger model result~\cite{Casher:1974vf}.} In momentum space\footnote{The Fourier transform is calculated on p.~284 in Ref.~\cite{Gelfand-Shilov}.}
\be
i\Delta_\un(p)
= \frac{iA(a)}{2(\xi m)^{2a}\,\sin(\pi a)}(-p^2 - i\epsilon)^a
= \frac{A(a)}{2\pi(\xi m)^{2a}}\int_0^\infty
dM^2\left(M^2\right)^a\frac{i}{p^2 - M^2 + i\epsilon}
\label{un-prop}
\ee
where
\be
A(a) \equiv -\frac{\sin(\pi a)\,\Gamma(-a)}{2^{1+2a}\,\pi\Gamma(1+a)} = \frac{1}{2^{1+2a}\left[\Gamma(1+a)\right]^2}
\ee
Since
\be
\mbox{Im}\,\Delta_\un(p) = -\frac{A(a)}{2(\xi
m)^{2a}}\theta(p^2)\left(p^2\right)^a
\label{Im-un-prop}
\ee
the unparticle phase space is
\be
\Phi_\un(p) = \frac{A(a)}{(\xi m)^{2a}}\,
\left(p^2\right)^a
\theta(p^0)\,\theta(p^2)
\label{un-phase-space}
\ee
Note that in the Schwinger model limit the phase space vanishes since $A(-1) = 0$.

Because we have the exact solution, we can
write (\ref{O-prop}) for arbitrary $x$ as
\be
i\Delta_{\cO}(x)
= i\Delta_\un(x)\exp\left[-4\pi ia\Delta(x)\right]
= i\Delta_\un(x)\sum_{n=0}^\infty\frac{\left(-4\pi
a\right)^n}{n!}\left[i\Delta(x)\right]^n
\label{O-prop-exp}
\ee
At distances not large compared to $1/m$, the higher terms in the
sum in (\ref{O-prop-exp}) become relevant. This will be important for studying the transition between the unparticle and particle regime.

\subsection{Higher $n$-point functions}

In a similar way, for a general $2n$-point function of the operator $\cO$ we have
\be
\vev{\cO(x_1)\ldots\cO(x_n)\,\cO^\ast(y_1)\ldots\cO^\ast(y_n)} = \frac{\prod_{j,k}
C(x_j-y_k)^4}{\prod_{k>j}C(x_j-x_k)^4\,C(y_j-y_k)^4}\,G^{(4n)}_{\rm free}
\label{n-pt-C-factors}
\ee
where $G^{(4n)}_{\rm free}$ is the corresponding free fermionic $4n$-point function:\footnote{Note that this simple form arises from the sum over all possible contractions of the free fermion fields.}
\be
G^{(4n)}_{\rm free} = \frac{(-1)^n}{(4\pi^2)^n}
\frac{\prod_{k>j}(x_j-x_k)^2(y_j-y_k)^2}{\prod_{j,k}(x_j-y_k)^2}
\label{free-n-pt}
\ee
Thus, using (\ref{O-prop}),
\be
\vev{\cO(x_1)\ldots\cO(x_n)\,\cO^\ast(y_1)\ldots\cO^\ast(y_n)} = \frac{\prod_{j,k} i\Delta_{\cO}(x_j-y_k)}
{\prod_{k>j}i\Delta_{\cO}(x_j-x_k)\,
i\Delta_{\cO}(y_j-y_k)}
\label{n-pt}
\ee
The $2n$-point function involves the two-point function $i\Delta_{\cO}(x)$, and also its inverse
\be
i\tilde\Delta_{\cO}(x)\equiv \frac{1}{i\Delta_{\cO}(x)}
\label{inverse}
\ee
We can represent the $2n$-point functions diagrammatically  with $2n$ vertices connected by two kinds of lines: solid lines representing
$i\Delta_{\cO}(x_j-y_k)$; dotted lines representing
$i\tilde\Delta_{\cO}(x_j-x_k)$ or $i\tilde\Delta_{\cO}(y_j-y_k)$. For example, the $2$-point function is just a single solid line, and the $6$- and $10$-point functions are shown in figure~\ref{fig-6-10pt}.
{\figsize\begin{figure}[htb]
$${\epsfxsize=.45\hsize \epsfbox{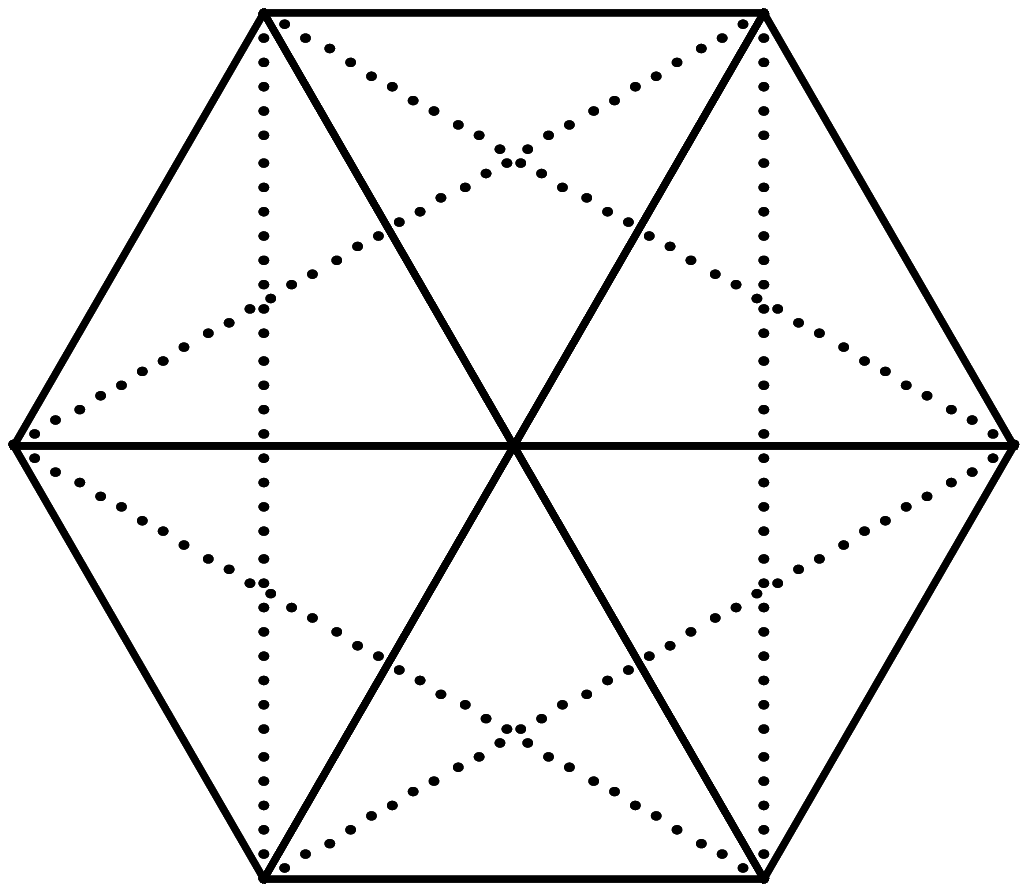}}\qquad\qquad
{\epsfxsize=.42\hsize \epsfbox{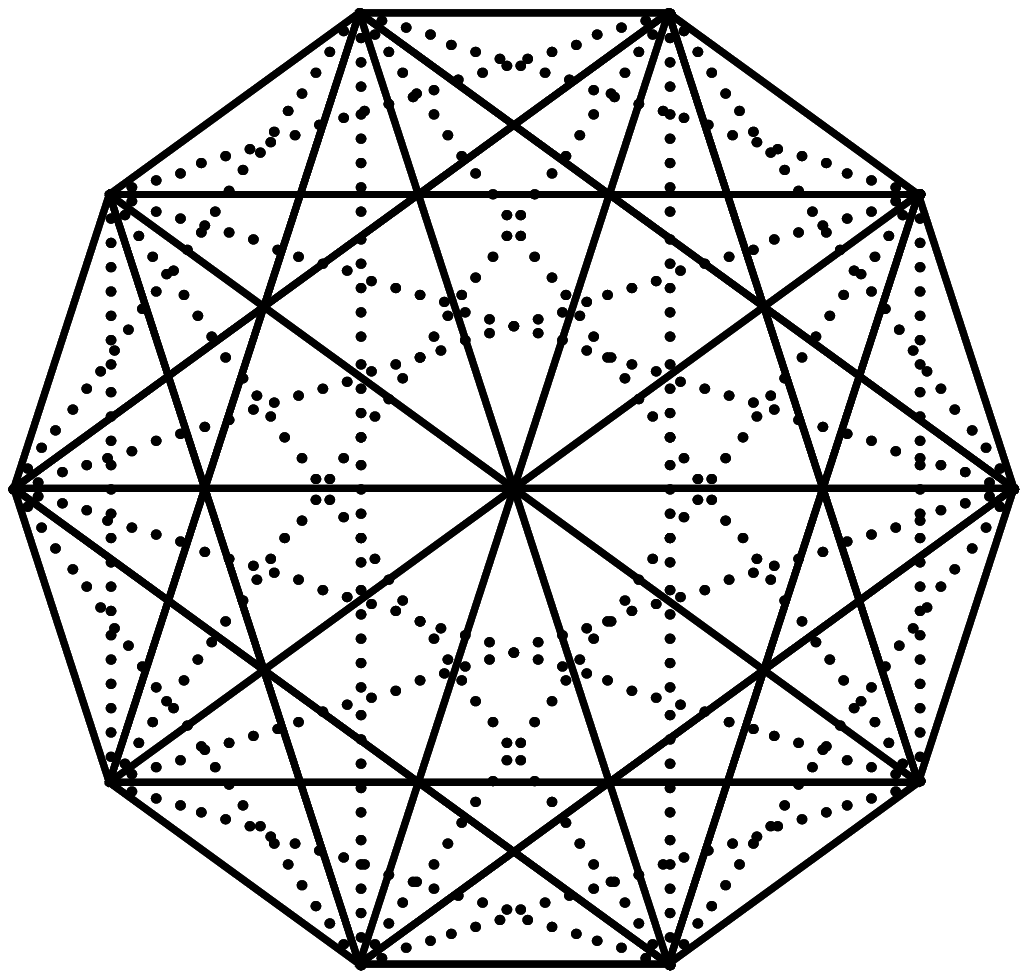}}$$
\caption{\figsize\sf\label{fig-6-10pt}The 6-point
function and the 10-point function of $\cO$.}\end{figure}}

It is important to remember that these diagrams, while useful as reminders of the structure of (\ref{n-pt}), are {\bf not} Feynman
diagrams. Eq. (\ref{n-pt}) is the sum of an infinite number of Feynman diagrams and describes the full $2n$-point function.\footnote{This includes also diagrams that can become disconnected in certain limits, since we included such diagrams in (\ref{free-n-pt}).}

Although we will not always indicate it explicitly, one should remember that the solid lines are actually directed, carrying the conserved chiral charge from an $\cO^\ast$ vertex to an $\cO$ vertex. The dotted lines are not directed. So for example we could represent the 6-point function as in figure~\ref{fig-6ptd}. Curiously, the charge carried by the solid lines depends on $n$. In the $2n$-point functions, each solid line carries $2/n$ units of charge. It is obvious that this conserves charge, because $n$ lines emerge from each $\cO^\ast$ vertex and $n$ flow into each $\cO$ vertex.
{\figsize\begin{figure}[htb]
$${\epsfxsize=.45\hsize \epsfbox{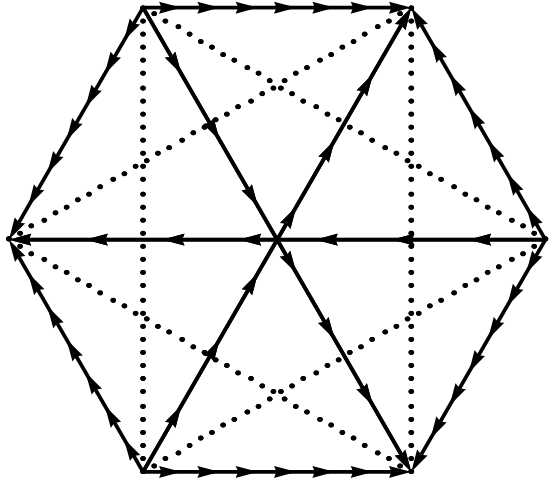}}$$
\caption{\figsize\sf\label{fig-6ptd}The 6-point
function with the direction of the charge flow indicated.}\end{figure}}

In this paper we will do most of our analysis on the $\cO$ 4-point function:
\begin{equation}
\begin{array}{c}
\displaystyle
\vev{\cO(x_1)\cO(x_2)\,\cO^\ast(y_1)\cO^\ast(y_2)} \\ \displaystyle
= \frac{i\Delta_\cO(x_1-y_1)\,
i\Delta_\cO(x_1-y_2)\, i\Delta_\cO(x_2-y_1)\,
i\Delta_\cO(x_2-y_2)}{i\Delta_\cO(x_1-x_2)\,
i\Delta_\cO(y_1-y_2)}
\end{array}
\label{4-pt}
\end{equation}
While its form may look peculiar, especially because of the propagators in the denominator that lead to IR divergences that we will analyze in section~\ref{momentum}, it is a typical form for conformal field theories, in any number of spacetime dimensions. More specifically, conformal invariance restricts the 4-point function of any four primary scalar operators $\cO_i$ with scaling dimensions $d_i$ to have the form~\cite{Ginsparg:1988ui}
\be
\vev{\cO_1(x_1)\cO_2(x_2)\cO_3(x_3)\cO_4(x_4)} = F(u,v)\prod_{i<j}\left(x_{ij}^2\right)^{d_T/6-(d_i+d_j)/2}
\ee
where $d_T = d_1 + d_2 + d_3 + d_4$ and $F(u,v)$ is an arbitrary function of the conformal-invariant cross-ratios
\be
u = \frac{x_{12}^2\, x_{34}^2}{x_{13}^2\, x_{24}^2}\,,\qquad
v = \frac{x_{12}^2\, x_{34}^2}{x_{14}^2\, x_{23}^2}
\label{cross-ratios}
\ee
When all $d_i$ are equal, like in our case of interest, this can be written more simply as
\be
\vev{\cO_1(x_1)\cO_2(x_2)\cO_3(x_3)\cO_4(x_4)} = \frac{f(u,v)}{(x_{13}^2)^d(x_{24}^2)^d}
\ee
In the unparticle limit of our model, i.e., with $\Delta_\cO \to \Delta_\un$ in (\ref{4-pt}), our 4-point function matches this form with $f(u,v) \propto v^d$. The function $f(u,v)$ can be different for other operators and in other conformal theories. However, because of the form of $u$ and $v$ in (\ref{cross-ratios}), when we expand $f(u,v)$ in powers of $u$ and $v$ each term will generically have powers of some $x_{ij}^2$ in the numerator and some in the denominator, similarly to what we have in our model.

\subsection{Momentum space and IR divergences\label{momentum}}

In momentum space the $2$-point function (\ref{O-prop}) [recall also (\ref{O-prop-exp})] becomes
\be
i\Delta_{\cO}(P)
= \sum_{n=0}^\infty\frac{\left(-4\pi
a\right)^n}{n!}\int\frac{d^2p_\un}{(2\pi)^2}
\,i\Delta_\un(p_\un)\left[\prod_{i=1}^n\frac{d^2p_i}{(2\pi)^2}
i\Delta(p_i)\right](2\pi)^2\delta^2\left(P-p_\un-\sum_{j=1}^n
p_j\right)
\label{solid-prop}
\ee
This describes a sum of two-point diagrams in which the incoming momentum $P$ splits between the unparticle propagator and $n$ massive scalar propagators.\footnote{Similar behavior in the Schwinger model, where the unparticle stuff is absent, has been discussed in~\cite{Casher:1973uf,Casher:1974vf}.} The discontinuity gives the phase space
\be
\begin{array}{c}
\displaystyle
\Phi(P) = \frac{A(a)}{(\xi m)^{2a}}\sum_{n=0}^\infty\frac{\left(-4\pi
a\right)^n}{n!}
\int\frac{d^2p_\un}{(2\pi)^2}\,\theta(p_\un^0)\,
\theta(p_\un^2)\left(p_\un^2\right)^a
\qquad\qquad\qquad\\ \displaystyle
\qquad\qquad\times\left[\prod_{i=1}^n\frac{d^2p_i}{(2\pi)^2}\,
2\pi\delta(p_i^2-m^2)\theta(p_i^0)\right]
(2\pi)^2\delta^2\left(P - p_\un - \sum_{j=1}^n p_j\right)
\end{array}
\label{full-phase-space}
\ee
This is the form that we used in \cite{Georgi:2008pq} to discuss the
transition from the low-energy unparticle regime and the high-energy free-fermion regime in our toy model. We will review this in section~\ref{sec-disappearance}. But to go beyond this simplest process, we need to understand the meaning of the higher $n$-point functions (\ref{n-pt}) in more detail.

The two point function $\Delta_{\cO}(x)$ goes to zero as
$-x^2\to\infty$. But that means that its inverse (\ref{inverse}) goes to $\infty$ as $-x^2\to\infty$. At first (and perhaps second) sight, this looks like a recipe for infrared divergences. Indeed, the Fourier transform of $\tilde\Delta_{\cO}(x)$ is very singular as $p^2\to0$. Formally, the dotted line propagator in momentum space can be written
as an expansion similar to (\ref{solid-prop}):
\bea
i\tilde\Delta_{\cO}(k) &\equiv& \int\,d^2x\,\frac{e^{ikx}}{i\Delta_{\cO}(x)}
= \int\,d^2x\,\frac{e^{ikx}}{i\Delta_\un(x)}\exp\left[4\pi ia\Delta(x)\right]\nn\\
&=& -i\frac{8\pi^4 A(-2-a)(\xi m)^{2a}}{\sin(\pi a)}
\int\frac{d^2p}{(2\pi)^2}\frac{1}{(-p^2 - i\epsilon)^{2+a}}\nn\\
&&\times \sum_{n=0}^\infty\frac{(4\pi a)^n}{n!}
\left[\prod_{i=1}^n\frac{d^2p_i}{(2\pi)^2}
i\Delta(p_i)\right]
(2\pi)^2\delta^2\left(k-p-\sum_{j=1}^n p_j\right)
\label{dotted-prop}
\eea
The inverse of the unparticle propagator, which gave the
\be
\frac{1}{(-p^2 - i\epsilon)^{2+a}}
\label{dotted-prop-up}
\ee
factor in (\ref{dotted-prop}), will lead to infrared divergences. The Fourier transform
\be
\frac{2^{-2(1+\gamma)}}{\pi}\frac{\Gamma(-\gamma)}{\Gamma(1+\gamma)}
\int\,d^2x\,{e^{ipx}}\,{(-x^2+i\epsilon)^\gamma}
=-\frac{1}{(-p^2-i\epsilon)^{1+\gamma}}
\label{ft}
\ee
has a well-behaved K\"all\'en-Lehmann representation for $-1< \gamma <0$:
\be
-\frac{\sin(\pi \gamma)}{\pi}\,
\int_0^\infty
\frac{ds}{s^{1+\gamma}}\,\frac{1}{p^2 - s + i\epsilon}
\label{kl}
\ee
The spectral function is positive and the singularity at $s=0$ is
integrable.  We used this implicitly in (\ref{full-phase-space}) with $\gamma = -1-a$. However, for the dotted line we needed to take $\gamma = 1 + a$ to obtain (\ref{dotted-prop}): this means $0 < \gamma < 1$, and then the integral in (\ref{kl}) does not converge near $s = 0$. More practically, the factor (\ref{dotted-prop-up}) will lead to divergences in cross-sections.

A better approach for doing the Fourier transform of the dotted line is to use (\ref{ft}) with $\gamma = a$ and get an extra factor of $x^2$ by differentiating both sides as
\begin{equation}
\frac{2^{-2(2+a)}}{\pi} \frac{\Gamma(-a)}{\Gamma(1+a)} \int\,d^2x\,{e^{ipx}}\,(-x^2+i\epsilon)^{1+a}
=-\frac{\pd^2}{\pd p^+\pd p^-}\frac{1}{(-p^2-i\epsilon)^{1+a}}
\label{ft2}
\end{equation}
where we use the lightcone components (\ref{lightcone-def}). It looks naively as if the derivatives just reproduce (\ref{dotted-prop-up}). Indeed, for $p^2\neq0$
\begin{equation}
\frac{\pd^2}{\pd p^+\pd p^-}\frac{1}{(-p^2-i\epsilon)^{1+a}}
=-(1+a)^2\,\frac{1}{(-p^2-i\epsilon)^{2+a}}
\label{naive}
\end{equation}
But it would be too naive to assume (\ref{naive}) for all $p^2$
because of the singularities of these generalized functions at
$p^2=0$. The derivative operator clearly produces a well-defined generalized function, but the singularities at $p^2=0$ would require special care.

We can think of two ways to deal with these infrared issues. In some situations, we can use (\ref{ft2}) directly by routing external momenta through the diagram in such a way that we can take the derivatives outside the loop. Then we can use (\ref{ft}) and (\ref{kl}). The other alternative is to use (\ref{ft2}) with an explicit infrared cut-off. This second alternative is instructive as well as useful, so we will outline it here. We will discuss both methods in more detail in section~\ref{sec-me-general} and appendix~\ref{infrared}.

Let the function $f_\lambda(s)$ be an IR-regulated version of $1/s^{1+a}$:
\begin{equation}
f_\lambda(s)\to\left\{
\begin{array}{l}
\displaystyle\frac{1}{s^{1+a}}\;\mbox{ for $s\gg\lambda$}\\
0\;\mbox{ as $s\to0$}
\end{array}\right.
\end{equation}
(for example, we could take $f_\lambda(s)=s^{-a}/(s+\lambda)$). Then an IR-regulated version of
\be
-\frac{\pi(1+a)^2}{\sin(\pi a)}\,\frac{1}{(-p^2-i\epsilon)^{2+a}}
\ee
is
\begin{equation}
{\renewcommand{\arraystretch}{2.2}\begin{array}{c}
\displaystyle
\frac{\pd^2}{\pd p^+\pd p^-}
\int_0^\infty
{ds}\,f_\lambda(s)\,\frac{1}{p^2 - s + i\epsilon}
\\ \displaystyle
= \int_0^\infty
{ds}\,f_\lambda(s)\,
\left(
\frac{1}{(p^2 - s + i\epsilon)^2}
+\frac{2s}{(p^2 - s + i\epsilon)^3}
\right)
\\ \displaystyle
= \int_0^\infty
{ds}\,f_\lambda(s)\,\left(\frac{\pd}{\pd s}
+s\,\frac{\pd^2}{\pd s^2}\right)
\,\frac{1}{p^2 - s + i\epsilon}
\\ \displaystyle
= \int_0^\infty
{ds}\,\Bigl(f'_\lambda(s)
+s\,f''_\lambda(s)\Bigr)
\,\frac{1}{p^2 - s + i\epsilon}
\end{array}}
\label{un-prop-ir-general}
\end{equation}
The spectral function in (\ref{un-prop-ir-general}) does not have a uniform sign. In fact, it is a total derivative, so the integral over $s$ vanishes. As $\lambda\to0$, the deviation of
\begin{equation}
\frac{1}{(1+a)^2}\,\Bigl(f'_\lambda(s)
+s\,f''_\lambda(s)\Bigr)
\end{equation}
from $1/s^{2+a}$ is squeezed down to $s=0$, but the integral over $s$
continues to vanish.

Because we now have a  K\"all\'en-Lehmann representation for the dotted line, we can safely write down the corresponding phase space from the discontinuity across the cut:
\bea
\tilde\Phi(k)
&=& \frac{8\pi^4 A(-2-a)(\xi m)^{2a}}{(1+a)^2}
\int\frac{d^2p}{(2\pi)^2}
\Bigl(f'_\lambda(p^2)
+p^2\,f''_\lambda(p^2)\Bigr)
\theta(p^2)
\label{full-dotted-phase-space}\\
&&\times \sum_{n=0}^\infty\frac{(4\pi a)^n}{n!}
\left[\prod_{i=1}^n\frac{d^2p_i}{(2\pi)^2}
2\pi\delta(p_i^2-m^2)\theta(p_i^0)\right]
(2\pi)^2\delta^2\left(k-p-\sum_{j=1}^n p_j\right)\nn
\eea
We will see in appendix~\ref{infrared} how (\ref{un-prop-ir-general})
and (\ref{full-dotted-phase-space}) work in a simple example.

\section{Disappearance process: $\phi + \phi \to \un$\label{sec-disappearance}}
\setcounter{equation}{0}

We now review the physical process described in figure~\ref{fig-1}a:
\be
\phi + \phi \to \mbox{Sommerfield stuff}
\ee
Because $\phi^2$ couples to $\cO^\ast$ at low energy (from
(\ref{lowenergy})),
we can obtain the total cross-section for this process from the discontinuity across the physical cut in the $\cO$ 2-point function. The cross-section for a given initial state $I$ to scatter into any possible final state $F$ can be obtained from the amplitude of $I$ going back to $I$ using the relation
\be
\sum_F \int d\Phi_F \left|\cM\left(I\to F\right)\right|^2
= -\,\mbox{Disc}\; i\cM\left(I\to I\right)
\label{disc}
\ee
where $\Phi_F$ is the phase space of the final state $F$, and $\mbox{Disc}$ refers to the discontinuity across the branch cut, $\cM(s+i\epsilon) - \cM(s-i\epsilon)$, where $s = E_{\rm cm}^2$. This is also known as the optical theorem, and the right-hand side of (\ref{disc}) is often written as $2\,\mbox{Im}\,\cM\left(I\to I\right)$. In our particular process, for $\phi$ momenta $P_1$ and $P_2$, this gives the cross-section
\be
\sigma = \frac{\mbox{Im}\,\cM(P_1,P_2\to P_1,P_2)}{s}
= -\frac{h^2}{s}\,\mbox{Im}\,\Delta_\cO(P)
\label{optical}
\ee
where $P = P_1 + P_2$ and $s = P^2$. In the unparticle limit ($\sqrt{s} \ll m$), using (\ref{Im-un-prop}), or directly the phase space (\ref{un-phase-space}), we find the fractional power behavior expected with unparticle production:
\be
\sigma = \frac{A(a)}{2}\frac{h^2}{(\xi m)^{2a}}\frac{1}{s^{1-a}}
\ee
On the other hand, in the free-particle limit that appears at high energies $\sqrt{s} \gg m$, we have $C(x) \to
1$ in (\ref{O-prop}) and then
\be
\sigma = \frac{h^2}{4}\frac{1}{s}
\label{sigma-short-dist}
\ee
which is the cross-section for $\phi + \phi \to \bar\psi_2 +
\psi_1$ without the Sommerfield interaction.

{\figsize\begin{figure}[htb]
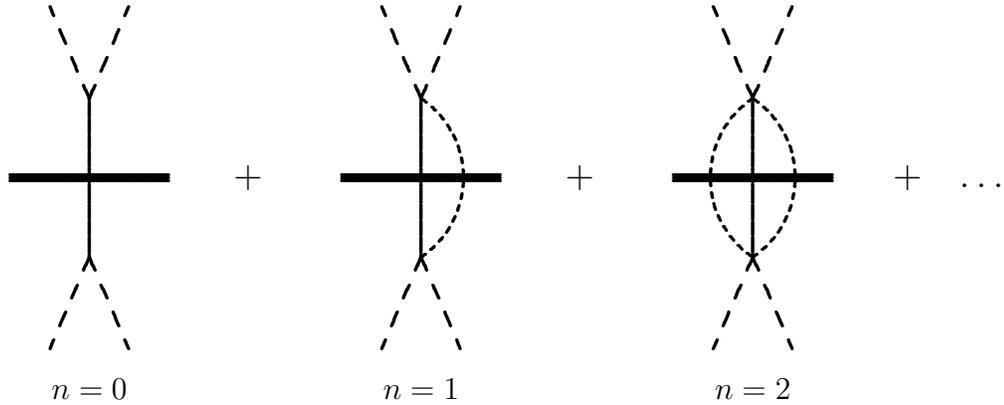

$$\beginpicture
\setcoordinatesystem units <1\tdim,1\tdim>
\stpltsmbl
\plot 0 -30 0 30 /
\plot -30 0 30 0 /
\plot -30 1 30 1 /
\plot -30 -1 30 -1 /
\setdashes <2mm>
\plot 0 30 15 65 /
\plot 0 30 -15 65 /
\plot 0 -30 15 -65 /
\plot 0 -30 -15 -65 /
\linethickness=0pt
\putrule from 0 -85 to 0 70
\putrule from -60 0 to 60 0
\put{$n=0$} at 0 -80
\put{\large{+}} at 60 0
\endpicture
\beginpicture
\setcoordinatesystem units <1\tdim,1\tdim>
\stpltsmbl
\plot 0 -30 0 30 /
\plot -30 0 30 0 /
\plot -30 1 30 1 /
\plot -30 -1 30 -1 /
\setdashes <2mm>
\plot 0 30 15 65 /
\plot 0 30 -15 65 /
\plot 0 -30 15 -65 /
\plot 0 -30 -15 -65 /
\setdashes <0.8mm>
\circulararc 115 degrees from 0 -30 center at -20 0
\linethickness=0pt
\putrule from 0 -85 to 0 70
\putrule from -60 0 to 60 0
\put{$n=1$} at 0 -80
\put{\large{+}} at 60 0
\endpicture
\beginpicture
\setcoordinatesystem units <1\tdim,1\tdim>
\stpltsmbl
\plot 0 -30 0 30 /
\plot -30 0 30 0 /
\plot -30 1 30 1 /
\plot -30 -1 30 -1 /
\setdashes <2mm>
\plot 0 30 15 65 /
\plot 0 30 -15 65 /
\plot 0 -30 15 -65 /
\plot 0 -30 -15 -65 /
\setdashes <0.8mm>
\circulararc 115 degrees from 0 30 center at 20 0
\circulararc 115 degrees from 0 -30 center at -20 0
\linethickness=0pt
\putrule from 0 -85 to 0 70
\putrule from -60 0 to 75 0
\put{$n=2$} at 0 -80
\put{\large{+\quad\ldots}} at 75 0
\endpicture$$
\caption{\figsize\sf\label{fig-massive-bosons}The cross-section for the disappearance process, based on the expansion (\ref{full-phase-space}). The long-dashed lines are the standard model particles $\phi$, the solid line represents the unparticle propagator while the short-dashed lines represent massive boson propagators. The thick line is a cut for computing the contribution to the cross-section.}\end{figure}}

The transition between the two limits can be studied using the expansion in (\ref{full-phase-space}) which shows that the general result can be described as the production of unparticle stuff along with an arbitrary number $n$ of massive bosons (to the extent that this is allowed energetically), see figure~\ref{fig-massive-bosons}. Along with an additional massive boson, each subsequent term is proportional to an additional power of $a$. One can easily obtain explicit results in the case of small $a$, when only the first few terms in the expansion contribute. For example, the $n=1$ term contributes
\be
\Phi^{(1)} = -a\,\theta(\sqrt{s}-m)\ln\frac{\sqrt{s}}{m} + \cO(a^2)
\ee
which gives the total phase space at the leading non-trivial order in $a$ as
\bea
\Phi = \frac{1}{2} - a\left[\ln\left(\frac{2}{e^{\gamma_E}}\frac{\xi
m}{\sqrt{s}}\right) + \theta(\sqrt{s}-m)\,\ln\frac{\sqrt{s}}{m}\right] + \cO(a^2)
\label{transition}
\eea
For energies $\sqrt{s} > m$, this expression reduces to
\be
\Phi = \frac{1}{2} + \cO(a^2)
\ee
that is the free-fermion result (\ref{sigma-short-dist}). Thus, for $|a|\ll 1$ there is a discontinuity in $d\Phi/d\sqrt{s}$ at $\sqrt{s} = m$, where a transition occurs from pure unparticle behavior below energy $m$ to pure free-fermion behavior above $m$ (see figure~\ref{fig-transition}).\footnote{The linear approximation (\ref{transition}) is not valid for $\sqrt{s} \ll m$
due to large $\ln\sqrt{s}$, but we have the exact expression
(\ref{un-phase-space}).}
Interestingly, the free-fermion behavior is obtained here as a sum of the $n = 0$ and $n = 1$ terms of figure~\ref{fig-massive-bosons}.
{\figsize\begin{figure}[htb]
$$\epsfxsize=2in \epsfbox[70 250 540 720]{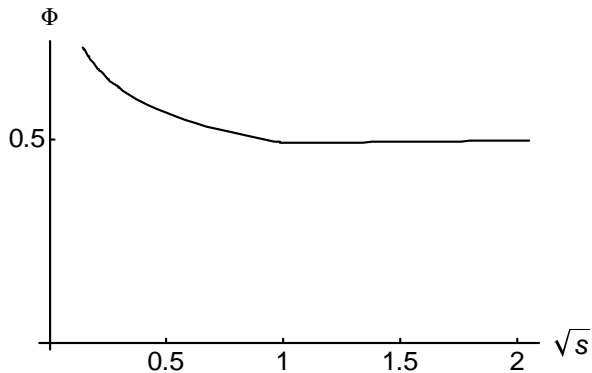}$$
\caption{\figsize\sf\label{fig-transition}Phase space $\Phi$ for the disappearance process in figure~\ref{fig-1}a as a function of the energy $\sqrt{s}$ (in units of $m$) for $a = -0.1$.}
\end{figure}}

For larger values of $|a|$, higher powers of $a$ will need to be taken into account in the contributions with extra bosons in order to reach the free-fermion regime at sufficiently high energies. Since each additional massive boson that we include gives a contribution with one extra power of $a$, we will need to include $N$ massive bosons to cancel all $a$-dependent terms up to $\cO(a^N)$. Then the free-fermion behavior will only appear at $\sqrt{s} > Nm$ (while in the range $m < \sqrt{s} < Nm$ mixed behavior will be observed, with discontinuities in $d\Phi/d\sqrt{s}$ for each multiple of $m$). In the limit $a \to -1$, the required value of $N$ becomes infinitely large: the condition $|a^N| \ll 1$ implies $N \gtrsim \frac{1}{-\ln(-a)}$. Note also that for $a = -1$ the unparticle contribution disappears since $A(-1)=0$. This limit is the Schwinger model; it has been studied in~\cite{Casher:1973uf,Casher:1974vf}.

\section{Missing charge process: $\phi + \phi \to \bar\phi + \bar\phi + \un$\label{sec-mc}}
\setcounter{equation}{0}

\subsection{Inclusive treatment: general\label{sec-mc-general}}

In this section we show that it is possible to go beyond the
simple calculation of section~\ref{sec-disappearance} to study processes in which standard model particles are radiated from the unparticle stuff. We begin with a process in which infrared issues do not intrude at low energies. In section~\ref{sec-me} we will consider another process in which the IR properties of the dotted line are more immediately important.

{\figsize\begin{figure}[htb]
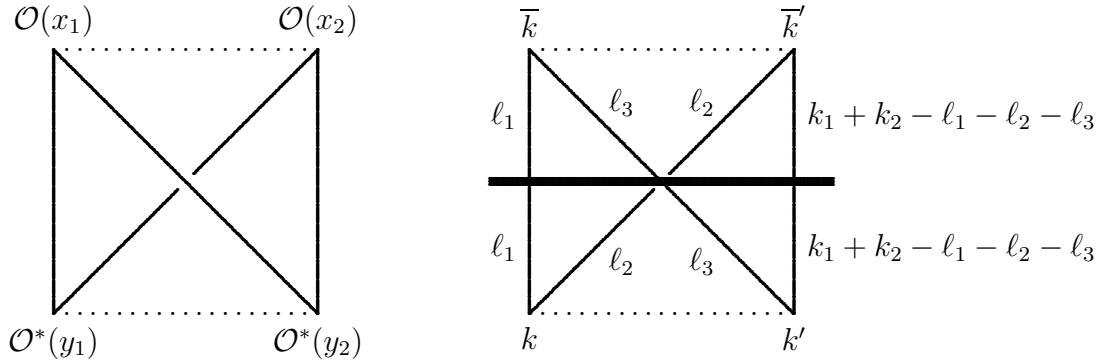

$$\beginpicture
\setcoordinatesystem units <1\tdim,1\tdim>
\stpltsmbl
\plot -50 -50 -50 50 50 -50 50 50 3 3 /
\plot -3 -3 -50 -50 /
\setdots
\plot 50 -50 -50 -50 /
\plot 50 50 -50 50 /
\put {$\cO^\ast(y_2)$} [t] at 50 -55
\put {$\cO(x_2)$} [b] at 50 55
\put {$\cO^\ast(y_1)$} [t] at -50 -55
\put {$\cO(x_1)$} [b] at -50 55
\linethickness=0pt
\putrule from 0 -60 to 0 60
\putrule from -80 0 to 80 0
\endpicture
\beginpicture
\setcoordinatesystem units <1\tdim,1\tdim>
\stpltsmbl
\plot -50 -50 -50 50 50 -50 50 50 3 3 /
\plot -3 -3 -50 -50 /
\plot -65 0 65 0 /
\plot -65 1 65 1 /
\plot -65 -1 65 -1 /
\setdots
\plot 50 -50 -50 -50 /
\plot 50 50 -50 50 /
\put {$k'$} [t] at 50 -55
\put {$\bar k'$} [b] at 50 55
\put {$k$} [t] at -50 -55
\put {$\bar k$} [b] at -50 55
\put {$k_1+k_2-\ell_1-\ell_2-\ell_3$} [l] at 55 -25
\put {$\ell_3$} [tr] at 20 -25
\put {$\ell_2$} [tl] at -20 -25
\put {$\ell_1$} [r] at -55 -25
\put {$k_1+k_2-\ell_1-\ell_2-\ell_3$} [l] at 55 25
\put {$\ell_2$} [br] at 20 25
\put {$\ell_3$} [bl] at -20 25
\put {$\ell_1$} [r] at -55 25
\linethickness=0pt
\putrule from 0 -60 to 0 60
\putrule from -100 0 to 180 0
\endpicture$$
\caption{\figsize\sf\label{fig-mc}The 4-point function~(\ref{4-pt}), in position space on the left and momentum space on the right. Momenta $k$ and $k'$ enter the diagram and $\bar k$ and $\bar k'$ leave it. The thick line is a cut for computing the cross-section of the process (\ref{missingcharge}).}\end{figure}}

Consider the missing charge process of figure~\ref{fig-1}b,
\be
\phi(p_1) + \phi(p_2) \to \bar\phi(q_1) + \bar\phi(q_2)
+ \mbox{Sommerfield stuff}
\label{missingcharge}
\ee
Because of (\ref{lowenergy}), and the fact that four units of $U(1)$ charge must be carried by the Sommerfield stuff in this process, the Sommerfield dynamics that contributes is associated with the discontinuity across the cut in the $\cO$ 4-point function (\ref{4-pt}) that is represented by figure~\ref{fig-mc}.
We obtain one contribution to the total cross-section of (\ref{missingcharge}) by annihilating two $\phi\,$s at $y_1$ (and injecting $k = p_1 + p_2$) and creating two $\bar\phi\,$s at $y_2$ (and injecting $k' = -q_1 - q_2$), and analogously for $x_1$ and $x_2$. More generally, $k$ can be taken to be the sum of any two out of $p_1$, $p_2$, $-q_1$, $-q_2$, and $k'$ the sum of the remaining two. Independently, $\bar k$ and $\bar k'$ can also be each assigned two of these four momenta. All the possible combinations are summed over in the cross-section. (In other words, we are taking into account the interference of all the possible ways of attaching the standard model particles.) Let's define the cross-section factor $\bar\sigma$ as
\be
\bar\sigma \equiv \sum_i \left|\cM_i\right|^2 \Phi_i
\label{bar-sigma}
\ee
where $\cM_i$ are the amplitudes of the various processes that contribute to our inclusive calculation and $\Phi_i$ are the phase spaces of the Sommerfield stuff that is produced in these processes. In order to obtain the actual cross-section one would still need to multiply $\bar\sigma$ by the couplings to the standard model $h^4$, divide it by the usual factors involving the momenta of the incoming particles, multiply by the phase space factors of the outgoing particles and integrate over the phase space. Based on (\ref{disc}), we have
\bea
&&\bar\sigma =
-\sum_{i,j}\int \Phi(Q-\ell_1-\ell_2-\ell_3)\,
\Phi(\ell_1)\frac{d^2\ell_1}{(2\pi)^2}\,
\Phi(\ell_2)\frac{d^2\ell_2}{(2\pi)^2}\,
\Phi(\ell_3)\,\frac{d^2\ell_3}{(2\pi)^2}\nn\\
&&\qquad\qquad\qquad\times\,
i\tilde\Delta_{\cO}(k_i-\ell_1-\ell_2)\, i\tilde\Delta_{\cO}(\bar k_j-\ell_1-\ell_3)
\label{sigma-mc-gen}
\eea
where $\tilde\Delta_{\cO}(k)$  is the dotted line propagator in momentum space given by (\ref{dotted-prop}) (with appropriate IR caveats), $\Phi(p)$ is the phase space of the Sommerfield stuff (along with its massive bosons) as given in (\ref{full-phase-space}), the momenta $k_i$ and $\bar k_j$ take the values
\be
k_i,\, \bar k_j \in \{p_1+p_2 \,,\; -q_1-q_2 \,,\; p_1-q_1 \,,\; p_1-q_2 \,,\; p_2-q_1 \,,\; p_2-q_2\}
\label{k's}
\ee
and we define
\be
P \equiv p_1 + p_2\,,\qquad
q \equiv q_1 + q_2\,,\qquad
Q \equiv P - q
\label{PqQ}
\ee
In the unparticle limit (i.e., without the massive bosons), (\ref{sigma-mc-gen}) reduces to
\bea
&&\bar\sigma = \frac{1}{2^3(2\pi)^6}\left[\frac{A(a)}{(\xi m)^{2a}}\right]^4 \sum_{i,j}
\int_0^\infty\,d\ell_1^+\,d\ell_2^+\,d\ell_3^+\,d\ell_1^-\,d\ell_2^-\,d\ell_3^-\,\nn\\
&&\qquad\qquad\qquad\qquad\qquad\times\,\theta\left((Q-\ell_1-\ell_2-\ell_3)^+\right)\,
\theta\left((Q-\ell_1-\ell_2-\ell_3)^-\right)\nn\\
&&\qquad\qquad\qquad\qquad\qquad\times\left[\ell_1^+\ell_2^+\ell_3^+(Q-\ell_1-\ell_2-\ell_3)^+ \ell_1^-\ell_2^-\ell_3^-(Q-\ell_1-\ell_2-\ell_3)^-\right]^a
\nn\\
&&\qquad\qquad\qquad\qquad\qquad\times\,\tilde\Delta_{\cO}(k_i-\ell_1-\ell_2)\, \tilde\Delta_{\cO}(\bar k_j-\ell_1-\ell_3)
\label{mc-un-cs}
\eea

\subsection{Results for small $a$}

It is easy to compute (\ref{mc-un-cs}) for $|a|\ll1$, since then
\be
i\tilde\Delta_\cO(k) =  -\frac{64i\pi^3 a}{\left(-k^+k^- - i\epsilon\right)^2} + \cO(a^2)
\label{ho-1}
\ee
and
\bea
\bar\sigma &=& \frac{1}{2^7(2\pi)^6}
\sum_{i,j}\int_0^\infty\,d\ell_1^+\,d\ell_2^+\,d\ell_3^+\,d\ell_1^-\,d\ell_2^-\,d\ell_3^-\,
\nn\\
&&\qquad\qquad\qquad \times\,\theta\left((Q-\ell_1-\ell_2-\ell_3)^+\right)
\theta\left((Q-\ell_1-\ell_2-\ell_3)^-\right)\nn\\
&&\qquad\qquad\qquad
\times\,\tilde\Delta_{\cO}(k_i-\ell_1-\ell_2)\, \tilde\Delta_{\cO}(\bar k_j-\ell_1-\ell_3) \nn\\
&=& \frac{a^2 h^4}{2} \sum_{i,j} I(Q^+,k_i^+,\bar k_j\,^+)\, I(Q^-,k_i^-,\bar k_j\,^-)
\label{sigma-missing-charge}
\eea
where\footnote{The following integral would be divergent for $0 < k_i < Q$ or $0 < \bar k_j < Q$, but the kinematics ensures that $k_i$ and $\bar k_j$ never fall in these ranges.}
\bea
I(Q, k, \bar k) &\equiv& \int_0^\infty d\ell_1\,d\ell_2\,d\ell_3\, \frac{\theta(Q-\ell_1-\ell_2-\ell_3)}{\left(k - \ell_1 - \ell_2\right)^2 \left(\bar k - \ell_1 - \ell_3\right)^2}\nn\\
&=& \frac{(Q-2\bar k)\,\ln\left(1-Q/\bar k\right) - (Q-2k)\,\ln\left(1 - Q/k\right)}{(\bar k - k)(Q - k - \bar k)}\,\theta(Q)
\label{I}
\eea

To analyze (\ref{sigma-missing-charge}) in more detail, we can consider, for simplicity, only the terms\footnote{In particular, these would be the only terms present if instead of (\ref{lowenergy}) we considered a model with two flavors of the standard model scalars, $\phi_A$ and $\phi_B$, with
$
\cL_{\rm int} =
\frac{h}{2}
\left[\cO\left({\phi_A^\ast}^2 + {\phi_B^\ast}^2\right) + \cO^\ast\left(\phi_A^2 + \phi_B^2\right)\right]
$
and asked about the process
$
\phi_A(p_1) + \phi_A(p_2) \to \bar\phi_B(q_1) + \bar\phi_B(q_2) + \mbox{unparticle stuff}
$.}
\be
k_i,\, \bar k_j \in \{p_1+p_2 \,,\; -q_1-q_2\}
\label{no-interference}
\ee
while omitting all the other interference terms in (\ref{k's}). Notice that for terms with $\bar k = k$ or $\bar k = Q - k$, (\ref{I}) reduces to
\be
I(Q, k, \bar k) = \left[\frac{Q}{k(k-Q)} + \frac{2\ln\left(1 - Q/k\right)}{2k-Q}\right]\theta(Q)
\ee
which gives, with the definitions (\ref{PqQ}),
\be
\bar\sigma = 2\,a^2\,\theta(Q^0)\,\theta(Q^2) \left(\frac{P^+-q^+}{P^+q^+} + \frac{2\ln(q^+/P^+)}{P^++q^+}\right)\left(\frac{P^--q^-}{P^-q^-} + \frac{2\ln(q^-/P^-)}{P^-+q^-}\right)
\label{mc-small-a-cs}
\ee
For $q \ll P$, that is when most of the energy goes into the unparticle stuff (which is possible if $P \gg m_\phi$), $\bar\sigma$ is dominated by the term
\be
\bar\sigma \propto \frac{1}{q^2}
\ee
This small-$q$ enhancement occurs because the momentum flowing through the dotted line (whose propagator is $i\tilde\Delta_\cO(k) \sim 1/(k^2)^2$) is allowed to be as small as $q$. On the other hand, in the limit when only a small fraction of the momentum goes into the unparticle stuff, $Q \ll P$, the cross-section behaves as
\be
\bar\sigma \propto \frac{\left(Q^2\right)^3}{(P^2)^4}
\label{sigma-small-un}
\ee
This result can be understood by observing that in this regime the points $x_1$ and $x_2$ in figure~\ref{fig-mc} are typically very separated from $y_1$ and $y_2$ (compared to the separation between $x_1$ and $x_2$ or $y_1$ and $y_2$). Then the four solid lines are essentially connected between the same two points and given by $[i\Delta_\un(X)]^4 \propto (-X^2)^{-4(1+a)}$ (where $X \equiv x-y$). In momentum space this becomes $\propto \left(-Q^2\right)^{3+4a}$. Each of the dotted lines has $i\tilde\Delta_\un(P) \propto (-P^2)^{-2-a}$, so we expect
\be
\bar\sigma \propto \frac{\left(Q^2\right)^{3+4a}}{(P^2)^{4+2a}}
\ee
This indeed agrees with (\ref{sigma-small-un}) (in the limit $|a|\ll 1$ that we assumed in the derivation of (\ref{sigma-small-un})). We will discuss this point of view in more generality in subsection~\ref{sec-mc-series}.

The missing charge process in the unparticle limit does not require us to use the IR-regulated form of the dotted line because the kinematics keeps the momentum carried by the dotted lines away from the light cone. At higher energies, above threshold for the production of the massive bosons, we do have to worry about these IR issues because while the total momentum carried by the dotted line cannot be lightlike, the momentum carried by its unparticle part can be. We will not discuss this here, because we will see a related issue already at low energy in the process analyzed in section~\ref{sec-me}.

\subsection{Inclusive treatment: series expansion\label{sec-mc-series}}

{\figsize\begin{figure}[htb]
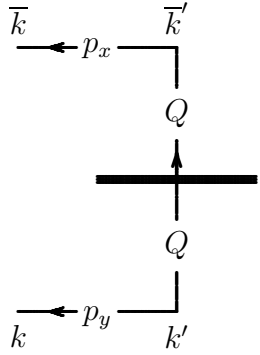

$$\beginpicture
\setcoordinatesystem units <1\tdim,1\tdim>
\stpltsmbl
\plot 30 -50 30 -35 /
\plot 30 -15 30 15 /
\plot 30 35 30 50 /
\tarrow from 30 0 to 30 10
\plot 8 50 30 50 /
\plot -30 50 -8 50 /
\tarrow from -10 50 to -18 50
\plot 30 -50 8 -50 /
\plot -8 -50 -30 -50 /
\tarrow from -10 -50 to -18 -50
\plot 0 0 60 0 /
\plot 0 1 60 1 /
\plot 0 -1 60 -1 /
\put {$k$} [t] at -30 -55
\put {$k'$} [t] at 30 -55
\put {$p_y$} [t] at 0 -48
\put {$\bar k$} [b] at -30 55
\put {$\bar k'$} [b] at 30 55
\put {$p_x$} [b] at 0 46
\put {$Q$} [c] at 30 25
\put {$Q$} [c] at 30 -25
\linethickness=0pt
\putrule from 0 -60 to 0 60
\putrule from -80 0 to 100 0
\endpicture$$
\caption{\figsize\sf\label{fig-mc-small-un-mom}A single term in (\ref{mc-4pt-expansion}). The $p_x$ and $p_y$ lines are described by (\ref{mc-zeta-FT}) and the $Q$ line by (\ref{mc-X-FT}). The arrows indicate the conventional direction of positive momentum. The thick line is a cut for computing the contribution to the cross-section.}\end{figure}}

We can also obtain analytic results, in the unparticle limit, for an arbitrary $a$, by expanding around the limit where the fraction of the momentum that goes into the unparticle stuff is small: $Q \ll P$. In this limit the contribution should come mainly from configurations of figure~\ref{fig-mc} in which $\zeta_x, \zeta_y \ll X$, where
\be
\zeta_x \equiv x_1 - x_2\,,\qquad
\zeta_y \equiv y_1 - y_2\,,\qquad
X \equiv x_2 - y_2
\label{zeta-x}
\ee
Expanding the 4-point function~(\ref{4-pt}) in $\zeta_x/X$ and $\zeta_y/X$ we have
\bea
&&\vev{\cO(x_1)\cO(x_2)\cO^\ast(y_1)\cO^\ast(y_2)} = i\tilde\Delta_\un(\zeta_x)\,i\tilde\Delta_\un(\zeta_y)\,[i\Delta_\un(X)]^4\nn\\
&&\qquad\times \left[1 - 2(1+a)\frac{\zeta_x^- - \zeta_y^-}{X^-} + 4(1+a)^2\frac{(\zeta_x - \zeta_y)^2}{X^2} \right.\nn\\
&&\qquad\qquad
+ (1+a)\frac{\left(3+2a\right)\left({\zeta_x^-}^2 + {\zeta_y^-}^2\right) - \left(5+4a\right)\zeta_x^-\zeta_y^-}{\left(X^-\right)^2} \nn\\
&&\qquad\qquad - (1+a)\left(1 + \frac{2a}{3}\right) \left(\zeta_x^- - \zeta_y^-\right) \frac{2\left(2+a\right)\left({\zeta_x^-}^2+{\zeta_y^-}^2\right) - \left(5+4a\right)\zeta_x^-\zeta_y^-}{\left(X^-\right)^3}
\nn\\
&&\qquad\qquad - 2(1+a)^2\left(\zeta_x^+ - \zeta_y^+\right)\frac{\left(3+2a\right)\left({\zeta_x^-}^2+{\zeta_y^-}^2\right)- \left(5+4a\right)\zeta_x^-\zeta_y^-}{\left(X^-\right)^2 X^+}\nn\\
&&\qquad\qquad\left. +\; \{-\leftrightarrow + \mbox{ for all asymmetric terms}\}
+ \ldots\tall\right]
\label{mc-4pt-expansion}
\eea
Each term in (\ref{mc-4pt-expansion}) describes a diagram as in figure~\ref{fig-mc-small-un-mom} in which momentum $Q$ flows through $X$, momentum $p_x \equiv \bar k$ through $\zeta_x$, and momentum $p_y \equiv -k$ through $\zeta_y$. To take the $\zeta$-dependent factors into momentum space, note that the Fourier transform of $i\tilde\Delta_\un(\zeta)(\zeta^-)^m(\zeta^+)^n$ is
\bea
\left(\frac{2}{i}\frac{\pd}{\pd p^+}\right)^m \left(\frac{2}{i}\frac{\pd}{\pd p^-}\right)^n
i\tilde\Delta_\un(p) &=& \left(2i\right)^{m+n}\,
\prod_{k=0}^{m-1}(2+a+k) \prod_{k=0}^{n-1}(2+a+k)
\frac{i\tilde\Delta_\un(p)}{(p^+)^m(p^-)^n}\nn\\
&=& \left(2i\right)^{m+n}\, \frac{\Gamma(2+a+m)\,\Gamma(2+a+n)}{[\Gamma(2+a)]^2}\frac{i\tilde\Delta_\un(p)}{(p^+)^m(p^-)^n}
\qquad\qquad
\label{mc-zeta-FT}
\eea
where, as in (\ref{dotted-prop}),
\be
i\tilde\Delta_\un(p) = -\frac{2^{3+2a}(2\pi)^3\,\Gamma(2+a)(\xi m)^{2a}}{\Gamma(-1-a)}\frac{i}{(-p^2-i\epsilon)^{2+a}}
\ee
To obtain the contribution of a term from (\ref{mc-4pt-expansion}) to the cross-section, we take the product of two factors of the form (\ref{mc-zeta-FT}) coming from the $\zeta_x$ and $\zeta_y$ lines and multiply it by the discontinuity across the cut through $X$. To find the latter, we compute the Fourier transform of
\be
\frac{[i\Delta_\un(X)]^4}{(X^-)^M(X^+)^N}
\label{mc-term}
\ee
(suppose, for definiteness, that $M \geq N$) as
\bea
&&\int d^2X\,e^{iQX} \frac{(X^+)^{M-N}[i\Delta_\un(X)]^4}{(X^2-i\epsilon)^M}
= \left(\frac{2}{i}\frac{\pd}{\pd Q^-}\right)^{M-N} \int d^2X\,\frac{e^{iQX}\,[i\Delta_\un(X)]^4}{(X^2-i\epsilon)^M}\nn\\
&&= -\frac{(-1)^M i}{2^{7+8a+2M}(2\pi)^7(\xi m)^{8a}}
\frac{\Gamma(-3-4a-M)}{\Gamma(4+4a+M)}
\left(\frac{2}{i}\frac{\pd}{\pd Q^-}\right)^{M-N}\left(-Q^2-i\epsilon\right)^{3+4a+M}\nn\\
&&= -\frac{i^{1+M+N}}{2^{7+8a+M+N}(2\pi)^7(\xi m)^{8a}}
\frac{\Gamma(-3-4a-N)}{\Gamma(4+4a+M)}\left(Q^+\right)^{M-N}
\left(-Q^2-i\epsilon\right)^{3+4a+N}
\label{mc-X-FT}
\eea
which gives the phase space factor
\bea
&&\frac{i^{M+N}(-1)^{N}\sin(4\pi a)\, \Gamma(-3-4a-N)}{2^{6+8a+M+N}(2\pi)^7(\xi m)^{8a}\,\Gamma(4+4a+M)}
\left(Q^+\right)^M \left(Q^-\right)^N
\left(Q^2\right)^{3+4a}\theta(Q^0)\,\theta(Q^2) \nn\\
&& = \frac{i^{M+N}\,\left(Q^+\right)^M \left(Q^-\right)^N
\left(Q^2\right)^{3+4a}\theta(Q^0)\,\theta(Q^2)}{2^{7+8a+M+N}(2\pi)^6(\xi m)^{8a}\,\Gamma(4+4a+M)\,\Gamma(4+4a+N)}
\label{mc-phase-space-factor}
\eea
Summing the contributions of all the terms in (\ref{mc-4pt-expansion}) and adding an overall minus sign from (\ref{disc}), we obtain
\bea
\bar\sigma &=& \frac{\tilde\Delta_\un(p_x)\tilde\Delta_\un(p_y)
\left(Q^2\right)^{3+4a}\theta(Q^0)\,\theta(Q^2)}{2^{7+8a}\,(2\pi)^6(\xi m)^{8a}\,[\Gamma(4+4a)]^2} \nn\\
&&\times
\left[1 + \frac{2+a}{2}\,Q^+\left(\frac{1}{p_x^+} - \frac{1}{p_y^+}\right)
+ \frac{(2+a)(3+a)(3+2a)}{4(5+4a)}\,{Q^+}^2\left(\frac{1}{{p_x^+}^2} + \frac{1}{{p_y^+}^2}\right) \right.\nn\\
&&\qquad - \frac{(2+a)^2}{4}\frac{{Q^+}^2}{p_x^+ p_y^+}
+ \frac{(2+a)^2}{4}\,Q^2\left(\frac{1}{p_x^2} + \frac{1}{p_y^2} - \frac{1}{p_x^+ p_y^-} - \frac{1}{p_y^+ p_y^-}\right)\nn\\
&&\qquad + \frac{(2+a)^2(3+a)}{24(5+4a)}\,{Q^+}^3 \left(\frac{1}{p_x^+} - \frac{1}{p_y^+}\right)\left[2(4+a)\left(\frac{1}{{p_x^+}^2}+\frac{1}{{p_y^+}^2}\right) - \frac{1+4a}{p_x^+p_y^+}\right] \nn\\
&&\qquad + \frac{(2+a)^2}{8}\,{Q^+}^2Q^- \left(\frac{1}{p_x^-} - \frac{1}{p_y^-}\right)\left[\frac{(3+a)(3+2a)}{5+4a}\left(\frac{1}{{p_x^+}^2} + \frac{1}{{p_y^+}^2}\right) - \frac{2+a}{p_x^+p_y^+}\right]\nn\\
&&\qquad\left. +\; \{+ \leftrightarrow - \mbox{ for all asymmetric terms}\} + \ldots\tall\right]
\eea
Summing over the 4 possibilities in (\ref{no-interference}) we get
\bea
\bar\sigma &=& \frac{[\tilde\Delta_\un(P)]^2
\left(Q^2\right)^{3+4a}\theta(Q^0)\,\theta(Q^2)}{2^{5+8a}\,(2\pi)^6(\xi m)^{8a}\,[\Gamma(4+4a)]^2} \nn\\
&&\times
\left[1 + (2+a)\left(\frac{Q^+}{P^+} + \frac{Q^-}{P^-}\right)
+ (2+a)^2\frac{Q^2}{P^2}\right.\nn\\
&&\qquad
+ \frac{(2+a)(28+31a+8a^2)}{4(5+4a)} \left[\left(\frac{Q^+}{P^+}\right)^2 + \left(\frac{Q^-}{P^-}\right)^2 + (2+a)\left(\frac{Q^2Q^+}{P^2P^+} + \frac{Q^2Q^-}{P^2P^-}\right)\right]\nn\\
&&\qquad\left. + \frac{(2+a)^2(3+a)(17+8a)}{12(5+4a)}\left(\left(\frac{Q^+}{P^+}\right)^3 + \left(\frac{Q^-}{P^-}\right)^3\right)
+ \ldots\right]
\label{mc-cs-series}
\eea
Note that at the leading order in $a$ this becomes
\bea
&&\bar\sigma = \frac{a^2}{18} \frac{\left(Q^2\right)^3}{\left(P^2\right)^4}
\left[1 + 2\left(\frac{Q^+}{P^+} + \frac{Q^-}{P^-}\right)
+ 4\frac{Q^2}{P^2}
+ \frac{14}{5}\left(\left(\frac{Q^+}{P^+}\right)^2 + \left(\frac{Q^-}{P^-}\right)^2\right)\right.\\
&&\qquad\qquad\qquad\;\left. + \frac{28}{5}\left(\frac{Q^2Q^+}{P^2P^+} + \frac{Q^2Q^-}{P^2P^-}\right) + \frac{17}{5}\left(\left(\frac{Q^+}{P^+}\right)^3 + \left(\frac{Q^-}{P^-}\right)^3\right)
+ \ldots\right] \theta(Q^0)\,\theta(Q^2)\nn
\eea
which is precisely (\ref{mc-small-a-cs}) expanded in powers of $Q/P$.

\subsection{Interpretation in terms of exclusive processes\label{sec-mc-interpretation}}

What physical states of the conformal sector are produced in the missing charge process whose inclusive cross-section we have computed here? Can the inclusive sum in (\ref{bar-sigma}) be decomposed into distinct well-defined contributions?

To understand the answer to these questions, let's look at our calculation from a different point of view. Consider the series expansion that we did in subsection~\ref{sec-mc-series}. Since we took the limit $\zeta_x, \zeta_y \ll X$, we can replace the products $\cO(x_1)\cO(x_2)$ and $\cO^\ast(y_1)\cO^\ast(y_2)$ in~(\ref{mc-4pt-expansion}) by their OPEs. In particular, using the shorthand notation
\be
\langle a|b\rangle \equiv \vev{\,a(X)\,b(0)}
\label{(a|b)}
\ee
we can write (\ref{mc-4pt-expansion}) as
\bea
&&\frac{\vev{\cO(x_1)\cO(x_2)\cO^\ast(y_1)\cO^\ast(y_2)}}{i\tilde\Delta_\un(\zeta_x)\,i\tilde\Delta_\un(\zeta_y)} = \nn\\
&&\qquad
\langle\cO^2|\cO^{\ast2}\rangle
+ \frac{1}{2}\left[\zeta_x^-\langle \pd_-\cO^2|\cO^{\ast 2}\rangle + \zeta_y^-\langle\cO^2|\pd_-\cO^{\ast2}\rangle\right]
\nn\\
&&\qquad + \frac{3+2a}{4(5+4a)}\left[{\zeta_x^-}^2\langle\pd_-^2\cO^2|\cO^{\ast2}\rangle + {\zeta_y^-}^2\langle\cO^2|\pd_-^2\cO^{\ast2}\rangle\right]
+ \frac{1}{4}\zeta_x^-\zeta_y^-\langle\pd_-\cO^2|\pd_-\cO^{\ast2}\rangle \nn\\
&&\qquad + \frac{1}{4}\left[\zeta_x^+\zeta_y^-\langle\pd_+\cO^2|\pd_-\cO^{\ast2}\rangle
+ \zeta_x^+\zeta_x^-\langle\pd_+\pd_-\cO^2|\cO^{\ast2}\rangle
+ \zeta_y^+\zeta_y^-\langle\cO^2|\pd_+\pd_-\cO^{\ast2}\rangle
\right]\nn\\
&&\qquad + \frac{2+a}{12(5+4a)} \left[{\zeta_x^-}^3\langle\pd_-^3\cO^2|\cO^{\ast 2}\rangle
+ {\zeta_y^-}^3\langle\cO^2|\pd_-^3\cO^{\ast 2}\rangle\right]
\nn\\
&&\qquad + \frac{3+2a}{8(5+4a)} \left[{\zeta_x^-}^2\zeta_y^-\langle\pd_-^2\cO^2|\pd_-\cO^{\ast 2}\rangle
+ \zeta_x^-{\zeta_y^-}^2\langle\pd_-\cO^2|\pd_-^2\cO^{\ast 2}\rangle \right.\nn\\
&&\qquad\qquad\qquad\qquad + {\zeta_x^-}^2\zeta_y^+\langle\pd_-^2\cO^2|\pd_+\cO^{\ast 2}\rangle
+ \zeta_x^+{\zeta_y^-}^2\langle\pd_+\cO^2|\pd_-^2\cO^{\ast 2}\rangle \nn\\
&&\qquad\qquad\qquad\qquad\left. + \zeta_x^+{\zeta_x^-}^2\langle\pd_+\pd_-^2\cO^2|\cO^{\ast 2}\rangle
+ {\zeta_y^-}^2\zeta_y^+\langle\cO^2|\pd_+\pd_-^2\cO^{\ast 2}\rangle \right] \nn\\
&&\qquad + \frac{1}{8}\left[\zeta_x^-\zeta_x^+\zeta_y^-\langle\pd_+\pd_-\cO^2|\pd_-\cO^{\ast 2}\rangle + \zeta_x^-\zeta_y^+\zeta_y^-\langle\pd_-\cO^2|\pd_+\pd_-\cO^{\ast 2}\rangle\right] \nn\\
&&\qquad
+ \;\{- \leftrightarrow + \mbox{ for all asymmetric terms}\}
+ \ldots
\label{mc-4pt-expansion-OPE}
\eea
where we defined $\cO^2$ to be the leading operator in the OPE for $\cO(x_1)\cO(x_2)$ with
\be
\langle\cO^2|\cO^{\ast2}\rangle = [i\Delta_\un(X)]^4
\label{O^2-2pt}
\ee
and the leading terms in the OPE are
\bea
&&\mbox{T}\cO(\zeta)\cO(0) = i\tilde\Delta(\zeta)\left[
1 + \frac{1}{2}\left(\zeta^-\pd_- + \zeta^+\pd_+\right)
+ \frac{3+2a}{4(5+4a)}\left({\zeta^-}^2\pd_-^2 + {\zeta^+}^2\pd_+^2\right)\right. \nn\\
&&\qquad\qquad\qquad\qquad\qquad
+ \frac{1}{4}\zeta^2\pd_+\pd_-
+ \frac{2+a}{12(5+4a)}\left({\zeta^-}^3\pd_-^3 + {\zeta^+}^3\pd_+^3\right) \label{OO-OPE}\\
&&\qquad\qquad\qquad\qquad\qquad\left.
+ \frac{3+2a}{8(5+4a)}\,\zeta^2\left(\zeta^-\pd_- + \zeta^+\pd_+\right)\pd_+\pd_-
+ \ldots\right]\cO^2(0) + \ldots\nn
\eea
The coefficients of the derivative terms in (\ref{OO-OPE}) must in fact have this precise form in order for the 3-point function $\vev{\cO(x)\cO(y)\cO^{\ast2}(z)}$ to be consistent with the conformal symmetry. In general~\cite{Petkou:1994ad} (see also~\cite{Ferrara:1971vh,Ferrara:1972xe,Ferrara:1973yt}), the contribution of all the derivatives of a scalar operator of dimension $\Delta$ (in our case, the operator $\cO^2$ has dimension $\Delta = 4(1+a)$) to the OPE of two scalar operators of dimension $d$ (in our case $d = 1+a$), in $D$ spacetime dimensions, should be proportional to
\bea
&& \frac{1}{\left(\zeta^2\right)^{d-\Delta/2}}\int_0^1 dt \left[t(1-t)\right]^{\Delta/2-1} \,_0F_1\left(\Delta+1-\frac{D}{2};\, -\frac{1}{4}t(1-t)\,\zeta^2\,\Box\right) e^{t\,\zeta\cdot\pd}\nn\\
&&= \frac{1}{\left(\zeta^2\right)^{d-\Delta/2}}
\sum_{m=0}^\infty \frac{\Gamma(\Delta+1-D/2)\left(-\zeta^2\,\Box\right)^m}{4^m\,m!\,\Gamma(\Delta+1-D/2+m)}
\int_0^1 dt \left[t(1-t)\right]^{\Delta/2+m-1} e^{t\,\zeta\cdot\pd}\nn\\
&&\propto \frac{1}{\left(\zeta^2\right)^{d-\Delta/2}}
\left[1 + \frac{\zeta\cdot\pd}{2}
+ \frac{2+\Delta}{8\,(1+\Delta)}(\zeta\cdot\pd)^2
- \frac{\Delta}{16\,(1+\Delta)\,(1+\Delta-D/2)}\,\zeta^2\,\Box \right. \nn\\
&&\qquad\qquad\qquad\left.
+ \frac{4+\Delta}{48\,(1+\Delta)}(\zeta\cdot\pd)^3
- \frac{\Delta}{32\,(1+\Delta)\,(1+\Delta-D/2)}\, \zeta^2\,(\zeta\cdot\pd)\,\Box
+ \ldots \right]\qquad\quad
\label{c-block-scalar}
\eea

Furthermore, there exists an explicit expression for the contribution that each rank-$\ell$ tensor $\cO_k$ in (\ref{4pt-CPWE}) (together with all its derivatives) can make to the 4-point function of 4 scalars~\cite{Dolan:2000ut}. The contribution to
\be
\vev{\cO_1(x_1)\cO_2(x_2)\cO_1(x_3)\cO_2(x_4)}
\ee
where $\cO_1$ and $\cO_2$ have the same dimension $d$ is proportional to\footnote{We present the expression that is relevant to 2 spacetime dimensions, but an analogous expression for 4 dimensions is given in~\cite{Dolan:2000ut} as well (see also the explanations in~\cite{Rattazzi:2008pe}).}
\bea
&&K_{\ell,\Delta} \equiv
\frac{u^{\frac{1}{2}\left(\Delta - \ell\right)} \left(\eta^-\right)^\ell}{\left(x_{12}^2\, x_{34}^2\right)^d}\,
_2F_1\left(\frac{\Delta+\ell}{2}, \frac{\Delta+\ell}{2}, \Delta+\ell; \eta^-\right)\,
_2F_1\left(\frac{\Delta-\ell}{2}, \frac{\Delta-\ell}{2}, \Delta-\ell; \eta^+\right)\nn\\
&&\qquad\quad\, +\, \{\eta^- \leftrightarrow \eta^+ \}
\label{Dolan-Osborn}
\eea
where $_2F_1$ is the ordinary hypergeometric function,
\be
\eta^\pm \equiv \frac{x_{12}^\pm x_{34}^\pm}{x_{13}^\pm x_{24}^\pm}
\ee
and $u = \eta^2$ as in (\ref{cross-ratios}). In our case
\be
x_{12} = \zeta_x\,,\qquad
x_{34}=\zeta_y\,,\qquad
x_{24} = X\,,\qquad
x_{13} = X + \zeta_x - \zeta_y
\ee
Expanding (\ref{Dolan-Osborn}) in small $\zeta_x$ and $\zeta_y$, with $\ell = 0$ and $\Delta = 4(1+a)$, we obtain precisely the expression (\ref{mc-4pt-expansion}) (up to an overall prefactor that is not fixed by the conformal symmetry). No other primary operators besides $\cO^2$ are required for reproducing the terms shown in (\ref{mc-4pt-expansion}), and this confirms the interpretation (\ref{mc-4pt-expansion-OPE}). It is then obvious from (\ref{mc-4pt-expansion-OPE}) that taking the cut in the 4-point function as we did describes the production of $\cO^2$ stuff:
\be
\phi + \phi \to \bar\phi + \bar\phi + \{\mbox{$\cO^2$ stuff}\}
\label{mc-O^2-stuff}
\ee

Additional operators do appear in higher-order terms that are not written in~(\ref{mc-4pt-expansion}). In particular, at order $(\zeta/X)^4$ there appears the extra contribution
\be
i\tilde\Delta_\un(\zeta_x)\,i\tilde\Delta_\un(\zeta_y)\,[i\Delta_\un(X)]^4\;
\frac{(1+a)^2}{2(5+4a)}\left(\frac{{\zeta_x^-}^2{\zeta_y^-}^2}{(X^-)^4} + \frac{{\zeta_x^+}^2{\zeta_y^+}^2}{(X^+)^4}\right)
\ee
that must be accounted for by the two-point functions of some new operator(s) of dimension $\Delta = 4(1+a)+2$ and rank $\ell = 2$.\footnote{In fact, it can be shown by more advanced methods that all operators of vector charge $0$ and axial charge $4$ in this theory have dimensions of the form $\Delta = 4(1+a)+n$ where $n$ is an integer~\cite{Luscher:1975js}.}

In conformal theories that are not solvable, which is typically the case in four dimensions, there is no easy way to determine which operators contribute to the OPE. However, the conserved currents of the CFT are the usual suspects to appear in OPEs and they always have their canonical dimensions. In our model, they will indeed appear in the OPE for $\cO^\ast(\zeta)\,\cO(0)$ that will be relevant for the missing energy process in section~\ref{sec-me}. The conserved current $j^\mu$ in our model could not appear in the OPE for $\cO(\zeta)\,\cO(0)$ because $\cO$ is charged under the axial $U(1)$ symmetry while $j^\mu$ is not. Yet further information can be obtained based on the conformal symmetry alone. For example, an upper bound on the dimension of the leading scalar operator in the OPE of two identical real scalars was derived in~\cite{Rattazzi:2008pe}. In our case, considering the operator $\cO + \cO^\ast$ (we are taking this combination in order to have a real operator) this bound implies that the OPEs for $\cO(\zeta)\,\cO(0)$ or $\cO^\ast(\zeta)\,\cO(0)$ must contain a scalar operator with dimension below $0.53 + 4(1+a)$. This condition is satisfied by the operator $\cO^2$ (whose dimension is $4(1+a)$) that appears in $\cO(\zeta)\,\cO(0)$. For $a > -\frac{1}{2}$, the condition is satisfied even earlier by the dimension-$2$ operator $j^2$ that appears in $\cO^\ast(\zeta)\,\cO(0)$ (we will analyze this OPE in section~\ref{sec-me-interpretation}).

\subsection{Exclusive treatment: amputated 3-point functions\label{sec-mc-amputated}}

We can also compute the exclusive cross-section for (\ref{mc-O^2-stuff}) directly, using the amputated 3-point function as discussed in section~\ref{sec-self-int}. For $D = 2$ and $d = 1+a$, (\ref{sigma-contrib-int}) reduces to
\be
|\cM|^2\,\Phi = \frac{2^{3-4a}\,\pi^2\,[\Gamma(\Delta/2+1-a)]^2\,C_3^2} {[\Gamma(1-\Delta)]^2\, [\Gamma(\Delta/2+a)]^2\,[\Gamma(\Delta/2)]^4\,C_2} \left(Q^2\right)^{\Delta-1} \theta(Q^0)\,\theta(Q^2)
\left|\tilde I\left(P,Q\right)\right|^2
\ee
with
\bea
\tilde I(P,Q) &=& \int_0^1 d\alpha_1\, d\alpha_2\, d\alpha_3\, \delta\left(\sum_j\alpha_j - 1\right)
\alpha_1^{-a-\Delta/2}\,\alpha_2^{\Delta/2-1}\,\alpha_3^{\Delta/2-1}
\nn\\
&&\times \int \frac{d^2k}{\left(k^2 + g(P,Q) + i\epsilon\right)^{1-a+\Delta/2}}
\eea
Wick rotating $k^1$, the integral over $k$ becomes
\be
2\pi i\int_0^\infty \frac{k\,dk}{\left(k^2 + g(P,Q)\right)^{1-a+\Delta/2}}
= \frac{2\pi i}{(\Delta - 2a) \left[g(P,Q)\right]^{\Delta/2-a}}
\ee
and then
\bea
|\cM|^2\,\Phi &=& \frac{2^{3-4a}\,\pi^4\,[\Gamma(\Delta/2-a)]^2\,C_3^2} {[\Gamma(1-\Delta)]^2\, [\Gamma(\Delta/2+a)]^2\,[\Gamma(\Delta/2)]^4\,C_2}\; \left(Q^2\right)^{\Delta-1} \theta(Q^0)\,\theta(Q^2)
\label{sigma-contrib-gen}\\
&&\times
\left|\int_0^1 d\alpha_1\, d\alpha_2\, d\alpha_3\,
\frac{\delta\left(\sum_j\alpha_j - 1\right)\, \alpha_1^{-a-\Delta/2}\,\alpha_2^{\Delta/2-1}\,\alpha_3^{\Delta/2-1}} {\left[\alpha_1(1-\alpha_1)P^2 - 2\alpha_1\alpha_3 QP + \alpha_3(1-\alpha_3)Q^2\right]^{\Delta/2-a}}\right|^2\nn
\eea
The integral in (\ref{sigma-contrib-gen}) is hard to solve analytically. However, expanding the integrand in powers of $Q/P$ it can be computed~\cite{Mathematica} (assuming for the purpose of calculation the unphysical range $-2a<\Delta<0$) to give
\bea
&&\frac{\Gamma(1-\Delta)\, [\Gamma(\Delta/2)]^2\,\Gamma(\Delta/2+a)} {\Gamma(\Delta)\,\Gamma(1-\Delta/2+a)} \frac{1}{\left(P^2\right)^{\Delta/2-a}} \nn\\
&&\times\left[1 + \frac{\Delta - 2a}{4} \left(\frac{Q^+}{P^+} + \frac{Q^-}{P^-}\right) + \frac{(\Delta-2a)^2}{16}\frac{Q^2}{P^2}\right. \\
&&\qquad
+ \frac{(\Delta-2a)(2+\Delta)(2+\Delta-2a)}{32\,(\Delta+1)} \left(\left(\frac{Q^+}{P^+}\right)^2 + \left(\frac{Q^-}{P^-}\right)^2 + \frac{\Delta-2a}{4}\left(\frac{Q^2Q^+}{P^2P^+} + \frac{Q^2Q^-}{P^2P^-}\right)\right)\nn\\
&&\qquad\left. + \frac{(\Delta-2a)\,(2+\Delta-2a)\,(4+\Delta)\,(4+\Delta-2a)}{384\,(\Delta+1)}\left(\left(\frac{Q^+}{P^+}\right)^3 + \left(\frac{Q^-}{P^-}\right)^3\right)
+ \ldots\right]\nn
\eea
so
\bea
|\cM|^2\,\Phi &=& \frac{2^{3-4a}\,\pi^4\,[\Gamma(\Delta/2-a)]^2\,C_3^2} {[\Gamma(\Delta)]^2\,[\Gamma(1-\Delta/2+a)]^2\,C_2}\; \frac{\left(Q^2\right)^{\Delta-1}}{\left(P^2\right)^{\Delta-2a}}\, \theta(Q^0)\,\theta(Q^2)\nn\\
&&\times\left[1 + \left(\frac{\Delta}{2}-a\right)\left(\frac{Q^+}{P^+} + \frac{Q^-}{P^-}\right) + \left(\frac{\Delta}{2}-a\right)^2\frac{Q^2}{P^2} \right.\nn\\
&&\quad + \left(\frac{\Delta}{2}-a\right)\frac{4+2\Delta^2+5\Delta-4a\Delta-6a} {8\,(1+\Delta)} \left[\left(\frac{Q^+}{P^+}\right)^2 + \left(\frac{Q^-}{P^-}\right)^2 \right.\nn\\
&&\qquad\qquad\qquad\qquad\qquad\qquad\qquad\qquad\qquad\quad\left.
+ \left(\frac{\Delta}{2}-a\right) \left(\frac{Q^2Q^+}{P^2P^+} + \frac{Q^2Q^-}{P^2P^-}\right)\right] \nn\\
&&\quad + \left(\frac{\Delta}{2}-a\right) \frac{\left(2+\Delta-2a\right) \left(8 + (7+2\Delta)\Delta - 2(5+2\Delta)a\right)}{48(1+\Delta)}\nn\\
&&\qquad\qquad\qquad\qquad\qquad\qquad\qquad\qquad\left.\times
\left(\left(\frac{Q^+}{P^+}\right)^3 + \left(\frac{Q^-}{P^-}\right)^3\right)
+ \ldots\right]
\label{sigma-contrib-gen-leading}
\eea
For the process (\ref{mc-O^2-stuff}) we have
\be
\Delta = 4(1+a)\,,\qquad
C_2 = \frac{1}{(2\pi)^8(\xi m)^{8a}}\,,\qquad
C_3 = \frac{1}{(2\pi)^6(\xi m)^{6a}}
\ee
where $\Delta$ and $C_2$ follow from (\ref{O^2-2pt}), and $C_3$ from (\ref{OO-OPE}) and conformal invariance. Then
\bea
|\cM_{\cO^2}|^2\,\Phi_{\cO^2} &=& \frac{[\Gamma(2+a)]^2} {2^{1+4a}\,[\Gamma(4+4a)]^2\,[\Gamma(-1-a)]^2\,(\xi m)^{4a}}\; \frac{\left(Q^2\right)^{3+4a}}{\left(P^2\right)^{4+2a}}\, \theta(Q^0)\,\theta(Q^2)\nn\\
&&\times\left[1 + \left(2+a\right)\left(\frac{Q^+}{P^+} + \frac{Q^-}{P^-}\right) + \left(2+a\right)^2\frac{Q^2}{P^2} \right.
\nn\\
&&\quad + \left(2+a\right)\frac{28+31a+8a^2} {4\,(5+4a)} \left[\left(\frac{Q^+}{P^+}\right)^2 + \left(\frac{Q^-}{P^-}\right)^2\right.\nn\\
&&\qquad\qquad\qquad\qquad\qquad\qquad\left.
 + (2+a)\left(\frac{Q^2Q^+}{P^2P^+} + \frac{Q^2Q^-}{P^2P^-}\right)\right]\nn\\
&&\left.\quad + \left(2+a\right)^2 \frac{\left(3+a\right) \left(17+8a\right)}{12(5+4a)} \left(\left(\frac{Q^+}{P^+}\right)^3 + \left(\frac{Q^-}{P^-}\right)^3\right)
 + \ldots\right]
\eea
After multiplying this by a factor of $4$ to account for the various ways of attaching the standard model particles (\ref{no-interference}), this agrees with (\ref{mc-cs-series}).

\newpage
\section{Missing energy process: $\phi + \phi \to \phi + \phi +
\un$\label{sec-me}}
\setcounter{equation}{0}

\subsection{Inclusive treatment: general\label{sec-me-general}}

{\figsize\begin{figure}[htb]
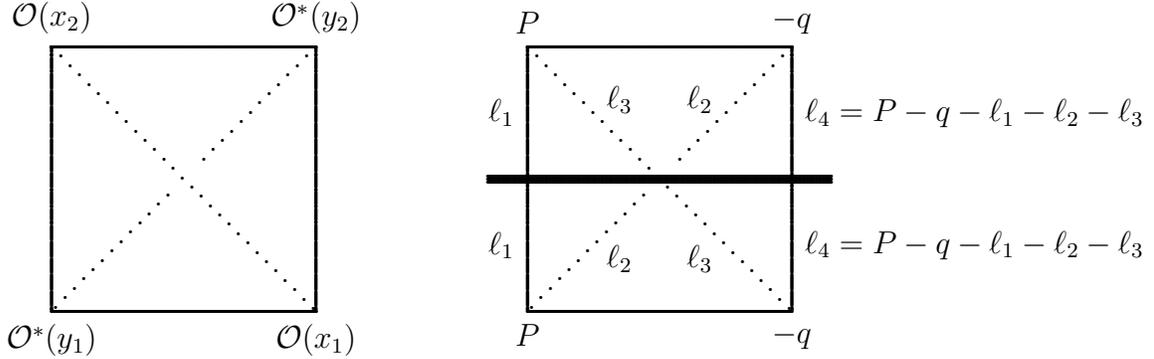

$$\beginpicture
\setcoordinatesystem units <1\tdim,1\tdim>
\stpltsmbl
\plot 50 50 -50 50 -50 -50 50 -50 50 50 /
\setdots
\plot -50 50 50 -50 /
\plot 50 50 6 6 /
\plot -6 -6 -50 -50 /
\put {$\cO(x_1)$} [t] at 50 -55
\put {$\cO^\ast(y_2)$} [b] at 50 55
\put {$\cO^\ast(y_1)$} [t] at -50 -55
\put {$\cO(x_2)$} [b] at -50 55
\linethickness=0pt
\putrule from 0 -60 to 0 60
\putrule from -80 0 to 80 0
\endpicture
\beginpicture
\setcoordinatesystem units <1\tdim,1\tdim>
\stpltsmbl
\plot 50 50 -50 50 -50 -50 50 -50 50 50 /
\plot -65 0 65 0 /
\plot -65 1 65 1 /
\plot -65 -1 65 -1 /
\setdots
\plot -50 50 50 -50 /
\plot 50 50 6 6 /
\plot -6 -6 -50 -50 /
\put {$-q$} [t] at 50 -55
\put {$-q$} [b] at 50 55
\put {$P$} [t] at -50 -55
\put {$P$} [b] at -50 55
\put {$\ell_4=P-q-\ell_1-\ell_2-\ell_3$} [l] at 55 -25
\put {$\ell_3$} [tr] at 20 -25
\put {$\ell_2$} [tl] at -20 -25
\put {$\ell_1$} [r] at -55 -25
\put {$\ell_4=P-q-\ell_1-\ell_2-\ell_3$} [l] at 55 25
\put {$\ell_2$} [br] at 20 25
\put {$\ell_3$} [bl] at -20 25
\put {$\ell_1$} [r] at -55 25
\linethickness=0pt
\putrule from 0 -60 to 0 60
\putrule from -100 0 to 180 0
\endpicture$$
\caption{\figsize\sf\label{fig-me}The 4-point function~(\ref{4-pt}), oriented in a way relevant to the process (\ref{missingenergy}). Momenta $P$ and $-q$ enter the diagram at the bottom and leave it at the top. The thick line is a cut for computing the cross-section.}\end{figure}}

Consider the missing energy process of figure~\ref{fig-1}c,
\be
\phi(p_1) + \phi(p_2) \to \phi(q_1) + \phi(q_2)
+ \mbox{Sommerfield stuff}
\label{missingenergy}
\ee
Because of (\ref{lowenergy}), the Sommerfield dynamics that contributes is associated with the discontinuity across the cut in the $\cO$ 4-point function (\ref{4-pt}) that we now orient as shown in figure~\ref{fig-me}, where we again use the definitions (\ref{PqQ}). The resulting cross-section factor (defined as in (\ref{bar-sigma})) can be computed by (\ref{disc}) as
\bea
\bar\sigma &=& -\int
\Phi(\ell_1)\frac{d^2\ell_1}{(2\pi)^2}\,
\tilde\Phi(\ell_2)\frac{d^2\ell_2}{(2\pi)^2}\,
\tilde\Phi(\ell_3)\,\frac{d^2\ell_3}{(2\pi)^2}
\Phi(\ell_4)\,\frac{d^2\ell_4}{(2\pi)^2}\,
(2\pi)^2\,\delta^2\Bigl(Q-\sum_j\ell_j\Bigr) \nn\\
&&\qquad\times\, i\Delta_{\cO}(P-\ell_1-\ell_2)\, i\Delta_{\cO}(P-\ell_1-\ell_3)
\label{sigma-me-gen}
\eea
where $\Phi(k)$ is the Sommerfield stuff phase space (\ref{full-phase-space}) and $\tilde\Phi(k)$ is the ``phase space''
corresponding to the dotted line (\ref{full-dotted-phase-space}).

In this process it is immediately clear that we have to worry about the IR dynamics of the dotted lines because the momentum carried by them can be light-like. We will not perform the full computation in practice, but will explain how it can be done in principle (numerically) using the procedures discussed in section~\ref{momentum} and exemplified in appendix~\ref{infrared}. However, in subsection~\ref{sec-me-series} we will be able to obtain detailed analytic results by expanding around the limit of small unparticle momentum.

Let us argue, in two different ways, that (\ref{sigma-me-gen}) gives a finite result in the limit that the IR cut-off $\lambda$ in (\ref{full-dotted-phase-space}) goes to zero. For simplicity, consider the low-energy unparticle limit where we can ignore the massive bosons and use the unparticle propagator (\ref{un-prop}) for $i\Delta_\cO$, and the phase spaces
\bea
&&\Phi(\ell) =
\frac{A(a)}{(\xi m)^{2a}}\theta(\ell^0)\,\theta(\ell^2)\left(\ell^2\right)^a \label{phase-space-solid} \\
&&\tilde\Phi(\ell) = \frac{16\pi^4 A(-2-a)(\xi m)^{2a}}{(1+a)^2}\,
 \theta(\ell^0)\,\theta(\ell^2)
\Bigl(f'_\lambda(\ell^2)
+\ell^2\,f''_\lambda(\ell^2)\Bigr)
\label{tilde}
\eea
Each of the integration measures in (\ref{sigma-me-gen}) can be written as
\be
d\ell^0d\ell^1=\frac{ds\,d\ell^1}{2\ell^0}
\ee
where $s=\ell^2$. Now the strategy is to do the integration over the spatial momenta $\ell_j^1$ with the $s_j$ held fixed. For fixed $s_j$, this is just like computing the cross-section for producing four particles with masses-squared $s_j$ with the amplitude-squared $|\cM|^2 \propto h^4\Delta_{\cO}(P-\ell_1-\ell_2) \Delta_{\cO}(P-\ell_1-\ell_3)$. Because this $|\cM|^2$ is a smooth function of the invariants, we expect no rapid dependence of the cross-section as a function of the $s_j$. The point is that while the phase space in (\ref{tilde}) is singular at $\ell^2=0$, the integral over $s$ is finite and well-defined. This remains true when the phase space is multiplied by the smooth matrix element.

There is also another way to see that our strategy should work and that we can remove the IR cut-off at the end, because there is another way to do the calculation. In figure~\ref{fig-ird}, we show a slightly modified diagram in which an additional momentum $k$ is flowing in such a way that the momenta $P$ and $q$ flow only through the two dotted lines. We know from the discussion in section~\ref{momentum} that the appropriate definition of the dotted ``propagator'' is as a derivative. For this momentum routing, we can do the differentiations with respect to $P$ and $q$ and take the derivatives outside the integrals over the loop momenta, and then none of the lines requires an IR cut-off, so the result is clearly independent of $\lambda$. After doing the loop integrations and doing the differentiations, we can set
\be
k=P+q
\ee
to obtain the result.
{\figsize\begin{figure}[htb]
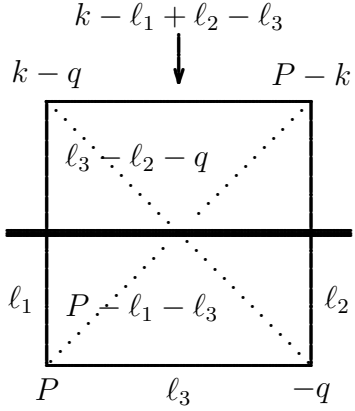

$$\beginpicture
\setcoordinatesystem units <1\tdim,1\tdim>
\stpltsmbl
\plot 50 50 -50 50 -50 -50 50 -50 50 50 /
\plot -65 0 65 0 /
\plot -65 1 65 1 /
\plot -65 -1 65 -1 /
\setdots
\plot -50 50 50 -50 /
\plot 50 50 6 6 /
\plot -6 -6 -50 -50 /
\put {$-q$} [t] at 50 -55
\put {$P-k$} [b] at 50 55
\put {$P$} [t] at -50 -55
\put {$\ell_3$} [t] at 0 -55
\put {$k-q$} [b] at -50 55
\put {$\ell_2$} [l] at 55 -25
\put {$P-\ell_1-\ell_3$} [tl] at -43 -23
\put {$\ell_1$} [r] at -55 -25
\put {$\ell_3-\ell_2-q$} [bl] at -43 23
\put {$k-\ell_1+\ell_2-\ell_3$} [b] at 0 77
\setsolid
\tarrow from 0 75 to 0 57
\linethickness=0pt
\putrule from 0 -60 to 0 60
\putrule from -100 0 to 100 0
\endpicture$$
\caption{\figsize\sf\label{fig-ird}Same as figure~\ref{fig-me} but with a different routing of momentum and an additional external momentum $k$ flowing through the diagram.}\end{figure}}

One may wonder what the result of this calculation will look like. When we took the limit $Q \ll P$ in the missing charge process (section~\ref{sec-mc}) we saw that the phase space reduced to four copies of the unparticle phase space (\ref{un-phase-space}) corresponding to the operator $\cO$. Na\"{\i}vely, this could be attributed to the fact that we took a cut across four solid lines. However, this cannot be interpreted as the production of four units of $\cO$ unparticle stuff because the $\cO$ stuff in the disappearance process of section~\ref{sec-disappearance} has charge $2$, while the putative ``four objects'' in the missing charge process carry a total charge of $4$, not $8$. Instead, it must be interpreted as the production of $\cO^2$ stuff (and stuff corresponding to higher-dimension primary operators in the OPE), that has no relation to $\cO$ stuff. Similarly, the missing energy process that we discuss here cannot be described as the production of two units of stuff corresponding to the solid lines $i\Delta_\cO(x)$ and two units corresponding to the dotted lines $i\tilde\Delta_\cO(x) = 1/[i\Delta_\cO(x)]$. The latter would be unsatisfactory also because the ``propagator'' $i\tilde\Delta_\cO(x)$ would describe an operator with dimension $\tilde d = -1-a < 0$ which is unacceptable from the physical point of view. In the next subsections we will see what actually happens.

\subsection{Inclusive treatment: series expansion\label{sec-me-series}}

{\figsize\begin{figure}[htb]
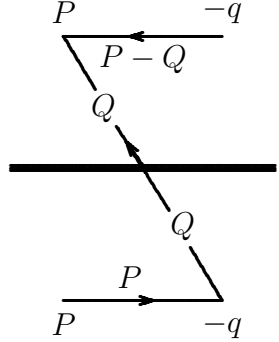

$$\beginpicture
\setcoordinatesystem units <1\tdim,1\tdim>
\stpltsmbl
\plot 30 -50 18 -30 /
\plot 10 -16 -10 16 /
\plot -18 30 -30 50 /
\tarrow from 0 1 to -6 10
\plot -30 50 30 50 /
\tarrow from -0 50 to -5 50
\plot 30 -50 -30 -50 /
\tarrow from -4 -50 to 4 -50
\plot -50 0 50 0 /
\plot -50 1 50 1 /
\plot -50 -1 50 -1 /
\put {$P$} [t] at -30 -55
\put {$-q$} [t] at 30 -55
\put {$P$} [t] at -5 -37
\put {$P$} [b] at -30 55
\put {$-q$} [b] at 30 55
\put {$P-Q$} [b] at 0 36
\put {$Q$} [c] at -15 23
\put {$Q$} [c] at 15 -22
\linethickness=0pt
\putrule from 0 -60 to 0 60
\putrule from -60 0 to 60 0
\endpicture$$
\caption{\figsize\sf\label{fig-me-small-un-mom}A single term in (\ref{me-4pt-expansion}). The $P$ and $P-Q$ lines are described by (\ref{FT-upzpzm}) while the cut through the $Q$ line gives the phase space factor (\ref{me-phase-space-factor}).}\end{figure}}
Similarly to what we did in subsection~\ref{sec-mc-series}, we can easily obtain analytic results by expanding around the limit where the fraction of the momentum that goes into the unparticle stuff is small: $Q \ll P$. In this limit the contribution should come mainly from configurations of figure~\ref{fig-me} in which points $x_i$, $y_j$ with $i = j$ are much closer to each other than points with $i\neq j$. Denoting
\be
\zeta_i \equiv y_i-x_i\,,\qquad
X \equiv x_2-x_1
\label{zeta-i}
\ee
we expand the 4-point function~(\ref{4-pt}) in small $\zeta_i/X$:
\bea
&&\vev{\cO(x_1)\cO(x_2)\cO^\ast(y_1)\cO^\ast(y_2)} \nn\\
&&= i\Delta_\un(\zeta_1)\,i\Delta_\un(\zeta_2)\left[1 + (1+a)\frac{\zeta_1^-\zeta_2^-}{(X^-)^2}
+ (1+a)\frac{\zeta_1^-\zeta_2^-(\zeta_1^- - \zeta_2^-)}{(X^-)^3}
\right.\nn\\
&&\quad\qquad\qquad\qquad\qquad
+ (1+a)\frac{\zeta_1^-\zeta_2^-\left(2{\zeta_1^-}^2 - (2-a)\,\zeta_1^-\zeta_2^- + 2{\zeta_2^-}^2\right)}{2(X^-)^4} \nn\\
&&\quad\qquad\qquad\qquad\qquad
+ (1+a)^2\frac{\zeta_1^2\zeta_2^2}{(X^2)^2}
+ (1+a)^2\frac{\zeta_1^2\zeta_2^2(\zeta_2^- - \zeta_1^-)}{(X^-)^3(X^+)^2}\nn\\
&&\quad\qquad\qquad\qquad\qquad
+ (1+a)\frac{{\zeta_1^-}^4\zeta_2^- - \zeta_1^-{\zeta_2^-}^4 - \left(1-a\right){\zeta_1^-}^2{\zeta_2^-}^2\left(\zeta_1^- - \zeta_2^-\right)}{(X^-)^5}\nn\\
&&\left.\quad\qquad\qquad\qquad\qquad
+\; \{-\to+ \mbox{ for all asymmetric terms}\}
+ \ldots\tall\right]
\label{me-4pt-expansion}
\eea
Written in this way, each term describes a diagram as in figure~\ref{fig-me-small-un-mom} in which momentum $Q$ flows through $X$, momentum $-P$ through $\zeta_1$, and momentum $-(P-Q)$ through $\zeta_2$. For the $\zeta$-dependent factors, note that the Fourier transform of $i\Delta_\un(\zeta)(\zeta^-)^m(\zeta^+)^n$ is
\bea
\left(\frac{2}{i}\frac{\pd}{\pd p^+}\right)^m \left(\frac{2}{i}\frac{\pd}{\pd p^-}\right)^n
i\Delta_\un(p)
&=& \left(\frac{2}{i}\right)^{m+n}\,
\prod_{k=0}^{m-1}(a-k) \prod_{k=0}^{n-1}(a-k)
\frac{i\Delta_\un(p)}{(p^+)^m(p^-)^n} \label{FT-upzpzm}\\
&=& \left(\frac{2}{i}\right)^{m+n}\,
\frac{[\Gamma(1+a)]^2}{\Gamma(1+a-m)\,\Gamma(1+a-n)}
\frac{i\Delta_\un(p)}{(p^+)^m(p^-)^n}\nn
\eea
To obtain the contribution of a term in (\ref{me-4pt-expansion}) to the cross-section we take the product of two factors of the form (\ref{FT-upzpzm}) coming from the $\zeta_1$ and $\zeta_2$ lines and multiply it by the discontinuity across the cut through $X$. Apart from the first term that describes a disconnected diagram,\footnote{As mentioned before, the expression we use, (\ref{free-n-pt}), does include diagrams that are disconnected in limits where the $C(x)$ factors in (\ref{n-pt-C-factors}) do not connect them. Here the solid and dotted lines that connect the top and the bottom parts of the diagram cancel each other in position space in the limit $y_1\to x_1, y_2\to x_2$. We will mention the disconnected term again later in this section.} all the terms in the expansion have the form
\be
\frac{1}{(X^-)^M(X^+)^N}
\label{1/XmXp}
\ee
where $M$ and $N$ are integers. They describe the production of some massless stuff.\footnote{Kinematically this looks like $M$ right-moving and $N$ left-moving free massless fermions, see (\ref{free-fermion-1})--(\ref{free-fermion-2}). But since the $M$  ``particles'' are massless and collinear (and similarly the $N$ ``particles''), their exact number is uncertain. Furthermore, some of the $1/X^\pm$ factors come from the momentum dependence of the interactions rather than describe an additional particle (we have already seen this happening in the missing charge process in section~\ref{sec-mc}). Even though terms in which both $M$ and $N$ are non-zero will describe states with mass $Q^2 > 0$, there will be no dependence on a fractional power like we had for the unparticle stuff discussed in the previous sections.} For finding the discontinuities across the cut, consider, for example, the $1/(X^-)^2$ term. We compute its Fourier transform as
\be
\int d^2X\,e^{iQX}\frac{(X^+)^2}{(-X^2+i\epsilon)^2}
= \left(\frac{2}{i}\frac{\pd}{\pd Q^-}\right)^2 \int d^2X\,\frac{e^{iQX}}{(-X^2+i\epsilon)^2}
= -\pi\frac{i(Q^+)^2}{Q^2 + i\epsilon}
\ee
where we used
\bea
\lim_{\delta\to0} \int d^2X\,\frac{e^{iQX}}{(-X^2+i\epsilon)^{2+\delta}}
&=& -i\frac{\pi}{4}\, \lim_{\delta\to0}\, \frac{\Gamma(-1-\delta)}{2^{2\delta}\,\Gamma(2+\delta)}(-Q^2-i\epsilon)^{1+\delta} \nn\\
&=& -i\frac{\pi}{4}\, (-Q^2-i\epsilon)\ln\left(\frac{-Q^2-i\epsilon}{c}\right)
\label{FT-2}
\eea
where $c$ is a ($\delta$-dependent) number. This gives the phase space factor
\be
-2\pi^2\,\theta(Q^+)(Q^+)^2\delta(Q^2)
\ee
In the same way, we can analyze terms (\ref{1/XmXp}) with arbitrary $M$ and $N$. Similarly to (\ref{FT-2}),
\be
\int d^2X\,\frac{e^{iQX}}{(X^2-i\epsilon)^M} = -\frac{i\pi}{\left(2^{M-1}(M-1)!\right)^2} \left(-Q^2-i\epsilon\right)^{M-1}\ln\left(\frac{-Q^2-i\epsilon}{c_M}\right)
\ee
where $c_M$ is a regulator-dependent number, so the Fourier transform of terms with $N = 0$ (and analogously terms with $M = 0$) is
\be
\left(\frac{2}{i}\frac{\pd}{\pd Q^-}\right)^M \int d^2X\,\frac{e^{iQX}}{(X^2-i\epsilon)^M}
= \frac{i^{M+1}\,\pi}{2^{M-2}(M-1)!}\; \frac{(Q^+)^M}{Q^2 + i\epsilon}
\label{FT-integer}
\ee
and the resulting phase space factor is
\be
\frac{i^M\,\pi^2}{2^{M-3}(M-1)!}\,\left(Q^+\right)^M\theta(Q^+)\,\delta(Q^2)
\label{me-phase-space-factor-N=0}
\ee
For terms with $M \geq N > 0$ (and similarly $N \geq M > 0$), the Fourier transform is
\bea
&&\left(\frac{2}{i}\frac{\pd}{\pd Q^-}\right)^{M-N} \int d^2X\,\frac{e^{iQX}}{(X^2-i\epsilon)^M} = \\
&&\frac{i^{M+N+1}\,\pi}{2^{M+N-2}(M-1)!(N-1)!} \left(Q^+\right)^{M-1}\left(Q^-\right)^{N-1} \ln\left(\frac{-Q^2-i\epsilon}{c_M}\right) + d_{M,N}\frac{\left(Q^+\right)^M \left(Q^-\right)^N}{Q^2 - i\epsilon}\nn
\eea
where the term with the unimportant prefactor $d_{M,N}$ will not contribute to the imaginary part. The resulting phase space factor is\footnote{Alternatively, notice that (\ref{1/XmXp}) can be represented as the $a\to -1$ limit of (\ref{mc-term}) with $\xi m = 2\pi$, and then (\ref{mc-phase-space-factor}) reduces to (\ref{me-phase-space-factor}).}
\be
\frac{i^{M+N}\pi^2}{2^{M+N-3}(M-1)!(N-1)!} \left(Q^+\right)^{M-1}\left(Q^-\right)^{N-1}\theta(Q^0)\,\theta(Q^2)
\label{me-phase-space-factor}
\ee
Summing the contributions of all the terms in (\ref{me-4pt-expansion}) and adding an overall minus sign from (\ref{disc}), we obtain
\bea
&&\bar\sigma = 8\pi^2 a^2(1+a)\left[\Delta_\un(P)\right]^2 \nn\\
&&\qquad\times
\left[\left(\frac{Q^+}{P^+}\right)^2
+ (1-a)\left(\frac{Q^+}{P^+}\right)^3
+ (1-a)\left(1 - \frac{a(7+a)}{12}\right)\left(\frac{Q^+}{P^+}\right)^4\right.\nn\\
&&\qquad\qquad\left.
+ (1-a)\left(1-\frac{a}{2}\right)\left(1-\frac{a(3+a)}{6}\right) \left(\frac{Q^+}{P^+}\right)^5
+ \ldots\right]\theta(Q^+)\,\delta(Q^2)
\label{missing-energy-result}\\
&&\qquad\quad + \,\{+\to-\}\nn\\
&&\qquad +\, 8\pi^2 a^4(1+a)^2\frac{\left[\Delta_\un(P)\right]^2}{P^2}\frac{Q^2}{P^2}\left[1 + (1-a)\left(\frac{Q^+}{P^+} + \frac{Q^-}{P^-}\right) + \ldots \right] \theta(Q^0)\,\theta(Q^2)\nn
\eea
Nicely, the IR divergences do not appear in this approach.

{\figsize\begin{figure}[htb]
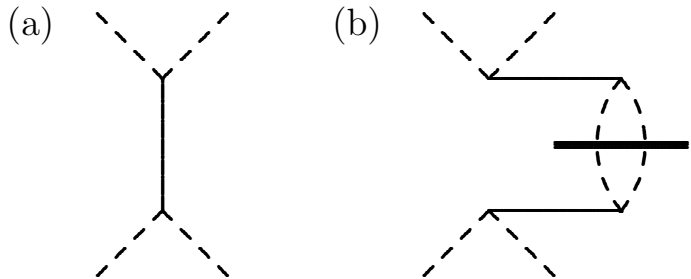

$$\beginpicture
\setcoordinatesystem units <1\tdim,1\tdim>
\stpltsmbl
\plot 0 -25 0 25 /
\setdashes
\plot 0 -25 -25 -50 /
\plot 0 -25 25 -50 /
\plot 0 25 -25 50 /
\plot 0 25 25 50 /
\put {\large{(a)}} at -50 45
\linethickness=0pt
\putrule from 0 -50 to 0 50
\putrule from -70 0 to 30 0
\endpicture
\qquad
\beginpicture
\setcoordinatesystem units <1\tdim,1\tdim>
\stpltsmbl
\plot 0 25 50 25 /
\plot 0 -25 50 -25 /
\plot 25 0 75 0 /
\plot 25 1 75 1 /
\plot 25 -1 75 -1 /
\setdashes
\plot 0 -25 -25 -50 /
\plot 0 -25 25 -50 /
\plot 0 25 -25 50 /
\plot 0 25 25 50 /
\circulararc 85 degrees from 50 25 center at 80 0
\circulararc -85 degrees from 50 25 center at 20 0
\put {\large{(b)}} at -50 45
\linethickness=0pt
\putrule from 0 -50 to 0 50
\putrule from -70 0 to 30 0
\endpicture$$
\caption{\figsize\sf\label{fig-no-me}The amplitude (a) and the cross-section (b) of the process (\ref{no-me}). The solid lines are the unparticle propagators.}\end{figure}}

We would like to note that while the cross-section (\ref{missing-energy-result}) vanishes in the limit of either $a \to 0$ or $Q \to 0$, there is also another process, without missing energy, whose amplitude contains just the unparticle 2-point function as shown in figure~\ref{fig-no-me}a:
\be
\phi + \phi \to \phi + \phi
\label{no-me}
\ee
with
\be
\bar\sigma = 2(2\pi)^2 \left[\Delta_\un(P)\right]^2 \theta(Q^0)\,\delta^2(Q)
\label{no-me-cs}
\ee
Figure~\ref{fig-no-me}b shows the diagram that is related to the cross-section of this process by the optical theorem. In fact, the 4-point function knows about this diagram: it is described by the disconnected term in (\ref{me-4pt-expansion}). The contribution (\ref{no-me-cs}) is hiding behind the IR divergences in section~\ref{sec-me-general}, similarly to the example analyzed in appendix~\ref{infrared}.

{\figsize\begin{figure}[htb]
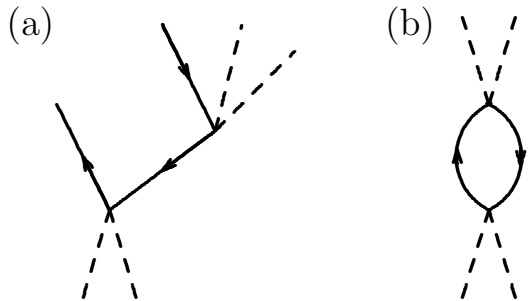

$$\beginpicture
\setcoordinatesystem units <1\tdim,1\tdim>
\stpltsmbl
\setdashes
\plot -10 -33 0 0 /
\plot 10 -33 0 0 /
\setsolid
\plot 0 0 -20 40 /
\tarrow from 0 0 to -10 20
\plot 0 0 40 30 /
\tarrow from 40 30 to 20 15
\plot 40 30 20 70 /
\tarrow from 20 70 to 30 50
\setdashes
\plot 40 30 50 70 /
\plot 40 30 70 60 /
\put {\large{(a)}} at -30 70
\linethickness=0pt
\putrule from 0 -35 to 0 75
\putrule from -40 0 to 80 0
\endpicture
\qquad
\beginpicture
\setcoordinatesystem units <1\tdim,1\tdim>
\stpltsmbl
\setdashes
\plot -10 -33 0 0 /
\plot 10 -33 0 0 /
\plot 0 40 -10 73 /
\plot 0 40 10 73 /
\setsolid
\circulararc 130 degrees from 0 0 center at -10 20
\circulararc -130 degrees from 0 0 center at 10 20
\tarrow from -12 17 to -12 23
\tarrow from 12 23 to 12 17
\put {\large{(b)}} at -30 70
\linethickness=0pt
\putrule from 0 -35 to 0 80
\putrule from -40 0 to 20 0
\endpicture$$
\caption{\figsize\sf\label{fig-me-free}(a) The amplitude for the missing energy processes in the free-fermion limit, eq.~(\ref{free-fermion-process}). The solid lines are $\psi$. (b) Similarly, the free-fermion limit of the process of figure~\ref{fig-no-me}a.}\end{figure}}

In the case $a=0$, Sommerfield model reduces to a theory of free fermions and the interaction (\ref{lowenergy}) describes the missing energy processes
\be
\phi + \phi \to \phi + \phi + \psi_1 + \bar\psi_1\,,\qquad
\phi + \phi \to \phi + \phi + \psi_2 + \bar\psi_2
\label{free-fermion-process}
\ee
as shown in figure~\ref{fig-me-free}a (while the process (\ref{no-me}) reduces to figure~\ref{fig-me-free}b). After integrating over the phase space of the fermions, for the first process we get
\be
|\cM|^2\,\Phi_{\psi_1\bar\psi_1} = \frac{(Q^+)^2}{2(P^+)^2}\frac{1}{1 - Q^+/P^+}\,\theta(Q^+)\,\delta(Q^2)
\ee
and the second process of (\ref{free-fermion-process}) is described by the same expression with $\{+\to-\}$. This is consistent with (\ref{missing-energy-result}) which for $a=0$ reduces to
\be
\bar\sigma = \frac{(Q^+)^2}{2(P^+)^2}\left[1 + \frac{Q^+}{P^+} + \left(\frac{Q^+}{P^+}\right)^2  + \left(\frac{Q^+}{P^+}\right)^3 + \ldots\right]\theta\left(Q^+\right) \delta\left(Q^2\right) + \{+\to-\}
\ee
On the other hand, the missing charge process from section~\ref{sec-mc} does not have a free-theory counterpart, which is consistent with the fact that the cross-section (\ref{sigma-missing-charge}) vanishes for $a=0$.

\subsection{Interpretation in terms of exclusive processes\label{sec-me-interpretation}}

We can interpret the expansion~(\ref{me-4pt-expansion}) that we did in the $\zeta_i \ll X$ limit by replacing the products $\cO^\ast(y_1)\,\cO(x_1)$ and $\cO^\ast(y_2)\,\cO(x_2)$ in the 4-point function by their OPEs
\bea
&&\mbox{T}\,\cO^\ast(y_i)\,\cO(x_i)
= i \Delta_\un(\zeta_i) + \sum_k c_k(\zeta_i)\,\cO_k(x_i)
\label{OPE-OO*}
\eea
It can be shown that in the low-energy effective theory describing the unparticle limit, the leading operators $\cO_k(x)$ that appear in this OPE are the components of the current $j^\mu$, products of several such operators (which include the stress-energy tensor $T^{\mu\nu}$), and their derivatives (for more details, see appendix~\ref{sec-operators} where this current is denoted $j_T^\mu$). For example, the $1/(X^-)^2$ and $1/(X^+)^2$ terms in~(\ref{me-4pt-expansion}) come from the two-point functions of $j^\pm$:
\be
\vev{j^\pm(X)j^\pm(0)} = -\frac{1+a}{\pi^2}\frac{1}{\left(X^\mp\right)^2}
\ee
The resulting contribution to the cross-section corresponds to producing the stuff that is created out of the vacuum by the action of $j^\pm$:
\be
\phi + \phi \to \phi + \phi + \{\mbox{$j^\pm$ stuff}\}
\ee
This stuff looks massless because $j^\pm$ has an integer dimension. Using the notation (\ref{(a|b)}), the expansion~(\ref{me-4pt-expansion}) can be written as
\bea
&&\frac{\vev{\cO(x_1)\cO(x_2)\cO^\ast(y_1)\cO^\ast(y_2)}}{i\Delta_\un(\zeta_1)\,i\Delta_\un(\zeta_2)} = \nn\\
&& 1 - \pi^2\zeta_1^-\zeta_2^-\left[\langle j^+|j^+ \rangle
+ \frac{\zeta_1^-\langle j^+|\pd_-j^+\rangle + \zeta_2^-\langle \pd_-j^+|j^+\rangle}{2}
+ \frac{\zeta_1^-\zeta_2^-\langle\pd_-j^+|\pd_-j^+\rangle}{4} \right. \nn\\
&&\qquad\qquad\qquad
+ \frac{{\zeta_1^-}^2\langle j^+|\pd_-^2j^+\rangle + {\zeta_2^-}^2\langle \pd_-^2j^+|j^+\rangle}{6}
+ \frac{{\zeta_1^-}^3\langle j^+|\pd_-^3j^+\rangle + {\zeta_2^-}^3 \langle\pd_-^3j^+|j^+\rangle}{24} \nn\\
&&\qquad\qquad\qquad\left.
+ \frac{{\zeta_1^-}^2\zeta_2^-\langle \pd_-j^+|\pd_-^2j^+\rangle + \zeta_1^-{\zeta_2^-}^2 \langle\pd_-^2j^+|\pd_-j^+\rangle}{12} + \ldots\right]\nn\\
&&\quad + \frac{(1+a)^2\pi^2}{4}\,{\zeta_1^-}^2{\zeta_2^-}^2\left[ \langle T^{++}|T^{++}\rangle + \frac{\zeta_1^-\langle T^{++}|\pd_-T^{++}\rangle + \zeta_2^- \langle\pd_-T^{++}|T^{++}\rangle}{2} + \ldots\right]
\nn\\
&&\quad +\, \pi^4\zeta_1^2\zeta_2^2\left[\langle j^2|j^2\rangle
+ \frac{\zeta_1^-\langle j^2|\pd_-j^2\rangle + \zeta_2^-\langle \pd_-j^2|j^2\rangle}{2} + \ldots\right] \nn\\
&&\quad
+ \;\{ + \leftrightarrow - \mbox{ for all asymmetric terms}\} + \ldots
\label{me-4pt-expansion-OPE}
\eea
This is in agreement with the OPE
\bea
&&\mbox{T}\,\cO^\ast(\zeta)\,\cO(0)\nn\\
&&= i\Delta_\un(\zeta)\left[ 1
+ i\pi\left(\zeta^+ j^- - \zeta^- j^+
+ \frac{(\zeta^+)^2\pd_+ j^- - (\zeta^-)^2\pd_- j^+}{2} \right.\right.\nn\\
&&\qquad\qquad\qquad\qquad\left.
+\, \frac{(\zeta^+)^3\pd_+^2 j^- - (\zeta^-)^3\pd_-^2 j^+}{6}
+ \frac{(\zeta^+)^4\pd_+^3 j^- - (\zeta^-)^4\pd_-^3 j^+}{24}
+ \ldots\right) \nn\\
&&\qquad\qquad\quad
-\, (1+a)\frac{\pi}{2} \left(\tall(\zeta^+)^2\, T^{--} + (\zeta^-)^2\, T^{++}
+ \frac{(\zeta^+)^3\,\pd_+T^{--} + (\zeta^-)^3\,\pd_-T^{++}}{2} + \ldots\right)\nn\\
&&\qquad\qquad\quad\left.
+\, \pi^2 \zeta^+\zeta^- \left(j^2 + \frac{\zeta^+\pd_+j^2 + \zeta^-\pd_-j^2}{2} + \ldots\right)
+ \ldots\right]
\label{OO*OPE}
\eea
where the operators on the r.h.s. are evaluated at position $0$. We can verify that the contribution of the derivatives is exactly the way it must be to respect the conformal symmetry: the contribution of $j^2$ (a scalar with dimension $\Delta = 2$) follows (\ref{c-block-scalar}). An analogous expression for the conserved current $j^\mu$ (which always has dimension $D-1$ in $D$ spacetime dimensions) is~\cite{Lang:1992pp,Arutyunov:2000ku}
\bea
&& \frac{\zeta_\mu}{\left(\zeta^2\right)^{d-D/2+1}}
\sum_{m=0}^\infty \frac{\Gamma(D/2)\left(-\zeta^2\,\Box\right)^m}{4^m\,m!\,\Gamma(D/2+m)}
\int_0^1 dt \left[t(1-t)\right]^{D/2+m-1} e^{t\,\zeta\cdot\pd}\nn\\
&&\propto \frac{\zeta_\mu}{\left(\zeta^2\right)^{d-D/2+1}}
\left[1 + \frac{\zeta\cdot\pd}{2}
+ \frac{D+2}{8\,(D+1)}(\zeta\cdot\pd)^2
- \frac{\zeta^2\,\Box}{8\,(D+1)} \right. \nn\\
&&\qquad\qquad\qquad\quad\left.
+ \frac{D+4}{48\,(D+1)}(\zeta\cdot\pd)^3
- \frac{\zeta^2\,(\zeta\cdot\pd)\,\Box}{16\,(D+1)}
+ \ldots \right]
\label{c-block-current}
\eea
which agrees with the $j^\mu$ terms in (\ref{OO*OPE}), if we take into account that in 2D the conservation of the vector and axial currents implies $\pd_+ j^+ = \pd_- j^- = 0$.\footnote{More precisely, in order to account for the relative signs between the $j^+$ and $j^-$ terms in (\ref{OO*OPE}), we should say that (\ref{c-block-current}) describes the contribution of the axial current $j^{5\mu} = -\epsilon^{\mu\nu}j_\nu$ rather than $j^\mu$.} Similarly, for the stress-energy tensor $T^{\mu\nu}$ (which always has dimension $D$) we have the expression~\cite{Arutyunov:2000ku}
\bea
&& \frac{\zeta_\mu\zeta_\nu}{\left(\zeta^2\right)^{d-D/2+1}}
\sum_{m=0}^\infty \frac{\Gamma(D/2+1)\left(-\zeta^2\,\Box\right)^m}{4^m\,m!\,\Gamma(D/2+1+m)}
\int_0^1 dt \left[t(1-t)\right]^{D/2+m} e^{t\,\zeta\cdot\pd}\nn\\
&&\propto \frac{\zeta_\mu\zeta_\nu}{\left(\zeta^2\right)^{d-D/2+1}}
\left(1 + \frac{\zeta\cdot\pd}{2}
+ \ldots \right)
\eea
which is consistent with (\ref{OO*OPE}).

We can also use (\ref{Dolan-Osborn}) with
\be
x_{12} = \zeta_1\,,\qquad
x_{34}=\zeta_2\,,\qquad
x_{24} = -X\,,\qquad
x_{13} = -X + \zeta_1 - \zeta_2
\ee
to verify the contributions of the various operators directly to (\ref{me-4pt-expansion}). The contribution of $j^\mu$ ($\ell = 1$, $\Delta = 1$) should be proportional to
\bea
&&-i\Delta_\un(\zeta_1)\,i\Delta_\un(\zeta_2)
\left[\ln(1-\eta^-) + \ln(1-\eta^+)\right]\nn\\
&&= i\Delta_\un(\zeta_1)\,i\Delta_\un(\zeta_2)
\frac{\zeta_1^-\zeta_2^-}{\left(X^-\right)^2}\left[1
+ \frac{\zeta_1^--\zeta_2^-}{X^-}
+ \frac{{\zeta_1^-}^2 - \frac{3}{2}\zeta_1^-\zeta_2^- + {\zeta_2^-}^2}{\left(X^-\right)^2}\right. \\
&&\qquad\qquad\qquad\qquad\qquad\qquad\left.
+ \frac{\left(\zeta_1^--\zeta_2^-\right)\left({\zeta_1^-}^2 - \zeta_1^-\zeta_2^- + {\zeta_2^-}^2\right)}{\left(X^-\right)^3}
+ \ldots \right]
+ \{-\to+\}\nn
\eea
which exactly reproduces the terms that are attributed to $j^+$ or $j^-$ (and their derivatives) in (\ref{me-4pt-expansion-OPE}). The contribution of $T^{\mu\nu}$ ($\ell = 2$, $\Delta = 2$) is proportional to
\bea
&& -12\,i\Delta_\un(\zeta_1)\,i\Delta_\un(\zeta_2)
\left[1 + \left(\frac{1}{\eta^-} - \frac{1}{2}\right)\ln(1-\eta^-)\right] + \{-\to+\}\nn\\
&& =
i\Delta_\un(\zeta_1)\,i\Delta_\un(\zeta_2)\,
\frac{{\zeta_1^-}^2 {\zeta_2^-}^2}{(X^-)^4}
\left[1
+ \frac{2(\zeta_1^- - \zeta_2^-)}{X^-} + \ldots\right] + \{-\to+\}
\eea
which agrees with what comes from the $T^{\pm\pm}$ terms in (\ref{me-4pt-expansion-OPE}). The contribution of $j^2$ ($\ell = 0$, $\Delta = 2$) is proportional to
\bea
&&i\Delta_\un(\zeta_1)\,i\Delta_\un(\zeta_2)\,
\ln(1-\eta^-)\ln(1-\eta^+)\nn\\
&&= i\Delta_\un(\zeta_1)\,i\Delta_\un(\zeta_2)\,
\frac{\zeta_1^2\zeta_2^2}{(X^2)^2}\left[1 + \frac{\zeta_1^- - \zeta_2^-}{X^-} + \frac{\zeta_1^+ - \zeta_2^+}{X^+} + \ldots\right]
\eea
which agrees with the $j^2$ terms from (\ref{me-4pt-expansion-OPE}). To summarize, in terms of $K_{\ell,\Delta}$ from (\ref{Dolan-Osborn}), (\ref{me-4pt-expansion}) is exactly reproduced by
\be
1 + \left(1+a\right)K_{1,1} + \frac{(1+a)^2}{2}\,K_{2,2} + \frac{(1+a)^2}{2}\,K_{0,2}
\ee
up to order $(\zeta/X)^5$. In order to account for even higher order terms that are not written explicitly in (\ref{me-4pt-expansion}) more operators are required. For example, we find that in order to reach terms of order $(\zeta/X)^{11}$ we need to take
\bea
&& 1 + \left(1+a\right)K_{1,1} + \frac{(1+a)^2}{2}\,K_{2,2} + \frac{(1+a)^2}{2}\,K_{0,2} + \frac{(1+a)^3}{3!}\,K_{3,3} + \frac{(1+a)^3}{2}\,K_{1,3} \nn\\
&&\; + \left[\frac{(1+a)^4}{4!} + \frac{(1+a)^2}{5!}\right]K_{4,4} + \frac{(1+a)^4}{6}\,K_{2,4} + \frac{(1+a)^4}{8}K_{0,4} \\
&&\; + \left[\frac{(1+a)^5}{5!} + \frac{(1+a)^3}{7\cdot 4!}\right]K_{5,5} + \left[\frac{(1+a)^5}{4!} + \frac{(1+a)^3}{5!}\right]K_{3,5} + \frac{(1+a)^5}{12}K_{1,5} \nn
\eea

In the free-fermion limit discussed at the end of subsection~\ref{sec-me-series}, operators that have, besides the $i\Delta_\un(\zeta)$, both $\zeta^+$ and $\zeta^-$ in their OPE coefficient  (e.g., $j^2$), will not contribute to the cross-section because (\ref{FT-upzpzm}) vanishes for $a\to0$ unless $m=0$ or $n=0$. On the other hand, all the operators with $\Delta = \ell$ (including $j^\mu$, $T^{\mu\nu}$, and others) and their derivatives contribute to the production of the fermions in (\ref{free-fermion-process}). In the free theory we have $j^+(x) = 2\,\psi_1^\ast(x)\psi_1(x)$, $j^-(x) = 2\,\psi_2^\ast(x)\psi_2(x)$, so it is clear why the operators $j^\pm$ produce the fermion pair. However, these operators produce the two fermions at the same spacetime point. Processes in which the two internal vertices in figure~\ref{fig-me-free}a are at a finite distance from each other are described by the additional operators that include both derivatives of $j^\mu$ as well as other primary operators that appear in the OPE of two $\psi\,$s.

\subsection{Exclusive treatment: amputated 3-point functions\label{sec-me-amputated}}

Like in section~\ref{sec-mc-amputated}, we can verify that our inclusive results agree with exclusive computations that are based on amputated 3-point functions. For example, for the process
\be
\phi + \phi \to \phi + \phi + \{\mbox{$j^2$ stuff}\}
\ee
we know that
\be
\vev{j^2(x)j^2(0)} = \frac{(1+a)^2}{\pi^4(x^2)^2}
\ee
\be
\vev{j^2(x)\,\cO^\ast(y')\,\cO(y)}
= \frac{(1+a)^2}{\pi^2} \frac{(y'-y)^2i\Delta_\un(y'-y)}{(x-y)^2(x-y')^2}
\ee
i.e.,
\be
\Delta = 2\,,\qquad
C_2 = \frac{(1+a)^2}{\pi^4}\,,\qquad
C_3 = \frac{(1+a)^2}{4\pi^4 (\xi m)^{2a}}
\ee
and then (\ref{sigma-contrib-gen-leading}) gives
\be
|\cM_{j^2}|^2\,\Phi_{j^2} = \frac{(1+a)^2\,[\Gamma(1-a)]^2} {2^{1+4a}\, [\Gamma(a)]^2\,(\xi m)^{4a}}\; \frac{Q^2}{\left(P^2\right)^{2-2a}}\, \theta(Q^0)\,\theta(Q^2)
\left[1 + \left(1-a\right)\left(\frac{Q^+}{P^+} + \frac{Q^-}{P^-}\right) + \ldots\right]
\ee
in agreement with the $j^2$ contribution to (\ref{missing-energy-result}).

For the process
\be
\phi + \phi \to \phi + \phi + \{\mbox{$j^\mu$ stuff}\}
\ee
we need a generalization of the analysis in section~\ref{sec-self-int} to the case where the amputated operator is a vector. The 3-point function, that is the analog of (\ref{3pt-scalar-2same}), is
\be
\vev{\cO_1(x_1)\cO_2(x_2)\cO^\mu(x_3)}
\propto \frac{\displaystyle \frac{x_{13}^\mu}{-x_{13}^2+i\epsilon} - \frac{x_{23}^\mu}{-x_{23}^2+i\epsilon}}
{\left(-x_{12}^2+i\epsilon\right)^{d - (\Delta - 1)/2}
\left(-x_{13}^2+i\epsilon\right)^{(\Delta - 1)/2}
\left(-x_{23}^2+i\epsilon\right)^{(\Delta - 1)/2}}
\label{3pt-vector-2same}
\ee
The appropriate analog of the D'EPP formula (\ref{D'EPP}) says that for $\delta_1 + \delta_2 + \delta_3 + 1 = D$~\cite{Fradkin:1997df}
\bea
&&\int d^Dx_4\, \left(\delta^\mu_\nu - 2\frac{x_{43}^\mu x_{43\,\nu}}{x_{43}^2}\right) \left(\frac{x_{14}^\nu}{x_{14}^2} - \frac{x_{24}^\nu}{x_{24}^2}\right)\frac{\Gamma(\delta_1+1)\,\Gamma(\delta_2+1)\,\Gamma(\delta_3+1)} {\left(x_{14}^2\right)^{\delta_1} \left(x_{24}^2\right)^{\delta_2} \left(x_{43}^2\right)^{\delta_3}} \nn\\
&& = \pi^{D/2}\left(\frac{x_{13}^\mu}{x_{13}^2} - \frac{x_{23}^\mu}{x_{23}^2}\right) \frac{(\delta_1+\delta_2)\,\Gamma\left(\frac{D}{2}-\delta_3\right) \Gamma\left(\frac{D}{2}-\delta_2\right) \Gamma\left(\frac{D}{2}-\delta_1\right)}{\left(x_{12}^2\right)^{D/2 - \delta_3} \left(x_{13}^2\right)^{D/2 - \delta_2 - 1} \left(x_{23}^2\right)^{D/2 - \delta_1 - 1}}
\label{D'EPP-1}
\eea
For solving (\ref{amputated}) we use this formula as
\bea
&&\frac{1}{\left(x_{12}^2\right)^{d - (\tilde\Delta-1)/2}}
\int d^Dx_4\, \left(\delta^\mu_\nu - 2\frac{x_{43}^\mu x_{43\,\nu}}{x_{43}^2}\right) \left(\frac{x_{14}^\nu}{x_{14}^2} - \frac{x_{24}^\nu}{x_{24}^2}\right)\frac{\left[\Gamma\left(\frac{\tilde\Delta + 1}{2}\right)\right]^2\,\Gamma(\Delta+1)} {\left(x_{14}^2\right)^{(\tilde\Delta - 1)/2} \left(x_{24}^2\right)^{(\tilde\Delta - 1)/2} \left(x_{43}^2\right)^{\Delta}} \nn\\
&& = \pi^{D/2}\left(\frac{x_{13}^\mu}{x_{13}^2} - \frac{x_{23}^\mu}{x_{23}^2}\right) \frac{(\tilde\Delta-1)\,\Gamma\left(\frac{D}{2}-\Delta\right) \left[\Gamma\left(\frac{\Delta + 1}{2}\right)\right]^2}{\left(x_{12}^2\right)^{d - (\Delta - 1)/2} \left(x_{13}^2\right)^{(\Delta - 1)/2} \left(x_{23}^2\right)^{(\Delta - 1)/2}}
\eea
Since in our case $\tilde\Delta = \Delta$, the amputated 3-point function is proportional to the ordinary 3-point function, with the prefactor fixed by (\ref{j2pt}):
\be
Q^\mu(x_3|x_1,x_2) = -\frac{2\pi}{1+a}\,\vev{\cO^\ast(x_1)\,\cO(x_2)\,j^\mu(x_3)}
\label{Qmu}
\ee
The ordinary 3-point function (\ref{j3ptord}) in momentum space is
\be
\langle \cO^\ast(-P)\,\cO(P-Q)\,j^\pm(Q)\rangle = \pm 2(1+a)\frac{i Q^\pm}{Q^2 + i\epsilon}\left[i\Delta_\un(P) - i\Delta_\un(P-Q)\right]
\label{j3pt-mom}
\ee
Thus the amplitude for the production of $j^\pm$ stuff is
\be
\cM_{j^\pm} = \mp\, i \frac{2\pi}{Q^\pm}\left[i\Delta_\un(P) - i\Delta_\un(P-Q)\right]
\label{Mjpm}
\ee

Alternatively, instead of using the D'EPP formula, we could do the amputation of $j^\mu$ directly in momentum space. We obtain the amputated 3-point function $iQ_\pm$ by dividing the ordinary 3-point function (\ref{j3pt-mom}) by the propagator
\be
\langle j^\pm(-Q)\,j^\pm(Q)\rangle
= i\frac{1+a}{\pi}\,\frac{\left(Q^\pm\right)^2}{Q^2 + i\epsilon}
\ee
where we used (\ref{j2pt}) and (\ref{FT-integer}) with $M=2$. This again gives (\ref{Mjpm}). Yet another method for obtaining the same result is described in appendix~\ref{sec-alt-amp-jm}.

From (\ref{j2pt}) and (\ref{me-phase-space-factor-N=0}) the phase space of $j^\pm$ is
\be
\Phi_{j^\pm}(Q) = 2(1+a)\left(Q^\pm\right)^2 \theta(Q^\pm)\,\delta(Q^2)
\ee
Thus
\be
|\cM_{j^\pm}|^2\,\Phi_{j^\pm}
= 8\pi^2(1+a)\, [\Delta_\un(P)]^2
\left[1 - \left(1 - \frac{Q^\pm}{P^\pm}\right)^a\right]^2 \,\theta(Q^\pm)\,\delta(Q^2)
\label{j+cs}
\ee
To compare (\ref{j+cs}) with our previous results, we expand it in $Q/P$:
\bea
&&|\cM_{j^\pm}|^2\,\Phi_{j^\pm} = 8\pi^2(1+a)\, [\Delta_\un(P)]^2
\left[\left(\frac{Q^\pm}{P^\pm}\right)^2 + (1-a)\left(\frac{Q^\pm}{P^\pm}\right)^3 \right. \\
&&\qquad\qquad\qquad\qquad\qquad\qquad\qquad\qquad\left. + (1-a)\frac{11-7a}{12}\left(\frac{Q^\pm}{P^\pm}\right)^4 + \ldots\right] \,\theta(Q^+)\,\delta(Q^2)\nn
\eea
This is exactly what we obtain if we repeat the computation that led to (\ref{missing-energy-result}) but include only terms that are attributed to $j^\pm$ and its derivatives in (\ref{me-4pt-expansion-OPE}).

\section{Comments about the massive bosons\label{sec-comments-massive}}
\setcounter{equation}{0}

When we go to energies above the conformal region, the massive boson propagators in $i\Delta_\cO(x)$ [see~(\ref{O-prop-exp})], and consequently in the higher $n$-point functions~(\ref{n-pt}), start playing a role. As we have shown in~\cite{Georgi:2008pq} and reviewed in section~\ref{sec-disappearance}, the resulting total cross-section for the disappearance process can be interpreted as involving phase space integrals over the unparticle stuff with an arbitrary number of additional massive bosons, as shown in figure~\ref{fig-massive-bosons}. These massive bosons have the same origin as the massive boson of the Schwinger model~\cite{Casher:1973uf,Casher:1974vf} and are analogous to the hadrons of QCD.

This interpretation of the disappearance process motivates us to describe the higher order processes of sections~\ref{sec-mc} and~\ref{sec-me} analogously: production of the various types of unparticle stuff like in the conformal limit and some number of additional massive bosons. In the following we show that this interpretation indeed fits into a consistent picture.

\subsection{Processes with a single massive boson\label{sec-1boson}}

{\figsize\begin{figure}[htb]
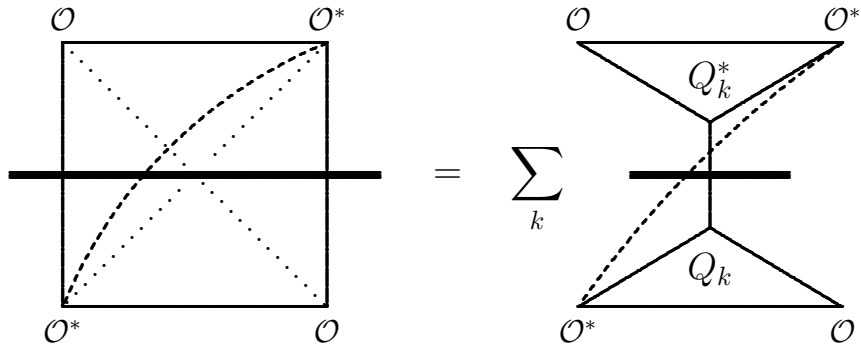

$$\beginpicture
\setcoordinatesystem units <1\tdim,1\tdim>
\stpltsmbl
\plot 50 50 -50 50 -50 -50 50 -50 50 50 /
\plot -70 0 70 0 /
\plot -70 1 70 1 /
\plot -70 -1 70 -1 /
\setdots
\plot -50 50 50 -50 /
\plot 50 50 6 6 /
\plot -6 -6 -50 -50 /
\setdashes <0.8mm>
\circulararc 45 degrees from 50 50 center at 120 -120
\put {$\cO$} [t] at 50 -55
\put {$\cO^\ast$} [b] at 50 55
\put {$\cO^\ast$} [t] at -50 -55
\put {$\cO$} [b] at -50 55
\linethickness=0pt
\putrule from 0 -60 to 0 60
\putrule from -90 0 to 90 0
\endpicture
\beginpicture
\setcoordinatesystem units <1\tdim,1\tdim>
\stpltsmbl
\plot 50 50 -50 50 0 20 50 50 /
\plot 50 -50 -50 -50 0 -20 50 -50 /
\plot 0 20 0 -20 /
\plot -30 0 30 0 /
\plot -30 1 30 1 /
\plot -30 -1 30 -1 /
\setdashes <0.8mm>
\circulararc 20 degrees from 50 50 center at 280 -280
\put {$\cO$} [t] at 50 -55
\put {$\cO^\ast$} [b] at 50 55
\put {$\cO^\ast$} [t] at -50 -55
\put {$\cO$} [b] at -50 55
\put {\large$Q_k^\ast$} [c] at 0 36
\put {\large$Q_k$} [c] at 0 -36
\put {\large$\displaystyle =\quad \sum_k$} [c] at -80 -5
\linethickness=0pt
\putrule from 0 -60 to 0 60
\putrule from -100 0 to 60 0
\endpicture$$
\caption{\figsize\sf\label{fig-me-w/boson-negative}A term with one massive boson (the short-dashed line) in the cross-section for the missing energy process.}\end{figure}}

Consider, for example, a term in which one massive boson line is attached to a dotted line of the missing energy process of figure~\ref{fig-me}, as shown in figure~\ref{fig-me-w/boson-negative}. We would expect this term to describe processes in which a massive boson $\cB$ is produced along with unparticle stuff of type $k$:
\be
\phi + \phi \to \phi + \phi + \{\mbox{$\cO_k$ stuff}\} + \cB
\label{me-w/boson}
\ee
Na\"{\i}vely this looks problematic because the contribution of figure~\ref{fig-me-w/boson-negative} to the cross-section is negative since similarly to (\ref{O-prop-exp}) we can write the full dotted line propagator as
\be
i\tilde\Delta_{\cO}(x) = \frac{1}{i\Delta_{\cO}(x)}
= \frac{1}{i\Delta_\un(x)}\exp\left[4\pi ia\Delta(x)\right]
= \frac{1}{i\Delta_\un(x)}\sum_{n=0}^\infty\frac{\left(4\pi
a\right)^n}{n!}\left[i\Delta(x)\right]^n
\label{dotted-prop-exp}
\ee
Since $a < 0$, the prefactors of terms that include an odd number $n$ of massive bosons are negative, unlike in (\ref{O-prop-exp}) where all the prefactors are positive. However, there actually exist several diagrams that contribute to the process (\ref{me-w/boson}), as shown in figure~\ref{fig-me-w/boson-interfere}, including diagrams in which the boson is attached to a solid line. Only the sum of the four diagrams should be non-negative, and this is indeed satisfied since the sum is proportional to\footnote{More precisely, besides the diagrams shown in figure~\ref{fig-me-w/boson-interfere}, there are other diagrams (of higher order in $a$) that contribute to the process (\ref{me-w/boson}), in which additional boson lines are attached between the bottom $\cO$ and $\cO^\ast$ and/or between the top $\cO$ and $\cO^\ast$ of the diagrams in figure~\ref{fig-me-w/boson-interfere}. They can be described by including the corresponding terms in (\ref{square}). The counting obviously works because it is the same as it would be with regular Feynman diagrams.}
\be
|Q_k(P) - Q_k(P-\ell)|^2
\label{square}
\ee
Here we use $Q_k(p)$ to denote the conformal 3-point function of $\cO^\ast$, $\cO$ and $\cO_k$ (with the $\cO_k$ leg amputated) in momentum space, with momentum $p$ entering the 3-point function at the $\cO^\ast$ point, momentum $Q$ leaving with the $\cO_k$ stuff (it is assumed to be the same in all the diagrams and therefore not specified explicitly as an argument of $Q_k$), and momentum $p - Q$ leaving at the $\cO$ point.

{\figsize\begin{figure}[htb]
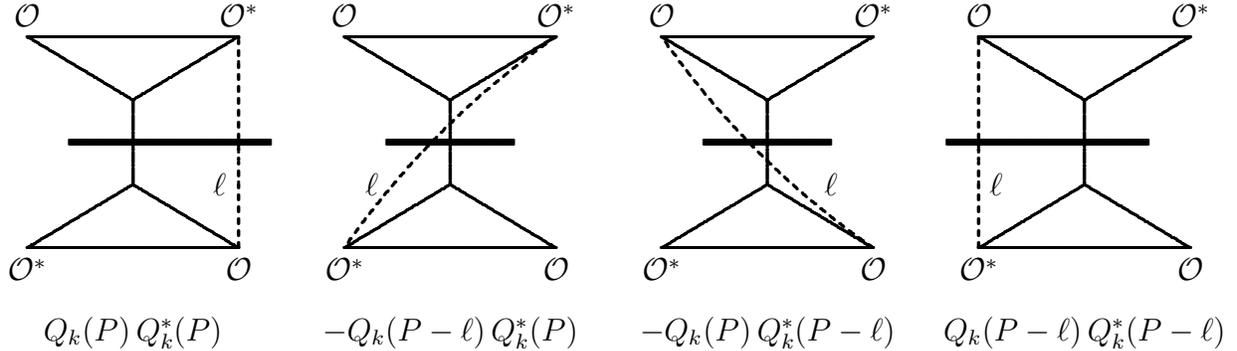

$$\beginpicture
\setcoordinatesystem units <0.8\tdim,0.8\tdim>
\stpltsmbl
\plot 50 50 -50 50 0 20 50 50 /
\plot 50 -50 -50 -50 0 -20 50 -50 /
\plot 0 20 0 -20 /
\plot -30 0 65 0 /
\plot -30 1 65 1 /
\plot -30 -1 65 -1 /
\setdashes <0.8mm>
\plot 50 -50 50 50 /
\put {$\cO$} [t] at 50 -55
\put {$\cO^\ast$} [b] at 50 55
\put {$\cO^\ast$} [t] at -50 -55
\put {$\cO$} [b] at -50 55
\put {$\ell$} at 41 -20
\linethickness=0pt
\putrule from 0 -95 to 0 60
\putrule from -60 0 to 75 0
\put {$Q_k(P)\,Q_k^\ast(P)$} at 0 -90
\endpicture
\beginpicture
\setcoordinatesystem units <0.8\tdim,0.8\tdim>
\stpltsmbl
\plot 50 50 -50 50 0 20 50 50 /
\plot 50 -50 -50 -50 0 -20 50 -50 /
\plot 0 20 0 -20 /
\plot -30 0 30 0 /
\plot -30 1 30 1 /
\plot -30 -1 30 -1 /
\setdashes <0.8mm>
\circulararc 20 degrees from 50 50 center at 280 -280
\put {$\cO$} [t] at 50 -55
\put {$\cO^\ast$} [b] at 50 55
\put {$\cO^\ast$} [t] at -50 -55
\put {$\cO$} [b] at -50 55
\put {$\ell$} at -37 -20
\linethickness=0pt
\putrule from 0 -95 to 0 60
\putrule from -75 0 to 75 0
\put {$-Q_k(P-\ell)\,Q_k^\ast(P)$} at 0 -90
\endpicture
\beginpicture
\setcoordinatesystem units <0.8\tdim,0.8\tdim>
\stpltsmbl
\plot 50 50 -50 50 0 20 50 50 /
\plot 50 -50 -50 -50 0 -20 50 -50 /
\plot 0 20 0 -20 /
\plot -30 0 30 0 /
\plot -30 1 30 1 /
\plot -30 -1 30 -1 /
\setdashes <0.8mm>
\circulararc -20 degrees from 50 -50 center at 280 280
\put {$\cO$} [t] at 50 -55
\put {$\cO^\ast$} [b] at 50 55
\put {$\cO^\ast$} [t] at -50 -55
\put {$\cO$} [b] at -50 55
\put {$\ell$} at 30 -20
\linethickness=0pt
\putrule from 0 -95 to 0 60
\putrule from -75 0 to 75 0
\put {$-Q_k(P)\,Q_k^\ast(P-\ell)$} at 0 -90
\endpicture
\beginpicture
\setcoordinatesystem units <0.8\tdim,0.8\tdim>
\stpltsmbl
\plot 50 50 -50 50 0 20 50 50 /
\plot 50 -50 -50 -50 0 -20 50 -50 /
\plot 0 20 0 -20 /
\plot -65 0 30 0 /
\plot -65 1 30 1 /
\plot -65 -1 30 -1 /
\setdashes <0.8mm>
\plot -50 -50 -50 50 /
\put {$\cO$} [t] at 50 -55
\put {$\cO^\ast$} [b] at 50 55
\put {$\cO^\ast$} [t] at -50 -55
\put {$\cO$} [b] at -50 55
\put {$\ell$} at -42 -20
\linethickness=0pt
\putrule from 0 -95 to 0 60
\putrule from -75 0 to 60 0
\put {$Q_k(P-\ell)\,Q_k^\ast(P-\ell)$} at 0 -90
\endpicture$$
\caption{\figsize\sf\label{fig-me-w/boson-interfere}The leading diagrams contributing to the cross-section of the process (\ref{me-w/boson}). Here $Q_k(p)$ denotes the unparticle amputated 3-point function with incoming momentum $p$ at the $\cO^\ast$ vertex. The momentum flowing through the boson line is denoted by $\ell$. The factor $-4\pi a$ that is common to all the diagrams, as well as the common factors of $h$ and the phase space factors, are not written explicitly.}\end{figure}}

{\figsize\begin{figure}[htb]
$$\beginpicture
\setcoordinatesystem units <1\tdim,1\tdim>
\stpltsmbl
\plot 0 0 0 60 /
\tarrow from 0 0 to 0 35
\setdashes <2mm>
\plot -40 -40 0 0 /
\tarrow from -40 -40 to -22 -22
\plot 40 -40 0 0 /
\tarrow from 40 -40 to 22 -22
\setdashes <0.8mm>
\plot 0 0 60 30 /
\linethickness=0pt
\putrule from 0 -40 to 0 65
\putrule from -40 0 to 60 0
\put {$V_1 = i\,\displaystyle\frac{2e}{m}\,h$} at -50 10
\endpicture$$
\caption{\figsize\sf\label{fig-1boson-vertex}A vertex described by (\ref{Lint-w/boson}). The long-dashed lines are the standard model particles $\phi$, the solid line is the unparticle leg, and the short-dashed line is the massive boson $\cB$. The arrows indicate the direction of flow of the chiral charge.}\end{figure}}

In the more common particle-physics language we can describe these processes as follows. We already know that in the unparticle limit (\ref{lowenergy}) gives the effective interaction
\be
\cL_{{\rm int},\,\un} =
\frac{h}{2}
\left(\cO_\un\,{\phi^\ast}^2 + \cO_\un^\ast\,\phi^2\right)
\label{Lint-U}
\ee
where $\cO_\un$ is the operator $\cO$ in the unparticle limit (which was often denoted simply by $\cO$ when we restricted ourselves to the unparticle limit in the previous sections). Similarly, as we will now verify, processes in which a single additional boson is produced can be described by including also the effective interaction
\be
\cL_{{\rm int},\,1} =
-i\,\frac{2e}{m}\,\frac{h}{2}
\left(\cB\,\cO_\un\,{\phi^\ast}^2 - \cB\,\cO_\un^\ast\,\phi^2\right)
\label{Lint-w/boson}
\ee
where we used $-4\pi a = (2e/m)^2$ and assumed that the massive boson field ${\cal B}$ has the standard kinetic term $\frac{1}{2}\pd_\mu\cB\pd^\mu\cB$.
This interaction gives the vertex $V_1$ shown in figure~\ref{fig-1boson-vertex} (and its hermitian conjugate $V_1^\ast = -V_1$). In the disappearance process with a single massive boson (the $n=1$ diagram in figure~\ref{fig-massive-bosons}), we have a $V_1$ vertex and a $V_1^\ast$ vertex, which give the extra factor of $(2e/m)^2 = -4\pi a$ relative to the unparticle-only disappearance process. This is consistent with (\ref{full-phase-space}). In the process (\ref{me-w/boson}), the four diagrams of figure~\ref{fig-me-w/boson-interfere} contain the following vertices involving the massive bosons (the first factor is the vertex from the bottom part of the diagram and the second from the top):
\be
V_1^\ast \cdot V_1\qquad\qquad
V_1 \cdot V_1 \qquad\qquad
V_1^\ast \cdot V_1^\ast \qquad\qquad
V_1 \cdot V_1^\ast
\ee
The products of the two vertices are positive for the first and fourth diagrams (which are precisely the diagrams where the boson is coming from the solid lines) and negative for the second and third diagrams (where the boson is coming from the dotted lines), which agrees with what we obtained above from our general approach.

{\figsize\begin{figure}[htb]
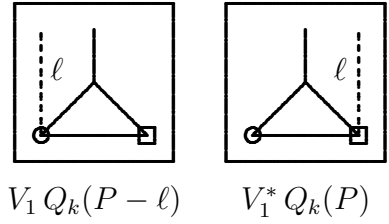

$$
\beginpicture
\setcoordinatesystem units <1\tdim,1\tdim>
\stpltsmbl
\plot -30 -30 -30 30 30 30 30 -30 -30 -30 /
\plot 0 20 0 0 -20 -20 20 -20 0 0 /
\circulararc 360 degrees from -22 -22 center at -20 -20
\plot 23 -23 23 -17 17 -17 17 -23 23 -23 /
\setdashes <0.8mm>
\plot -20 -20 -20 20 /
\put {$\ell$} at -14 5
\linethickness=0pt
\putrule from 0 -50 to 0 35
\putrule from -40 0 to 40 0
\put {$V_1\,Q_k(P-\ell)$} at 0 -45
\endpicture
\beginpicture
\setcoordinatesystem units <1\tdim,1\tdim>
\stpltsmbl
\plot -30 -30 -30 30 30 30 30 -30 -30 -30 /
\plot 0 20 0 0 -20 -20 20 -20 0 0 /
\circulararc 360 degrees from -22 -22 center at -20 -20
\plot 23 -23 23 -17 17 -17 17 -23 23 -23 /
\setdashes <0.8mm>
\plot 20 -20 20 20 /
\put {$\ell$} at 13 5
\linethickness=0pt
\putrule from 0 -50 to 0 35
\putrule from -40 0 to 40 0
\put {$V_1^\ast\,Q_k(P)$} at 0 -45
\endpicture
$$
\caption{\figsize\sf\label{fig-me-w/1bosons-amp}The leading diagrams for the \emph{amplitude} for producing one additional boson in the missing energy process. The circle marks the point at which the standard model particles inject the chiral charge into the diagram, while the square is the point at which the opposite process occurs. The massive boson lines are short-dashed and carry momentum $\ell$.}\end{figure}}

Even more simply, from the perspective of the effective interaction (\ref{Lint-w/boson}), the sum of the diagrams in figure~\ref{fig-me-w/boson-interfere} can be viewed as the standard calculation of $\int |\cM|^2 d\Phi$, where $\cM$ is the sum of the two interfering diagrams (figure~\ref{fig-me-w/1bosons-amp}) that one can construct for the amplitude of the process (\ref{me-w/boson}) using the vertex from figure~\ref{fig-1boson-vertex} and its conjugate. The relative minus sign in (\ref{square}) appears because one diagram in figure~\ref{fig-me-w/1bosons-amp} has a factor of $V_1$ while the other has $V_1^\ast = -V_1$.

It is interesting to note that there exists a missing energy process in which only the massive boson is produced, without the unparticle stuff:
\be
\phi + \phi \to \phi + \phi + \cB
\label{me-only-boson}
\ee
This arises similarly to (\ref{me-w/boson}), except that $\cO_k$ is the identity operator in (\ref{OO*OPE}) which gives rise to the disconnected term in (\ref{me-4pt-expansion-OPE}) and (\ref{me-4pt-expansion}). This is different from the disappearance process of section~\ref{sec-disappearance} where the massive bosons can only be produced together with unparticle stuff.

We can also approach the process (\ref{me-only-boson}), and similarly the processes (\ref{me-w/boson}), from a different point of view by considering the operator $j_S^\mu$ (or equivalently its axial counterpart $j_S^{5\mu} = -\epsilon^{\mu\nu}j_{S\nu}$) that is discussed appendix~\ref{sec-operators}. In particular, the OPE contains the term
\be
\mbox{T}\,\cO^\ast(\zeta)\,\cO(0)
\,\supset\, -2\pi i\,i\Delta_\un(\zeta)\,
\zeta_\mu\, j_S^{5\mu}(0)
\ee
The form of the two-point function of $j_S^\mu$, (\ref{jS+jS+})--(\ref{jS+jS-}), suggests the replacement
\be
j_S^{5\mu} \to -\sqrt{-\frac{a}{\pi}}\,\pd^\mu\cB
\ee
where we use
\be
(\Box + m^2)\,i\Delta(x) = -i\delta^2(x)
\ee
to reconcile (\ref{jS+jS-}) (up to a contact term, which is ambiguous anyway). Then we can write
\be
\mbox{T}\,\cO^\ast(\zeta)\,\cO(0)
\,\supset\, i\frac{2e}{m}\,i\Delta_\un(\zeta)\,\zeta^\mu\pd_\mu\cB(0) \ee
This means that for small $y-x$
\be
\mbox{T}\,\cO^\ast(y)\,\cO(x)
\,\supset\, i\frac{2e}{m}\,i\Delta_\un(y-x) \left[\cB(y)-\cB(x)\right]
\ee
which is precisely proportional to the amplitude we would write for the process (\ref{me-only-boson}) by summing the two diagrams in figure~\ref{fig-me-only-boson-amp}.

{\figsize\begin{figure}[htb]
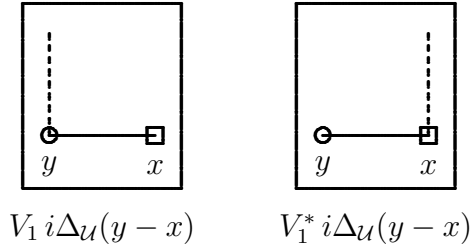

$$
\beginpicture
\setcoordinatesystem units <1\tdim,1\tdim>
\stpltsmbl
\plot -30 -40 -30 30 30 30 30 -40 -30 -40 /
\plot -20 -20 20 -20 /
\circulararc 360 degrees from -22 -22 center at -20 -20
\plot 23 -23 23 -17 17 -17 17 -23 23 -23 /
\setdashes <0.8mm>
\plot -20 -20 -20 20 /
\put {$y$} at -20 -32
\put {$x$} at 20 -32
\linethickness=0pt
\putrule from 0 -60 to 0 35
\putrule from -40 0 to 40 0
\put {$V_1\,i\Delta_\un(y-x)$} at 0 -55
\endpicture
\qquad
\beginpicture
\setcoordinatesystem units <1\tdim,1\tdim>
\stpltsmbl
\plot -30 -40 -30 30 30 30 30 -40 -30 -40 /
\plot -20 -20 20 -20 /
\circulararc 360 degrees from -22 -22 center at -20 -20
\plot 23 -23 23 -17 17 -17 17 -23 23 -23 /
\setdashes <0.8mm>
\plot 20 -20 20 20 /
\put {$y$} at -20 -32
\put {$x$} at 20 -32
\linethickness=0pt
\putrule from 0 -60 to 0 35
\putrule from -40 0 to 40 0
\put {$V_1^\ast\,i\Delta_\un(y-x)$} at 0 -55
\endpicture
$$
\caption{\figsize\sf\label{fig-me-only-boson-amp}The leading position-space diagrams for the amplitude for producing just a boson, without the unparticle stuff, in the missing energy process.}\end{figure}}

\subsection{Processes with multiple massive bosons}

{\figsize\begin{figure}[htb]
$$\beginpicture
\setcoordinatesystem units <1\tdim,1\tdim>
\stpltsmbl
\plot 0 0 0 60 /
\tarrow from 0 0 to 0 35
\setdashes <2mm>
\plot -40 -40 0 0 /
\tarrow from -40 -40 to -22 -22
\plot 40 -40 0 0 /
\tarrow from 40 -40 to 22 -22
\setdashes <0.8mm>
\plot 0 0 60 30 /
\plot 0 0 65 20 /
\plot 0 0 67 10 /
\plot 0 0 68 0 /
\plot 0 0 67 -10 /
\plot 0 0 65 -20 /
\linethickness=0pt
\putrule from 0 -40 to 0 60
\putrule from -40 0 to 60 0
\put {$V_n = \displaystyle\left(i\frac{2e}{m}\right)^n h$} at -55 10
\endpicture$$
\caption{\figsize\sf\label{fig-nboson-vertex}An $n$-boson vertex described by (\ref{Lint-gen}). The long-dashed lines are the standard model particles $\phi$, the solid line is the unparticle leg, and the short-dashed lines are the $n$ massive bosons $\cB$. The arrows indicate the direction of flow of the chiral charge.}\end{figure}}

The ideas discussed above can be generalized to processes with an arbitrary number of bosons by writing
\be
\cL_{\rm int} =
\frac{h}{2}
\left(e^{-i\left(2e/m\right)\cB}\,\cO_\un\,{\phi^\ast}^2 + e^{i\left(2e/m\right)\cB}\,\cO_\un^\ast\,\phi^2\right)
\label{Lint-gen}
\ee
This expression contains (\ref{Lint-U}), (\ref{Lint-w/boson}), and similar multiple-boson vertices, as shown in figure~\ref{fig-nboson-vertex}. One can see that (\ref{Lint-gen}) is equivalent to our original description by noticing that the interaction (\ref{lowenergy}) gives
\bea
&&\vev{\phi^2(x_1)\ldots\phi^2(x_n)\,\phi^{\ast2}(y_1)\ldots\phi^{\ast2}(y_n)} \nn\\
&&\propto\vev{\cO(x_1)\ldots\cO(x_n)\,\cO^\ast(y_1)\ldots\cO^\ast(y_n)} \nn\\
&&= \frac{\prod_{j,k} \exp\left[\left(2e/m\right)^2 i\Delta(x_j-y_k)\right]}
{\prod_{k>j} \exp\left[\left(2e/m\right)^2 i\Delta(x_j-x_k)\right]
\exp\left[\left(2e/m\right)^2 i\Delta(y_j-y_k)\right]} \nn\\
&&\quad\times\vev{\cO_\un(x_1)\ldots\cO_\un(x_n)\,\cO_\un^\ast(y_1)\ldots\cO_\un^\ast(y_n)}
\label{n-pt-bosons-1}
\eea
where we used (\ref{n-pt}) with (\ref{O-prop-exp}), while (\ref{Lint-gen}) gives
\bea
&&\vev{\phi^2(x_1)\ldots\phi^2(x_n)\,\phi^{\ast2}(y_1)\ldots\phi^{\ast2}(y_n)} \nn\\
&&\propto\vev{e^{-i(2e/m)\cB(x_1)}\ldots e^{-i(2e/m)\cB(x_n)}\, e^{i(2e/m)\cB(y_1)}\ldots e^{i(2e/m)\cB(y_n)}} \nn\\
&&\quad\times\vev{\cO_\un(x_1)\ldots\cO_\un(x_n)\,\cO_\un^\ast(y_1)\ldots\cO_\un^\ast(y_n)}
\label{n-pt-bosons-2}
\eea
After contracting the boson operators in (\ref{n-pt-bosons-2}) we indeed obtain the result (\ref{n-pt-bosons-1}).

{\figsize\begin{figure}[htb]
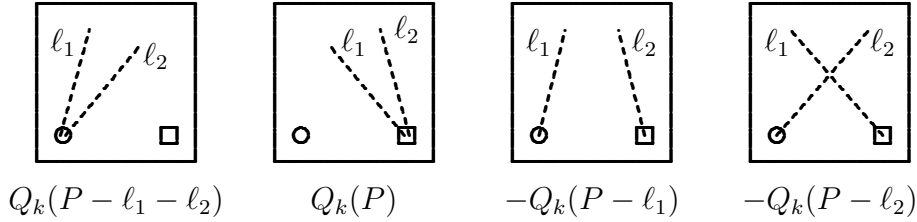

$$
\beginpicture
\setcoordinatesystem units <1\tdim,1\tdim>
\stpltsmbl
\plot -30 -30 -30 30 30 30 30 -30 -30 -30 /
\circulararc 360 degrees from -22 -22 center at -20 -20
\plot 23 -23 23 -17 17 -17 17 -23 23 -23 /
\setdashes <0.8mm>
\plot -21 -20 -10 20 /
\plot -19 -20 10 15 /
\put {$\ell_1$} at -20 15
\put {$\ell_2$} at 15 10
\put {$Q_k(P-\ell_1-\ell_2)$} at 0 -45
\linethickness=0pt
\putrule from 0 -50 to 0 35
\putrule from -45 0 to 45 0
\endpicture
\beginpicture
\setcoordinatesystem units <1\tdim,1\tdim>
\stpltsmbl
\plot -30 -30 -30 30 30 30 30 -30 -30 -30 /
\circulararc 360 degrees from -22 -22 center at -20 -20
\plot 23 -23 23 -17 17 -17 17 -23 23 -23 /
\setdashes <0.8mm>
\plot 21 -20 10 20 /
\plot 19 -20 -10 15 /
\put {$\ell_1$} at 0 15
\put {$\ell_2$} at 20 20
\put {$Q_k(P)$} at 0 -45
\linethickness=0pt
\putrule from 0 -50 to 0 35
\putrule from -45 0 to 45 0
\endpicture
\beginpicture
\setcoordinatesystem units <1\tdim,1\tdim>
\stpltsmbl
\plot -30 -30 -30 30 30 30 30 -30 -30 -30 /
\circulararc 360 degrees from -22 -22 center at -20 -20
\plot 23 -23 23 -17 17 -17 17 -23 23 -23 /
\setdashes <0.8mm>
\plot 20 -20 10 20 /
\plot -20 -20 -10 20 /
\put {$\ell_1$} at -20 15
\put {$\ell_2$} at 20 15
\put {$-Q_k(P-\ell_1)$} at 0 -45
\linethickness=0pt
\putrule from 0 -50 to 0 35
\putrule from -45 0 to 45 0
\endpicture
\beginpicture
\setcoordinatesystem units <1\tdim,1\tdim>
\stpltsmbl
\plot -30 -30 -30 30 30 30 30 -30 -30 -30 /
\circulararc 360 degrees from -22 -22 center at -20 -20
\plot 23 -23 23 -17 17 -17 17 -23 23 -23 /
\setdashes <0.8mm>
\plot 20 -20 -15 20 /
\plot -20 -20 15 20 /
\put {$\ell_1$} at -20 15
\put {$\ell_2$} at 20 15
\put {$-Q_k(P-\ell_2)$} at 0 -45
\linethickness=0pt
\putrule from 0 -50 to 0 35
\putrule from -45 0 to 45 0
\endpicture
$$
\caption{\figsize\sf\label{fig-me-w/2bosons-amp}The leading diagrams for the \emph{amplitude} for producing 2 additional bosons in the missing energy process. For clarity, only the boson lines are shown explicitly (but the unparticle 3-point function is also assumed to be present like in figure~\ref{fig-me-w/1bosons-amp}). Common prefactors are not written.}\end{figure}}

To see more explicitly how (\ref{Lint-gen}) works, let us look at missing energy processes that produce 2 bosons. Using the vertices $V_1$ and $V_2$ from figure~\ref{fig-nboson-vertex} (and their conjugates), the amplitude for such a process is given by the sum of the diagrams in figure~\ref{fig-me-w/2bosons-amp}. Squaring the amplitude and including a factor of $1/2$ due to the $2$ identical bosons in the final state, the cross-section is proportional to
\be
\frac{1}{2}\int d\Phi\, \left|Q_k\left(P-\ell_1-\ell_2\right) + Q_k\left(P\right) - Q_k\left(P-\ell_1\right) - Q_k\left(P-\ell_2\right)\right|^2
\label{2bosons-cs}
\ee
On the other hand, this is precisely the result we get from our general formalism by summing the diagrams in figure~\ref{fig-me-w/2bosons} where we used (\ref{O-prop-exp}) and (\ref{dotted-prop-exp}) for computing the effects of the bosons.\footnote{To make the agreement manifest we need to rename the integration variables $\ell_1 \leftrightarrow \ell_2$ in some of the terms that we obtain after expanding the square in (\ref{2bosons-cs}).}

{\figsize\begin{figure}[htb]
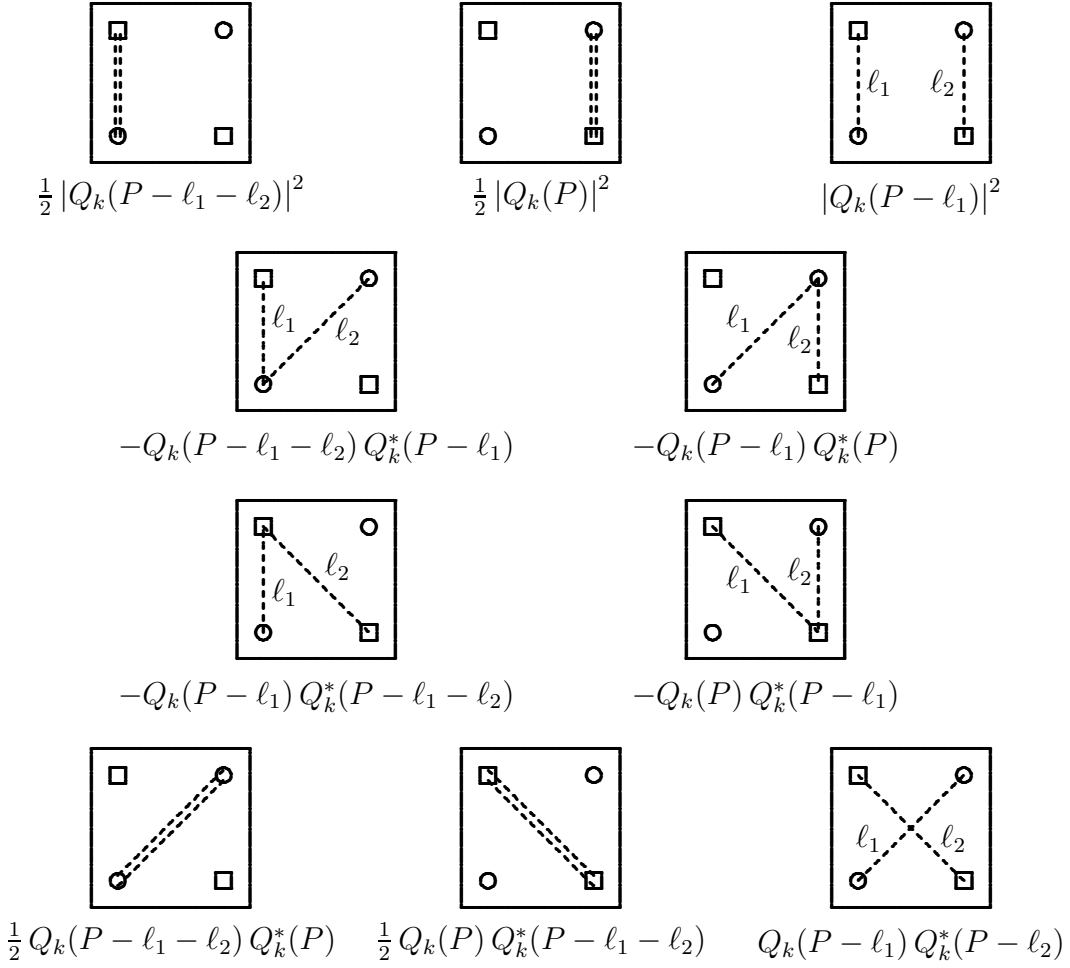

\bea
&\beginpicture
\setcoordinatesystem units <1\tdim,1\tdim>
\stpltsmbl
\plot -30 -30 -30 30 30 30 30 -30 -30 -30 /
\circulararc 360 degrees from 22 22 center at 20 20
\circulararc 360 degrees from -22 -22 center at -20 -20
\plot -23 23 -17 23 -17 17 -23 17 -23 23 /
\plot 23 -23 23 -17 17 -17 17 -23 23 -23 /
\setdashes <0.8mm>
\plot -21 -20 -21 20 /
\plot -19 -20 -19 20 /
\put {$\frac{1}{2}\left|Q_k(P-\ell_1-\ell_2)\right|^2$} [b] at 0 -50
\linethickness=0pt
\putrule from 0 -55 to 0 35
\putrule from -70 0 to 70 0
\endpicture
\beginpicture
\setcoordinatesystem units <1\tdim,1\tdim>
\stpltsmbl
\plot -30 -30 -30 30 30 30 30 -30 -30 -30 /
\circulararc 360 degrees from 22 22 center at 20 20
\circulararc 360 degrees from -22 -22 center at -20 -20
\plot -23 23 -17 23 -17 17 -23 17 -23 23 /
\plot 23 -23 23 -17 17 -17 17 -23 23 -23 /
\setdashes <0.8mm>
\plot 21 -20 21 20 /
\plot 19 -20 19 20 /
\put {$\frac{1}{2}\left|Q_k(P)\right|^2$} [b] at 0 -50
\linethickness=0pt
\putrule from 0 -55 to 0 35
\putrule from -70 0 to 70 0
\endpicture
\beginpicture
\setcoordinatesystem units <1\tdim,1\tdim>
\stpltsmbl
\plot -30 -30 -30 30 30 30 30 -30 -30 -30 /
\circulararc 360 degrees from 22 22 center at 20 20
\circulararc 360 degrees from -22 -22 center at -20 -20
\plot -23 23 -17 23 -17 17 -23 17 -23 23 /
\plot 23 -23 23 -17 17 -17 17 -23 23 -23 /
\setdashes <0.8mm>
\plot -20 -20 -20 20 /
\plot 20 -20 20 20 /
\put {$\ell_1$} at -12 0
\put {$\ell_2$} at 12 0
\put {$\left|Q_k(P-\ell_1)\right|^2$} [b] at 0 -50
\linethickness=0pt
\putrule from 0 -55 to 0 35
\putrule from -70 0 to 70 0
\endpicture&
\nn\\
&\beginpicture
\setcoordinatesystem units <1\tdim,1\tdim>
\stpltsmbl
\plot -30 -30 -30 30 30 30 30 -30 -30 -30 /
\circulararc 360 degrees from 22 22 center at 20 20
\circulararc 360 degrees from -22 -22 center at -20 -20
\plot -23 23 -17 23 -17 17 -23 17 -23 23 /
\plot 23 -23 23 -17 17 -17 17 -23 23 -23 /
\setdashes <0.8mm>
\plot -20 -20 20 20 /
\plot -20 -20 -20 20 /
\put {$\ell_1$} at -12 5
\put {$\ell_2$} at 12 0
\put {$-Q_k(P-\ell_1-\ell_2)\,Q_k^\ast(P-\ell_1)$} [b] at 0 -50
\linethickness=0pt
\putrule from 0 -55 to 0 35
\putrule from -85 0 to 85 0
\endpicture
\beginpicture
\setcoordinatesystem units <1\tdim,1\tdim>
\stpltsmbl
\plot -30 -30 -30 30 30 30 30 -30 -30 -30 /
\circulararc 360 degrees from 22 22 center at 20 20
\circulararc 360 degrees from -22 -22 center at -20 -20
\plot -23 23 -17 23 -17 17 -23 17 -23 23 /
\plot 23 -23 23 -17 17 -17 17 -23 23 -23 /
\setdashes <0.8mm>
\plot -20 -20 20 20 /
\plot 20 20 20 -20 /
\put {$\ell_1$} at -10 5
\put {$\ell_2$} at 13 -3
\put {$-Q_k(P-\ell_1)\,Q_k^\ast(P)$} [b] at 0 -50
\linethickness=0pt
\putrule from 0 -55 to 0 35
\putrule from -85 0 to 85 0
\endpicture&\nn\\
&\beginpicture
\setcoordinatesystem units <1\tdim,1\tdim>
\stpltsmbl
\plot -30 -30 -30 30 30 30 30 -30 -30 -30 /
\circulararc 360 degrees from 22 22 center at 20 20
\circulararc 360 degrees from -22 -22 center at -20 -20
\plot -23 23 -17 23 -17 17 -23 17 -23 23 /
\plot 23 -23 23 -17 17 -17 17 -23 23 -23 /
\setdashes <0.8mm>
\plot -20 20 20 -20 /
\plot -20 -20 -20 20 /
\put {$\ell_1$} at -12 -5
\put {$\ell_2$} at 8 5
\put {$-Q_k(P-\ell_1)\,Q_k^\ast(P-\ell_1-\ell_2)$} [b] at 0 -50
\linethickness=0pt
\putrule from 0 -55 to 0 35
\putrule from -85 0 to 85 0
\endpicture
\beginpicture
\setcoordinatesystem units <1\tdim,1\tdim>
\stpltsmbl
\plot -30 -30 -30 30 30 30 30 -30 -30 -30 /
\circulararc 360 degrees from 22 22 center at 20 20
\circulararc 360 degrees from -22 -22 center at -20 -20
\plot -23 23 -17 23 -17 17 -23 17 -23 23 /
\plot 23 -23 23 -17 17 -17 17 -23 23 -23 /
\setdashes <0.8mm>
\plot -20 20 20 -20 /
\plot 20 20 20 -20 /
\put {$\ell_1$} at -10 0
\put {$\ell_2$} at 13 3
\put {$-Q_k(P)\,Q_k^\ast(P-\ell_1)$} [b] at 0 -50
\linethickness=0pt
\putrule from 0 -55 to 0 35
\putrule from -85 0 to 85 0
\endpicture&
\nn\\
&\beginpicture
\setcoordinatesystem units <1\tdim,1\tdim>
\stpltsmbl
\plot -30 -30 -30 30 30 30 30 -30 -30 -30 /
\circulararc 360 degrees from 22 22 center at 20 20
\circulararc 360 degrees from -22 -22 center at -20 -20
\plot -23 23 -17 23 -17 17 -23 17 -23 23 /
\plot 23 -23 23 -17 17 -17 17 -23 23 -23 /
\setdashes <0.8mm>
\plot -20 -18 20 22 /
\plot -20 -22 20 18 /
\put {$\frac{1}{2}\,Q_k(P-\ell_1-\ell_2)\,Q_k^\ast(P)$} [b] at 0 -50
\linethickness=0pt
\putrule from 0 -55 to 0 35
\putrule from -70 0 to 70 0
\endpicture
\beginpicture
\setcoordinatesystem units <1\tdim,1\tdim>
\stpltsmbl
\plot -30 -30 -30 30 30 30 30 -30 -30 -30 /
\circulararc 360 degrees from 22 22 center at 20 20
\circulararc 360 degrees from -22 -22 center at -20 -20
\plot -23 23 -17 23 -17 17 -23 17 -23 23 /
\plot 23 -23 23 -17 17 -17 17 -23 23 -23 /
\setdashes <0.8mm>
\plot 20 -18 -20 22 /
\plot 20 -22 -20 18 /
\put {$\frac{1}{2}\,Q_k(P)\,Q_k^\ast(P-\ell_1-\ell_2)$} [b] at 0 -50
\linethickness=0pt
\putrule from 0 -55 to 0 35
\putrule from -70 0 to 70 0
\endpicture
\beginpicture
\setcoordinatesystem units <1\tdim,1\tdim>
\stpltsmbl
\plot -30 -30 -30 30 30 30 30 -30 -30 -30 /
\circulararc 360 degrees from 22 22 center at 20 20
\circulararc 360 degrees from -22 -22 center at -20 -20
\plot -23 23 -17 23 -17 17 -23 17 -23 23 /
\plot 23 -23 23 -17 17 -17 17 -23 23 -23 /
\setdashes <0.8mm>
\plot -20 -20 20 20 /
\plot 20 -20 -20 20 /
\put {$\ell_1$} at -16 -4
\put {$\ell_2$} at 16 -4
\put {$Q_k(P-\ell_1)\,Q_k^\ast(P-\ell_2)$} [b] at 0 -50
\linethickness=0pt
\putrule from 0 -55 to 0 35
\putrule from -70 0 to 70 0
\endpicture&
\nn
\eea
\caption{\figsize\sf\label{fig-me-w/2bosons}The leading diagrams for the \emph{cross-section} of producing 2 additional bosons in the missing energy process. For clarity, only the boson lines are shown explicitly. Common prefactors and phase space factors are not written.}\end{figure}}

As another check of (\ref{Lint-gen}), consider the case in which $n$ boson lines are attached between the two ends of an internal solid or dotted line (and for simplicity, no other bosons are attached to these vertices). From (\ref{O-prop-exp}) and (\ref{dotted-prop-exp}) we have a factor of
\be
\frac{(-4\pi a)^n}{n!}\qquad
\mbox{or}\qquad
\frac{(4\pi a)^n}{n!}\qquad
\label{nbosons-internal}
\ee
in the case of a solid or dotted line, respectively. On the other hand, taking the product of the two relevant vertices (using figure~\ref{fig-nboson-vertex} and/or its conjugate) gives
\be
\left(\pm i\frac{2e}{m}\right)^n \left(\mp i\frac{2e}{m}\right)^n = (-4\pi a)^n\qquad
\mbox{or}\qquad
\left(\pm i\frac{2e}{m}\right)^n \left(\pm i\frac{2e}{m}\right)^n = (4\pi a)^n
\label{nbosons-internal-prod-vert}
\ee
for these two cases. Dividing the expressions in (\ref{nbosons-internal-prod-vert}) by the symmetry factor $n!$ we obtain an agreement with (\ref{nbosons-internal}).

\subsection{Massive boson decay}

At the leading order in $h$, the massive boson $\cB$ decays as
\be
\cB \to \un_\cO + \bar\phi + \bar\phi\,, \qquad\qquad
\cB \to \bar\un_\cO + \phi + \phi
\ee
via the vertex in figure~\ref{fig-1boson-vertex} and its conjugate. The rate for each of these two modes is given by
\be
\Gamma = \frac{1}{2}\,\frac{1}{2m}\,|\cM|^2\,\Phi
\ee
where
\be
|\cM|^2 = \frac{4e^2}{m^2}h^2
\ee
and
\bea
\Phi &=& \frac{A(a)}{(\xi m)^{2a}}
\int\frac{d^2p_\un}{(2\pi)^2}\,\theta(p_\un^0)\,\theta(p_\un^2)
\left(p_\un^2\right)^a \nn\\
&&\qquad\qquad\times\left[\prod_{i=1}^2\int\frac{d^2p_i}{(2\pi)^2}\,
2\pi\,\delta(p_i^2-m_\phi^2)\,\theta(p_i^0)\right]
(2\pi)^2\,\delta^2\left(p_\cB - p_\un - p_1 - p_2\right) \nn\\
&=& \frac{A(a)}{4(2\pi)^2(\xi m)^{2a}}
\int_{\frac{m}{2}\left(1 - \sqrt{1 - 4m_\phi^2/m^2}\right)}^{\frac{m}{2}\left(1 + \sqrt{1 - 4m_\phi^2/m^2}\right)} \frac{dp_1^+}{p_1^+}
\int_{\frac{m_\phi^2}{m - m_\phi^2/p_1^+}}^{m-p_1^+} \frac{dp_2^+}{p_2^+}\nn\\
&&\qquad\qquad\qquad\quad\times
\left(m - p_1^+ - p_2^+\right)^a
\left(m - \frac{m_\phi^2}{p_1^+} - \frac{m_\phi^2}{p_2^+}\right)^a
\eea
At the leading order in $a$ and for $m_\phi \ll m$ we obtain
\be
\Gamma \simeq \frac{h^2}{8\pi^2}\frac{e^2}{m^3} \left[\ln^2\left(\frac{m_\phi}{m}\right) - \frac{\pi^2}{12}\right]
\ee

\section{Conclusions\label{conclusions}}
\setcounter{equation}{0}

The Sommerfield model is a useful toy model of the Banks-Zaks sector. It becomes scale-invariant in the infrared, with fractional anomalous dimensions. Since all of its correlation functions can be computed exactly, we can answer explicitly many of the questions regarding the physics of the Banks-Zaks sector as seen by the (toy) standard model observer, for an arbitrary coupling strength in the conformal sector. In particular, we were able to explore the behavior of the theory away from the low-energy scale-invariant regime and to incorporate the unparticle self-interactions. Most importantly, we believe, we used this toy model to provide consistency checks for the two new ideas in this paper: the extension of unparticle phase space calculations beyond the two point function to processes involving unparticle self-interactions; and the interpretation of the result using the conformal partial-wave expansion in terms of an ``amplitude'' for production of different types of unparticle stuff. These results are sufficiently general that they should apply to unparticle theories in 3+1 dimensions.

\section*{Acknowledgments}

We are grateful to Aqil Sajjad and Brian Shuve for discussions.
This research is supported in part by
the National Science Foundation under grant PHY-0244821.

\appendix

\section{Operators in the Sommerfield model\label{sec-operators}}
\setcounter{equation}{0}

The description of the Sommerfield model in terms of the free fields $\Psi$, $\cV$, $\cA$ in (\ref{Sommerfield-redefined}) immediately allows us to compute any correlation function involving these fields. Using this fact we were able to compute arbitrary correlation functions of the original interacting fermionic fields $\psi$ in section~\ref{sec-sommerfield}, and it is straightforward to extend these results to correlation functions that involve $A^\mu$ as well. We have also analyzed in section~\ref{correlation} the composite operator $\cO$ that appears in the OPE
\be
\mbox{T}\psi_2^\ast(\zeta)\psi_1(0)
= \frac{C_0(\zeta)}{C(\zeta)}\,\cO(0)+\cdots
\label{21OPEreminder}
\ee
In this appendix we would like to discuss the properties of additional operators that exist in the theory, obtained using similar methods. In particular, let's consider the vector current that is classically given by $j^\mu = \bar\psi\gamma^\mu\psi$. We expect to find $j^+$ in the OPE
\bea
&&\mbox{T}\psi_1(\zeta)\psi_1^\ast(0)
= \mbox{T} e^{-ie(\cV(\zeta)+\cA(\zeta))}\Psi_1(\zeta)\, e^{ie(\cV(0)+\cA(0))}\Psi_1^\ast(0)\nn\\
&&\quad = iS^1(\zeta) + \frac{C_0(\zeta)\,C(\zeta)}{2}\left[J^+(0) - \frac{e}{\pi}\left(\pd_- + \frac{\zeta^+}{\zeta^-}\pd_+\right)(\cV(0)+\cA(0))\right] + \ldots
\eea
where $J^+$ is the free current from
\be
\mbox{T}\Psi_1(\zeta)\Psi_1^\ast(0) = iS_0^1(\zeta) + \frac{1}{2}J^+(0) + \ldots
\ee
The auxiliary fields $\cV$ and $\cA$ are our own constructions. In the theory, they appear only in the combination $A^\mu$. Thus we see from here (and the analogous expression for $\mbox{T}\psi_2(\zeta)\psi_2^\ast(0)$) that we should define
\be
j^\pm \equiv J^\pm \mp \frac{2e}{\pi}\pd_\mp\cA
\ee
which gives
\be
\mbox{T}\psi_1(\zeta)\psi_1^\ast(0)
= iS^1(\zeta) + \frac{C_0(\zeta)\,C(\zeta)}{2}
\left[j^+(0) - \frac{e}{2\pi}\left(A^+(0) + \frac{\zeta^+}{\zeta^-}A^-(0)\right)\right] + \ldots
\ee
\be
\mbox{T}\psi_2(\zeta)\psi_2^\ast(0)
= iS^2(\zeta) + \frac{C_0(\zeta)\,C(\zeta)}{2}
\left[j^-(0) - \frac{e}{2\pi}\left(\frac{\zeta^-}{\zeta^+}A^+(0) + A^-(0)\right)\right] + \ldots
\ee
It can be shown by deriving Ward identities that both $j^\mu$ and $A^\mu$ are conserved vector currents, but they are not independent:
\be
\vev{j^\mu(x)A^\nu(x)}\neq 0
\ee
It is useful to replace $j^\mu$ and $A^\mu$ by two linear combinations that do not mix. These are
\be
j_T^\mu \equiv (1+a)\left(j^\mu - \frac{e}{\pi}A^\mu\right)
\qquad\mbox{and}\qquad
j_S^\mu \equiv -a\, j^\mu + (1+a)\frac{e}{\pi}A^\mu
\label{K,L}
\ee
Rewriting the OPEs in terms of $j_T^\mu$ and $j_S^\mu$ we have
\bea
&&\mbox{T} \psi_1(\zeta)\psi_1^\ast(0)
= iS^1(\zeta) + \frac{C_0(\zeta)\,C(\zeta)}{4}
\left[\frac{2+a}{1+a}\,j_T^+(0) + \frac{-a}{1+a}\frac{\zeta^+}{\zeta^-}\,j_T^-(0) \right.\nn\\
&&\qquad\qquad\qquad\quad\qquad\qquad\qquad\qquad\qquad\qquad\left.
+ j_S^+(0) - \frac{\zeta^+}{\zeta^-}\,j_S^-(0)\right] + \ldots
\eea
\bea
&&\mbox{T} \psi_2(\zeta)\psi_2^\ast(0)
= iS^2(\zeta) + \frac{C_0(\zeta)\,C(\zeta)}{4}
\left[\frac{2+a}{1+a}\,j_T^-(0) + \frac{-a}{1+a}\frac{\zeta^-}{\zeta^+}\,j_T^+(0) \right.\nn\\
&&\qquad\qquad\qquad\quad\qquad\qquad\qquad\qquad\qquad\qquad\left.
+ j_S^-(0) - \frac{\zeta^-}{\zeta^+}\,j_S^+(0)\right] + \ldots
\eea

The current $j_T^\mu$ happens to be exactly the combination whose axial counterpart $j_T^{5\mu} = -\epsilon^{\mu\nu}j_{T\,\nu}$ is anomaly-free (unlike $j^{5\mu} = -\epsilon^{\mu\nu}j_\nu$)~\cite{Georgi:1971iu}. It has the non-canonical ratio of $(1+a)$ between the axial and vector charge of $\psi_1$, exactly as in the Thirring model~\cite{Thirring:1958in,Wilson:1970pq,Georgi:1971sk,DellAntonio:1971zt}. The current $j_S^\mu$ can be dropped at low energies because its correlation functions always involve the massive propagator $\Delta(x)$ from (\ref{massive-propagator}). As a result, at low energies the OPEs reduce to those of the Thirring model where $j_T^\mu$ is the only current.\footnote{This is the reason for the name $j_T$. Similarly, $j_S$ gets its name because it is the only current that is present in the Schwinger model.} The OPE (\ref{21OPEreminder}) also agrees with the Thirring model at low energies~\cite{Wilson:1970pq}. In terms of our parameters, the Thirring model interaction is
\be
\cL_{\rm int} = \frac{\pi a}{2(1+a)}\,j_T^2 = -\frac{e^2}{2m_0^2}\,j_T^2
\ee
where $j_T^2 = j_T^\mu j_{T\,\mu} = j_T^+ j_T^-$.
A more general treatment of the operators in the Thirring model can be found in~\cite{Luscher:1975js}.

For the product $\cO(x)\cO^\ast(y)$ to which we refer in section~\ref{sec-me-interpretation}, we find the OPE
\bea
&&\mbox{T}\,\cO^\ast(\zeta)\,\cO(0)
= i\Delta_\cO(\zeta)\left[1 + 2\pi i\,\epsilon_{\mu\nu}\zeta^\mu j^\nu(0) - 2\pi^2 \left(\epsilon_{\mu\nu}\zeta^\mu j^\nu\right)^2(0)\right.\nn\\
&&\left.\qquad\qquad\qquad\qquad\qquad\quad
+ \pi i\, \epsilon^{\mu\nu}\left(\zeta^2\pd_\mu j_\nu(0) + \zeta_\mu\zeta_\rho\pd_\nu j^\rho(0)\right)
+ \ldots\right]
\eea
In the unparticle limit we can again drop $j_S^\mu$ to obtain
\bea
&&\mbox{T}\,\cO^\ast(\zeta)\,\cO(0)
= i\Delta_\un(\zeta)\left[\tall 1 + i\pi\left[\zeta^+ j_T^-(0) - \zeta^- j_T^+(0)\right]\right.\nn\\
&&\qquad\qquad\qquad\qquad\qquad\quad\;
+\, i\frac{\pi}{2}\left[(\zeta^+)^2\pd_+ j_T^-(0) - (\zeta^-)^2\pd_- j_T^+(0)\right] \\
&&\qquad\qquad\qquad\qquad\qquad\quad\;\left.
-\, \frac{\pi^2}{2} \left[(\zeta^+)^2 {j_T^-}^2(0) + (\zeta^-)^2{j_T^+}^2(0)\right]
+ \pi^2 \zeta^+\zeta^- j_T^2(0)
+ \ldots\right] \nn
\eea
where we used the fact that since both $j_T^\mu$ and $j_T^{5\mu}$ are conserved $\pd_\pm j_T^\pm(x) = 0$. The two-point functions of $j_T^\mu$ are\footnote{It is worth saying here that $\vev{j_T^+(x)j_T^-(0)}$ may contain a contact term which is ambiguous since current 2-point functions in $1+1$ dimensions are subject to subtractive renormalization. This is analog in $1+1$ of the ambiguity of current 3-point functions in $3+1$ that leads, for example, to different forms for the anomaly.}
\be
\vev{j_T^\pm(x)j_T^\pm(0)} = -\frac{1+a}{\pi^2} \frac{{x^\pm}^2}{\left(x^2-i\epsilon\right)^2}\qquad\qquad
\vev{j_T^+(x)j_T^-(0)} = 0
\label{j2pt}
\ee
The $x$-dependence is actually fixed by the requirement that $j_T^\mu$ cannot have an anomalous dimension since it generates a symmetry. The operators ${j_T^\pm}^2$ are the components $T^{\pm\pm}$ of the stress-energy tensor~\cite{Georgi:1971sk}:
\be
T^{\mu\nu} = \frac{\pi}{1+a}\left(j_T^\mu j_T^\nu - \frac{1}{2}g^{\mu\nu}j_T^\lambda j_{T\lambda}\right)\qquad\qquad
T^{\pm\pm} = \frac{\pi}{1+a}\,{j_T^\pm}^2
\ee
while $j_T^2$ is a scalar operator. The correlation functions of such products of currents can be obtained (apart from the contact terms) by contracting the currents like free fields. It is also useful to know the 3-point function
\be
\langle 0|\mbox{T}\cO^\ast(x_1)\,\cO(x_2)\,{j_T^\pm}(x_3)\,|0\rangle
= \mp (1+a)\frac{i}{\pi}
\left(\frac{x_{13}^\pm}{x_{13}^2 - i\epsilon} - \frac{x_{23}^\pm}{x_{23}^2 - i\epsilon}\right) i\Delta_\un(x_{12})
\label{j3ptord}
\ee
The two-point function of the current $j_S^\mu$, which we discuss further in section~\ref{sec-1boson}, is
\bea
\vev{j_S^\pm(x)j_S^\pm(0)} &=& \frac{4a}{\pi}\,\pd_\mp^2i\Delta(x)
\label{jS+jS+}\\
\vev{j_S^+(x)j_S^-(0)} &=& \frac{a}{\pi}m^2\,i\Delta(x) - a(1+a)\frac{i}{\pi}\,\delta^2(x)
\label{jS+jS-}
\eea

\section{Basic properties of lightcone coordinates\label{sec-lightcone}}
\setcounter{equation}{0}

The lightcone coordinates $x^\pm = x^0 \pm x^1$ in 2D have the properties
\be
x_\pm = \frac{x^\mp}{2} \qquad
x^2 = x^+x^- \qquad
\pd_\pm = \frac{1}{2}\left(\pd_0 \pm \pd_1\right) \qquad
\Box = 4\,\pd_+\pd_-
\ee
\be
g_{+-} = g_{-+} = \frac{1}{2} \qquad
g^{+-} = g^{-+} = 2 \qquad
\epsilon^{-+} = -\epsilon^{+-} = 2 \qquad
\epsilon_{-+} = -\epsilon_{+-} = -\frac{1}{2}
\ee
\be
d^2p = \frac{1}{2}dp^+dp^- \qquad
\theta(p^0)\,\theta(p^2) = \theta(p^+)\,\theta(p^-)
\ee

Under Lorentz transformation, components of vectors transform as $A^\pm \to e^{\mp\eta}A^\pm$, where $\eta$ is the rapidity, so objects of the form $A^+B^-$, $A^+/B^+$, $A^-/B^-$ are scalars. They can be written in the more conventional Lorentz-invariant form as
\be
A^+B^- = A_\mu B^\mu + \epsilon^{\mu\nu}A_\mu B_\nu \qquad\qquad
\frac{A^\pm}{B^\pm} = \frac{A^2}{A_\mu B^\mu \mp \epsilon^{\mu\nu} A_\mu B_\nu}
\ee

\section{Numerical treatment of IR divergences\label{infrared}}
\setcounter{equation}{0}

In this appendix we consider the simple (unphysical) diagram in figure~\ref{fig-ir} in order to show explicitly how to treat the IR divergences of the dotted line in momentum space.

{\figsize\begin{figure}[htb]
$$\beginpicture
\setcoordinatesystem units <\tdim,\tdim>
\put {{\epsfysize=100\tdim \epsfbox{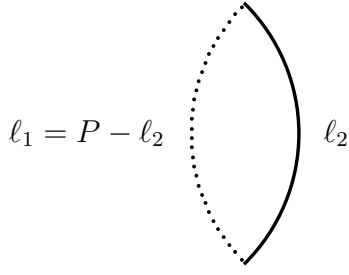}}} at 0 0
\put {$\ell_1 = P - \ell_2$} [rc] at -30 0
\put {$\ell_2$} [lc] at 30 0
\endpicture$$
\caption{\figsize\sf\label{fig-ir}As before, the solid line represents the propagator $i\Delta$ and the dotted line $i\tilde\Delta$. The momentum flowing into the diagram is $P = (M,0)$.}\end{figure}}

In position space, the diagram is simply\footnote{Since this is not a diagram that actually exists in our theory, we do not expect factors of $i$ to work out, so we added a minus sign in order for it to match with what we get using our algorithm.}
\be
i\cM = -\;i\Delta_\cO(x_1-x_2)
\,i\tilde\Delta_\cO(x_1-x_2) = -1
\ee
so in momentum space $\cM = i(2\pi)^2\delta^2(P)$, which gives the ``phase space''
\be
\Phi = 2(2\pi)^2\,\theta(P^0)\,\delta^2(P)
\label{phase-space-2lines}
\ee
On the other hand, if we want to do the calculation in momentum space, as we are suggesting to do for the diagrams of section~\ref{sec-me-general}, we need to compute the integral\footnote{One should further multiply this by $16\sin^2(\pi a)$ in order to obtain the phase space with the correct prefactor.}
\be
I =
\int d^2\ell_1\, d^2\ell_2\,
\,\theta(\ell_1^0)\,\theta(\ell_1^2)F(\ell_1)
\,\theta(\ell_2^0)\,\theta(\ell_2^2)\left(\ell_2^2\right)^a
\delta^2(P-\ell_1-\ell_2)
\label{nintegral}
\ee
According to one method of calculation, we take
\be
F(\ell_1) = f_\lambda(\ell_1^2)
\label{F-first}
\ee
This also needs to differentiated with
\be
\frac{\pd^2}{\pd\ell_1^+\pd\ell_1^-}
\ee
in order to properly account for the dotted line. But using the $\delta$-function to write $\ell_1 = P - \ell_2$ we can just differentiate with
\be
\frac{\pd^2}{\pd P^+\pd P^-} = \frac{\pd}{\pd(P^2)} + P^2\frac{\pd^2}{\pd(P^2)^2}
\ee
after computing the integral.\footnote{We used the regulated $f_\lambda(\ell_1^2)$ in (\ref{F-first}) for numerical convenience.} According to the second method we just take
\be
F(\ell_1) = f_\lambda'(\ell_1^2) + \ell_1^2 f_\lambda''(\ell_1^2)
\label{F-second}
\ee

The resulting $\Phi(P^2)$ should reproduce the $\delta$-function from (\ref{phase-space-2lines}) in the sense that
\be
\int_0^{P_{max}} dP^+ \int_0^{P_{max}} dP^-\;\frac{\Phi(P^2)}{2(2\pi)^2} \to 1
\label{numerical-delta}
\ee
for $P_{max} \gg \lambda$. The integral is finite despite the fact that $\Phi(P^2)$ is a function of $P^2$ alone because
\be
\int_0^{P_{max}^2} dP^2\, \Phi(P^2) \to 0
\label{tot}
\ee

{\figsize
\begin{figure}[htb]
$$\epsfxsize=2.5in \epsfbox[70 300 540 720]{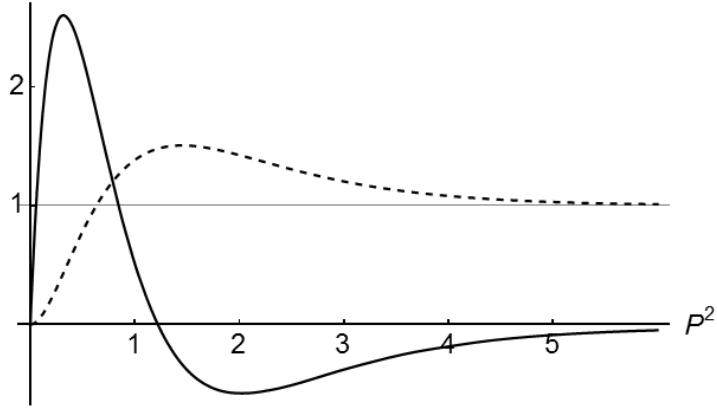}$$
\caption{\figsize\sf\label{fig-2lines-1}The dashed line shows the regulated $\theta$-function $\theta_{\rm reg}(P^2/\lambda^2)$ from (\ref{theta-reg}). The solid line is the numerical result for the integral (\ref{nintegral}) with (\ref{F-second}), that is proportional to $\Phi(P^2)$. (For $P^2$, we use units in which $\lambda=1$.)}
\end{figure}}

{\figsize
\begin{figure}[htb]
$$\epsfxsize=2.5in \epsfbox[70 280 540 720]{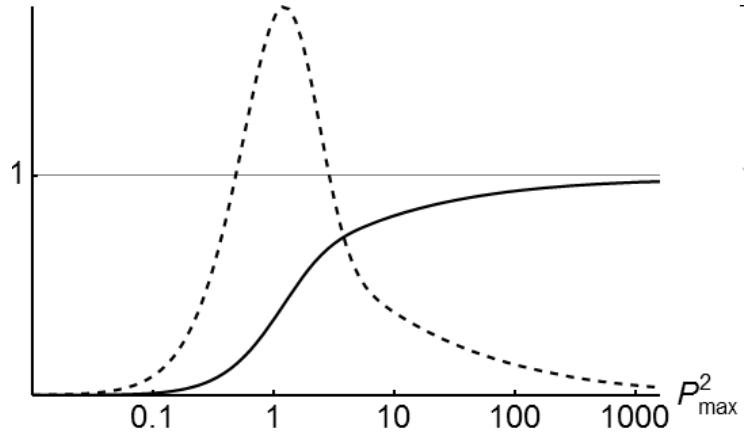}$$
\caption{\figsize\sf\label{fig-2lines-2}The solid line presents the two-dimensional integral over our numerical ``$\delta$-function'', as in (\ref{numerical-delta}). The dashed line (normalized arbitrarily) demonstrates (\ref{tot}). (For $P_{\rm max}^2$, we use units in which $\lambda=1$.)}
\end{figure}}

To confirm the viability of this procedure, we have performed the calculations described here numerically~\cite{Mathematica}. We used
\be
f_\lambda(s) = \frac{1}{s^{1+a}}\,\theta_{\rm reg}\left(\frac{s}{\lambda^2}\right)
\ee
with a regulated $\theta$-function
\be
\theta_{\rm reg}(x) = \sum_{n=0}^3 c_n\,\exp(-n x)
\label{theta-reg}
\ee
with the coefficients $c_n$ chosen such that $\theta_{\rm reg}(0) = \theta_{\rm reg}'(0) = 0$ and $\theta_{\rm reg}(x)\to 1$ for $x\gg1$. The results for $a = -1/2$ are presented in figures~\ref{fig-2lines-1} and~\ref{fig-2lines-2}. The specific data we show were obtained using (\ref{F-second}). A computation using (\ref{F-first}) yielded results that were essentially the same.

\section{An alternative method for amputating the $j^\mu$ leg\label{sec-alt-amp-jm}}
\setcounter{equation}{0}

Based on the observation that for $j^\mu$ we have $\tilde\Delta = \Delta$, we expect (and confirm in the following) that the amputated 3-point function is proportional to the ordinary 3-point function. The latter is given by (\ref{j3ptord}) and can be written as
\be
\vev{j^\mu(x)\cO^\ast(x_1)\cO(x_2)} = \epsilon^{\mu\nu}\pd_\nu f(x|x_1,x_2)
\label{jmu-3pt}
\ee
where the derivative is with respect to $x$ and
\be
f(x|x_1,x_2) = (1+a)\frac{i}{2\pi}\left[\ln(-(x-x_2)^2+i\epsilon) - \ln(-(x-x_1)^2+i\epsilon)\right]i\Delta_\un(x_1-x_2)
\ee
We also write (\ref{j2pt}) as
\be
\vev{j^\mu(x)j^\nu(0)} = \frac{1+a}{(2\pi)^2}\left(\pd^\mu\pd^\nu\ln(-x^2+i\epsilon) - 2\pi i\,g^{\mu\nu}\delta^2(x)\right)
\label{jmjn-2}
\ee
For using (\ref{amputated}), note that
\bea
I^\mu &\equiv& \int d^2x\, \pd^\mu\pd^\nu\ln(-(x'-x)^2+i\epsilon)\, \epsilon_{\nu\rho}\pd^\rho\ln(-(x-x_1)^2+i\epsilon)\nn\\
&=& 4\pi i\epsilon^{\mu\nu}\pd_\nu \ln(-(x'-x_1)^2+i\epsilon)
\eea
Then the effect of the first term of (\ref{jmjn-2}) is to multiply the 3-point function by
\be
(1+a)\frac{i}{\pi}
\ee
At the same time, the second term of (\ref{jmjn-2}) multiplies the 3-point function by
\be
-(1+a)\frac{i}{2\pi}
\ee
Thus the amputated 3-point function that would satisfy
\be
\int d^2x\, \vev{j^\mu(x')j^\nu(x)}\, iQ_\nu(x|x_1,x_2)\nn\\
= \vev{j^\mu(x')\cO^\ast(x_1)\cO(x_2)}
\ee
is
\be
Q^\mu(x|x_1,x_2) = -\frac{2\pi}{1+a}\, \vev{j^\mu(x)\cO^\ast(x_1)\cO(x_2)}
\ee
which agrees with (\ref{Qmu}).

\bibliography{up4}

\end{document}